\documentclass[twocolumn]{aastex63}
\submitjournal{ApJS}

\usepackage{color}
\usepackage{mathrsfs}
\usepackage{graphicx}
\usepackage{cases}
\usepackage{txfonts}
\usepackage{epsfig}
\usepackage{multirow}
\usepackage{bm}
\usepackage{aas_macro}
\def\be{\begin{equation}}
\def\ee{\end{equation}}

\def\bea{\begin{eqarray}}
\def\eea{\end{eqarray}}
\def\bs{\bm}

\begin{document}

\title{Fast scalar quadratic maximum likelihood estimators for the CMB $B$-mode power spectrum}
% \title{Fast and optimal estimators for CMB $B$-mode power spectrum}

\author{Jiming Chen}\email{chenjm94@mail.ustc.edu.cn}
\affiliation{CAS Key Laboratory for Researches in Galaxies and Cosmology, Department of Astronomy, University of Science and Technology of China, Chinese Academy of Sciences, Hefei, Anhui 230026, China}
\affiliation{School of Astronomy and Space Science, University of Science and Technology of China, Hefei, 230026, China}

\author{Shamik Ghosh}\email{shamik@ustc.edu.cn}
\affiliation{CAS Key Laboratory for Researches in Galaxies and Cosmology, Department of Astronomy, University of Science and Technology of China, Chinese Academy of Sciences, Hefei, Anhui 230026, China}
\affiliation{School of Astronomy and Space Science, University of Science and Technology of China, Hefei, 230026, China}

\author{Hao Liu}
\affiliation{School of Physics and Material Science, Anhui University, 111 Jiulong Road, Hefei, Anhui, 230601, China}

\author{Larissa Santos}
\affiliation{Center for Gravitation and Cosmology, College of Physical Science and Technology, Yangzhou University, Yangzhou, 225009, China}
\affiliation{School of Aeronautics and Astronautics, Shanghai Jiao Tong University, Shanghai 200240, China} 

\author{Wenjuan Fang}
\affiliation{CAS Key Laboratory for Researches in Galaxies and Cosmology, Department of Astronomy, University of Science and Technology of China, Chinese Academy of Sciences, Hefei, Anhui 230026, China}
\affiliation{School of Astronomy and Space Science, University of Science and Technology of China, Hefei, 230026, China}

\author{Siyu Li}
\affiliation{Institute of High Energy Physics, Chinese Academy of Sciences, Beijing 100049, China}

\author{Yang Liu}
\affiliation{Institute of High Energy Physics, Chinese Academy of Sciences, Beijing 100049, China}

\author{Hong Li}
\affiliation{Institute of High Energy Physics, Chinese Academy of Sciences, Beijing 100049, China}

\author{Jiaxin Wang}
\affiliation{Department of Astronomy, Shanghai Jiao Tong University, Shanghai, 200240, China}

\author{Le Zhang}
\affiliation{ School of Physics and Astronomy, Sun Yat-Sen University, 2 Daxue Road, Tangjia, Zhuhai, 519082, China}

\author{Bin Hu}
\affiliation{Department of Astronomy, Beijing Normal University, Beijing 100875, China}

\author{Wen Zhao}\email{wzhao7@ustc.edu.cn}
\affiliation{CAS Key Laboratory for Researches in Galaxies and Cosmology, Department of Astronomy, University of Science and Technology of China, Chinese Academy of Sciences, Hefei, Anhui 230026, China}
\affiliation{School of Astronomy and Space Science, University of Science and Technology of China, Hefei, 230026, China}

%\correspondingauthor{...}

\shorttitle{Fast scalar QML estimators for the CMB $B$-mode power spectrum}
\shortauthors{Chen, J., Ghosh, S., Liu, H., Santos, L., et al.,}

\begin{abstract}

Constructing a fast and efficient estimator for the $B$-mode power spectrum of cosmic microwave background (CMB) is of critical importance for CMB science. For a general CMB survey, the Quadratic Maximum Likelihood (QML) estimator for CMB polarization has been proved to be the optimal estimator with minimal uncertainties, but it is computationally very expensive. In this article, we propose two new QML methods for $B$-mode power spectrum estimation. We use the Smith-Zaldarriaga approach to prepare pure-$B$ mode map, and $E$-mode recycling method to obtain a leakage free $B$-mode map. We then use the scalar QML estimator to analyze the scalar pure-$B$ map (QML-SZ) or $B$-mode map (QML-TC). The QML-SZ and QML-TC estimators have similar error bars as the standard QML estimators but their computational cost is nearly one order of magnitude smaller. The basic idea is that one can construct the pure $B$-mode CMB map by using the $E$-$B$ separation method proposed by Smith \& Zaldarriaga (SZ) or the one considering the template cleaning (TC) technique, then apply QML estimator to these scalar fields. By simulating potential observations of space-based and ground-based detectors, we test the reliability of these estimators by comparing them with the corresponding results of the traditional QML estimator and the pure $B$-mode pseudo-$C_{\ell}$ estimator.

%The QML estimator is the minimum variance estimator for CMB power spectra estimation. The computational requirements for the classic QML estimator for polarization, in its minimal form, scales as $\mathcal{O}$($8N^3_{\rm pix, obs}$), where $N_{\rm pix, obs}$ is the number of observed pixels. This makes the QML method for polarization computationally prohibitive beyond the lowest resolution CMB maps, for the largest angular modes. In this work we demonstrate two new QML methods for $B$-mode power spectrum estimation. We use the Smith-Zaldarriaga approach to prepare pure-$B$ mode map, and $E$-mode recycling method to obtain leakage free $B$-mode map. We then use the scalar QML estimator to analyze the scalar pure-$B$ or $B$-mode map. The computation requirement scales as $\mathcal{O}$($N^3_{\rm pix, obs}$). We test our new QML estimators with CMB and noise simulation for satellite-based and ground-based experiment-like cases. We find that our new QML methods give rise to over an order of magnitude reduction in the computation, and a factor $> 7$ reduction in the memory requirements, while giving unbiased $B$-mode power spectrum estimates with near-optimal error bars. Only for the QML estimates with pure-$B$ map for a large sky coverage do we find sub-optimal error bars for the few of the largest scale modes. This work shows that with our new QML estimators we can extend the minimum variance estimator techniques to higher resolution maps making them very important in the search for the primordial $B$-mode signal.
%
\end{abstract}


\section{Introduction}
\label{intro}
The reliable characterization and scientific exploitation of the polarized cosmic microwave background (CMB) signal will provide a wealth of information of the dynamical evolution of the universe \citep{kamionkowski97, zaldarriaga-b-mode, zhao2009a, zhao2009b, zhao2009c,zhaogrishchuk2010} and the nature of both dark matter \citep{Gorski2005, 2001PhRvL..87s1301B} and dark energy \citep{Giovi2003}. Polarized anisotropies of the CMB radiation can be separated into $E$-mode and the $B$-mode components \citep{1997PhRvD..55.1830Z,kamionkowski97,kamionkowski2005,zhao2006,grishchuk2006,weinberg2007}. In the past two decades, many experiments e.g. DASI \citep{DASI}, WMAP \citep{WMAP1, WMAP2, WMAP3}, BOOMERanG \citep{2006ApJ...647..813M}, QUAD \citep{QUAD}, BICEP \citep{BICEP}, QUIET \citep{2012ApJ...760..145Q}, ACT \citep{2014JCAP...10..007N}, Planck \citep{2014A&A...571A..16P}, SPTpol \citep{2018ApJ...852...97H} have already detected the $E$-mode signal with a high confidence level. For instance, the Planck satellite experiment provides precise constraints on the $E$-mode polarization properties in a wide range of angular scales.
%, spanning from the largest scales down to a few arc minutes.

The detection of the large-scale $B$-mode polarization would indicate the presence of a stochastic background of gravitational waves, which may be a leftover from the inflationary epoch in the early universe \citep{grishchuk1974,1997PhRvD..55.3288H, 1999PhRvD..59l3522L, 2000cils.book.....L, kamionkowski97, zaldarriaga-b-mode, mayinzhe}. % \textcolor{red}{The amplitude of the background would allow us to directly deduce the energy scale of inflation, dramatically extending our understanding of the history of the Universe to its very beginnings.} 
%A detection of the $B$ component on large scales would thus indicate the presence of a background of gravity waves, a prediction of inflationary models \citep{kamionkowski97, zaldarriaga-b-mode}. 
%Such a detection would determine the energy scale of inflation and it could provide a stringent test of inflationary models \citep{1998PhRvD..58l3506K}. 
%On small scales, % \textcolor{red}{the B modes will most probably be dominated by secondary contributions produced after last scattering, the leading one being gravitational lensing \citep{zaldarriaga-lensing}. A detection of these contributions could provide information about the distribution of matter all the way up to the last-scattering surface. There are many proposals for how to detect and use this effect \citep{2000PhRvD..62d3517G, 2001PhRvD..63d3501B, 2002ApJ...574..566H}.  In standard models, however, the B component is likely to be quite difficult to detect \citep{2000PhRvD..61h3501J, 2001PhRvD..64f3001T, 2002PhRvD..65d3003B}}
However, the $B$-mode signal is dominated by lensed $B$-modes produced by the weak gravitational lensing effect of the CMB photons during their travel from the last scattering surface to us \citep{zaldarriaga-lensing}, which converts the $E$-modes into $B$-modes. Great effort has been made to observe the CMB $B$-modes with ongoing ground-based experiments, such as BICEP-3 \citep{2014SPIE.9153E..1NA}, AdvACTPol \citep{2016JLTP..184..772H} and SPT-3G \citep{2014SPIE.9153E..1PB}. Other upcoming experiments, like AliCPT-1 \citep{Hong2017, 2021arXiv210109608S}, Simons Observatory \citep{2019JCAP...02..056A}, LSPE \citep{2012arXiv1208.0281T}, QUIJOTE \citep{2014SPIE.9145E..4TP} and CMB-S4 \citep{2020arXiv200812619T}, will join the efforts to look for the primordial CMB $B$-mode from the ground, while the future LiteBIRD \citep{2019JLTP..194..443H} satellite aims to observe the CMB polarized signal from space. 

%It is regrettable that $B$-mode anisotropy, caused by primordial gravitational waves, has not been detected yet.
The CMB $B$-mode signal is much fainter than the $E$-mode. Its detection is then greatly limited by polarized astrophysical foregrounds, primarily dominated by dust and synchrotron emissions. Even if we are able to obtain clean maps by perfectly removing theses foregrounds, we still face the challenge of dealing with the $E$-$B$ leakage, which arises from $E$-$B$ decomposition on an incomplete sky \citep{2001PhRvD..64f3001T}.
 %$B$-mode signal is expected to be much smaller than the $E$ one. Moreover, the $B$-mode measurement is limited by several polarized foregrounds, mainly synchrotron and dust emissions. Even if we perfectly remove the polarized foreground from the observed maps and get cleaned maps, we will still face the challenge of $E$-$B$ leakage, which arises from $E$-$B$ decomposition on an incomplete sky \citep{2001PhRvD..64f3001T}.
In the partial sky case, the polarization field cannot be uniquely decomposed into $E$ and $B$ modes due to the ambiguity in the relationship between the Stokes parameters, and therefore in the $E$ and $B$ modes \citep{ebmixture0, bunn2008b}. In short, besides the ambiguous modes, the complete $B$-mode information consists of both the primordial and the lensed signal. Even though no bias is directly introduced, this leakage will play an important role in isolating the cosmological $B$-mode power spectrum from the input partial sky maps due to the increase of the overall uncertainty of the estimated signal.

%In short, the complete $B$-mode information consists of the primordial, the lensed and ambiguous modes. Even though no bias is directly introduced, this leakage will play an important role when trying to isolate the cosmological $B$-mode power spectrum from the polarized maps in a partial sky, potentially precluding its detection due to the increase of the overall uncertainty of the estimated signal.
 
 %which inflates the overall uncertainty of the estimated $B$-modes signal potentially precluding its detection.

Various methods have been proposed to alleviate the $E$-$B$ mixing problem and restore the cosmological $B$-mode information from an incomplete sky coverage, including several extensions of the standard pseudo-$C_{\ell}$ (PCL) methods \citep{2002ApJ...567....2H, 2003MNRAS.343..559H, ebmixture9, 2007PhRvD..76d3001S, 2010PhRvD..82b3001Z, 2010A&A...519A.104K, 2011A&A...531A..32K, grain2012, 2019PhRvD.100b3538L,Ghosh2020}, which solve for the power spectra by inverting the linear system relating the full sky power to the power from incomplete sky. These methods use fast spherical harmonic transforms with the advantage of speeding up their computation. The method proposed in \citep{ebmixture9, 2007PhRvD..76d3001S} (hereafter the SZ method) was shown to be the PCL estimator \citep{ferte2013} with the smallest errors. In our paper, we then consider the SZ method using an analytically apodized window function.  

Gibbs Sampling technique \citep{Jewell2004, Wandelt2004, Larson2007} may provide an unified way to jointly estimate pure $E$- and $B$-mode spectra and maps from a masked polarization sky, since this technique circumvents the $E$/$B$ decomposition by sampling full-sky realizations of the polarization in terms of their likelihood. However, the Gibbs sampling suffers a convergence issue at low signal to noise regime and would take a long convergence time for $B$-mode spectra to be estimated correctly. Generalized Wiener filtering methods can also been used for $E$/$B$ decomposition \citep{Bunn2017, KodiRamanah2018, KodiRamanah2019}.  

The Quadratic Maximum Likelihood (QML) method \citep{2001PhRvD..64f3001T}, which is a pixel-based estimator, provides another way to solve the $E$-$B$ mixing problem. It has the advantage of minimizing spectra uncertainties, but, at the same time, it involves matrix inversions and multiplications which significantly increases the calculation time and the demand for computational memory.

%The Quadratic Maximum Likelihood (QML) method \citep{2001PhRvD..64f3001T}, which is a pixel-based estimator, provides another way to solve the $E$-$B$ mixing problem with the advantage of minimizing spectra uncertainties, ie. optimal error.It involves matrix inversions and multiplications which scale as $\mathcal{O}(N_d^{3})$ \citep{2006MNRAS.370..343E}. So, the calculation time and the demand for computational memory dramatically increases with increase in the number of pixels. 
%Moreover, thanks to a series of experiments to measure the $E$- mode, as mentioned above, we have obtained high-precision $E$-mode information. 

Recently, another method was proposed to partly solve the $E$-$B$ mixture problem \citep{2019PhRvD.100b3538L}. Based on the available high-precision $E$-mode datasets, the template cleaning (TC) method can be used to estimate the $E$-$B$ leakage that will be later removed from the observed CMB maps. Finally, we can use an appropriate method to reconstruct the $B$-mode power spectrum from `real' $B$ maps.

%Based on the available high-precision $E$-mode datasets, we can use the template cleaning (TC) method to estimate $E$-$B$ leakage to be removed from the observed CMB maps, then we can use an appropriate method to reconstruct the $B$-mode power spectrum from `real' $B$ maps.

Combining the advantages of the above three methods, we propose two new methods to reconstruct the large-scale $B$-mode power spectrum: the QML-SZ method and the QML-TC method. The QML-SZ method uses the SZ-method to derive the pure $B$-mode map $\mathcal{B}(\hat{n})$ from Stokes $Q$ and $U$ maps, which can be ultimately treated as scalar fields, enabling us to use the QML method developed for CMB temperature maps. Similarly, in QML-TC method, we can first use the template cleaning method to get the leakage free $B$-mode map from $Q$ and $U$ maps, then use the scalar-mode QML method to estimate the $B$-mode power spectrum. Since we adopt the SZ and/or template cleaning methods to transform the Stokes $Q$ and $U$ maps into scalar $B$-mode maps, the number of pixels drops to $1/2$ of the standard QML method. This means that the computational running time will be 8 times shorter than standard QML estimator. We are able to precisely reconstruct an unbiased power spectrum with reasonably small errors, using the methodology described above, with the advantage of drastically reducing our computational requirements. 

%Since we adopt the SZ and/or template cleaning methods to transform the Stokes $Q$ and $U$ maps into scalar maps with the $B$-mode polarization information, the number of pixels drops to $1/2$ of the traditional QML method.

%then the pseudo $B$ maps as scalar maps can be treated as ‘temperature’ maps,  and we can use QML method for temperature to calculate it. Similarly, we can use template cleaning method to get `real' $B$ maps form $Q$, $U$ maps first and then use scalar-mode QML method estimate the $B$-mode signal. Because we adopt SZ-method and template cleaning method to transform polarized CMB $Q$, $U$ maps to scale maps, so the number of pixels became the $1/2$ of traditional QML method, which means the running time  just $1/8$ time of traditional QML estimator in theory. At the same time, the results of the three methods are all kept high precision, all get unbiased power spectrum and have the almost same variance.

This paper organized as follows. In Section~\ref{sec:power_spectrum_estimators}, we review the traditional QML method first. Next, we introduce the SZ method and combine it with scalar QML method to construct the QML-SZ estimator. After that, we briefly introduce the template cleaning method and adopt the scalar QML method to construct the QML-TC estimator. In Section~\ref{sec:tests_and_applications}, we  test the effects of various factors that may affect the uncertainties of these two estimators. In Section~\ref{sec:realistic_examples}, we apply the these methods to more realistic situations and make a comprehensive comparison of their performance. Conclusions and discussions are given in Section~\ref{sec:conclusion}.

%*****************************************************************%

% %*********************************************************************%
\section{POWER SPECTRUM ESTIMATORS} % (fold)
\label{sec:power_spectrum_estimators}

\subsection{$E$-$B$ mixture in partial-sky surveys}
\label{CMB $B$-mode decomposition}
The linear polarization of the CMB field can be completely described by the Stokes parameters $Q$ and $U$. Along the line-of-sight, $\hat n$, the polarization field can be written as:
\begin{equation}
\label{PQU}
	P_{\pm}(\hat{n})=Q(\hat{n})\pm iU(\hat{n}),
\end{equation}
which behaves as a spin-(2) and a spin-(-2) field. In the simplest full sky case, these fields are expanded over spin-weighted spherical harmonics, ${}_{\pm s}Y_{\ell m}$, as \citep{seljak1996}:
\begin{equation}
P_{\pm}(\hat{n})=\sum_{\ell m} a_{\pm2,\ell m}~{}_{\pm 2}Y_{\ell m}(\hat{n}).
\end{equation}
Detailed expressions are given in appendix \ref{apdx:imp_relations}.

%where ${}_{\pm s}Y_{\ell m}$ stands for the spin-weighted spherical harmonics, and the detailed expressions are given in appendix \ref{apdx:imp_relations}. 

However, describing the polarization field by means of the the Stokes parameters is frame dependent. Therefore, for convenience, we write the polarization field in terms of the rotationally invariant $E$ and $B$ components. These components are defined in the harmonic space in terms of spin-harmonic coefficients $a_{\pm 2,\ell m}$. 
\begin{eqnarray}
a_{E,\ell m}  &\equiv& -\frac{1}{2}[a_{2,\ell m}+a_{-2,\ell m}] \nonumber \\
& =& -\frac{1}{2}\left[\int P_+(\hat n){}_2Y^*_{\ell m}(\hat n)d\hat{n} +\int P_-(\hat n){}_{-2}Y^*_{\ell m}(\hat n)d\hat{n}\right], \label{eq_elm}\\%\right. \nonumber\\ \left. 
a_{B,\ell m} & \equiv& -\frac{1}{2i}[a_{2,\ell m}-a_{-2,\ell m}] \nonumber \\ 
&=& \frac{i}{2}\left[\int P_+(\hat n){}_2Y^*_{\ell m}(\hat n)d\hat{n} - \int P_-(\hat n){}_{-2}Y^*_{\ell m}(\hat n)d\hat{n}\right]. \label{eq_blm} %\right. \nonumber \left.
\end{eqnarray}
We can then define the $E(\hat{n})$ and  $B(\hat{n})$ sky maps as:
%One can now define the electric-type polarization (scalar) sky map, $E(\hat{n})$, and the magnetic-type polarization (pseudo-scalar) sky map, $B(\hat{n})$ as:
 \begin{eqnarray}
 E(\hat{n}) \equiv \sum_{\ell m}a_{E,\ell m} Y_{\ell m}(\hat{n}), \qquad
 B(\hat{n}) \equiv \sum_{\ell m}a_{B,\ell m} Y_{\ell m}(\hat{n}).
 \end{eqnarray}
Finally, the power spectra can be obtained as follows
%Finally, the power spectra are calculated as:
 \begin{eqnarray}
 C_{\ell}^{EE} \equiv \langle E_{\ell m} E_{\ell m}^* \rangle, \qquad %\frac{1}{2\ell+1}\sum_m
 C_{\ell}^{BB} \equiv \langle B_{\ell m} B_{\ell m}^* \rangle, %\frac{1}{2\ell+1}\sum_m
 \end{eqnarray}
where the angular brackets denote the average over realizations. 

For an incomplete sky observation, defined by the window function $W(\hat n)$, in principle, we can also define the partial sky $E$- and $B$-mode spherical harmonic coefficients (indicated by overhead tilde) as:
%$\tilde{}$
\begin{eqnarray}
    \tilde a_{E, \ell m} &=& -\frac{1}{2}\left[\int P_+\, W\, {}_2Y^*_{\ell m}d\hat{n} +\int P_-\, W\, {}_{-2}Y^*_{\ell m}d\hat{n}\right], \\
    \tilde a_{B, \ell m} &=& \frac{i}{2}\left[\int P_+\, W\, {}_2Y^*_{\ell m}d\hat{n} - \int P_-\, W\, {}_{-2}Y^*_{\ell m}d\hat{n}\right].
\end{eqnarray}
These coefficients $\tilde a_{E, \ell m}$ and $\tilde a_{B, \ell m}$ relate to the pure $E$ and $B$ coefficients $a_{E, \ell m}$ and $a_{B, \ell m}$  as follows:
\begin{eqnarray}
    \tilde a_{E, \ell m} &=& \sum_{\ell' m'}\left[ K_{\ell m \ell' m'}^{EE}a_{E, \ell' m'} + iK_{\ell m \ell' m'}^{EB}a_{B, \ell' m'}\right], \nonumber\\
    \tilde a_{B, \ell m} &=& \sum_{\ell' m'}\left[ -iK_{\ell m \ell' m'}^{BE}a_{E, \ell' m'} + K_{\ell m \ell' m'}^{BB}a_{B, \ell' m'}\right].
    \label{eq:EB_partsky}
\end{eqnarray}
The coupling matrices $K_{\ell m \ell' m'}^{XY}$ are the mixing kernels for the partial sky observation, and the full form of these matrices can be found in \citet{2009PhRvD..79l3515G}.Various methods have already been proposed to avoid the mixing by constructing the pure $E$-type and pure $B$-type fields, such as \cite{ebmixture0, 2011PhRvD..83h3003B, 2003PhRvD..68h3509L, 2009ApJ...706.1545C, 2013MNRAS.435.2040L, 2009PhRvD..79l3515G,ebmixture9, 2007PhRvD..76d3001S, 2010PhRvD..82b3001Z, 2010A&A...519A.104K,larissa2016,larissa2017}. 

%We can see that $K_{\ell m \ell' m'}^{EB}$ mixes the full sky $B$-mode signal to the partial sky $E$-mode signal and $K_{\ell m \ell' m'}^{BE}$ mixes the full sky $E$ modes with the partial sky $B$ modes. Various methods have already been proposed to avoid the mixing by constructing the pure $E$-type and $B$-type polarizations, such as \cite{ebmixture0, 2011PhRvD..83h3003B, 2003PhRvD..68h3509L, 2009ApJ...706.1545C, 2013MNRAS.435.2040L, 2009PhRvD..79l3515G,ebmixture9, 2007PhRvD..76d3001S, 2010PhRvD..82b3001Z, 2010A&A...519A.104K,larissa2017}. 

%Correcting for the $E$-$B$ leakage is critical in detection of primordial $B$ modes as the leakage of $E$-modes to $B$-modes can dominate the faint primordial $B$-mode signal. Even though various methods for eliminating the mixing have been proposed, residuals are still present and should be carefully taken into account (see for instance \cite{larissa2017}). 

% subsection convention_and_notation (end)

% %*********************************************************************%
% %*********************************************************************%

\subsection{Standard QML estimator} % (fold)
\label{sub:qml_estimator}
In this article, we mainly focus on how to construct the fast estimators of the CMB $B$-mode power spectrumwith small errors. For CMB polarization maps with any sky coverage, \citet{1997PhRvD..55.5895T} defines a QML estimator for the CMB temperature power spectrum, which is generalized for CMB polarization power spectra by \cite{2001PhRvD..64f3001T}. In this section, we briefly review the QML estimator for polarization. In pixel domain, we can define an input data vector, $\bm{x}$, which consists of the temperature fluctuation field and the Stokes $Q$ and $U$ (with respect to a fixed coordinate system) specified at the $i^{\rm th}$ pixel as
 \begin{equation}       %开始数学环境
\label{define_x}
\bm{x}_i=              %左括号
  \left(
\begin{array}{c}   %该矩阵一共3列，每一列都居中放置
    \Delta T_i  \\  %第一行元素
    Q_i  \\  %第二行元素
    U_i  %第二行元素
  \end{array}     \right)             %右括号
  +
  \left(
\begin{array}{c}   %该矩阵一共3列，每一列都居中放置
    n^T_i  \\  %第一行元素
    n^Q_i  \\  %第二行元素
    n^U_i  %第二行元素
  \end{array}     \right)  .
\end{equation}
Following \citet{2001PhRvD..64f3001T}, the optimal quadratic estimate of the power spectrum, $y^r_\ell$, is defined as :
\begin{equation} 	
 \label{QML_yrl}
	y^r_\ell = \bm{x}^t_i \bm{E}_{ij}^{r\ell} \bm{x}_j - b^r_\ell, \qquad r\in \left[T,TE,E,B\right],
\end{equation}
where $t$ indicates matrix transpose operation. Here, $r$ is the index denoting the power spectrum and $i$, $j$ are indices over pixels. The data $\bm{x}_i$ at a particular pixel is a $TQU$ component vector and $\bm{E}_{ij}^{r\ell}$ is $3 \times 3 $ matrix. The bias vector, $b^r_\ell$, corrects for the noise bias and is given by ${\rm Tr}[\bs E^r_\ell \bs N]$, when the noise, $n$, is uncorrelated between pixels. Note that we have assumed the summation convention. The $\bm{E}^{r\ell}$ matrices are computed as:
\begin{eqnarray}	
	\bm{E}^{r \ell}=\frac{1}{2} \bm{C}^{-1} \frac{\partial \bm{C}}{\partial \bm{C}^r_\ell} \bm{C}^{-1}
	\label{QML_Erl}
\end{eqnarray}
and the covariance matrix of the data $\bm{x}$, denoted by $\bm{C}$, is:
\begin{eqnarray}       %开始数学环境
 \label{QML_matrix_C}
\bm{C}_{ij}=\langle \bm{x}_i \bm{x}^t_j\rangle &= \bm{R}(\alpha_{ij})\bm{M}(\hat r_i \cdot \hat r_j ) \bm{R}(\alpha_{ji})^t + \bs N_{ij} \nonumber\\
&=
%   \left( \right. \matrix{
    \left(
\begin{array}{ccc}
    C^{TT}_{ij}  & \quad C^{TQ}_{ij} &\quad  C^{TU}_{ij}\\  %第一行元素
    \\
    C^{QT}_{ij}  & \quad C^{QQ}_{ij} &\quad  C^{QU}_{ij}\\  %第二行元素
    \\
    C^{UT}_{ij}  & \quad C^{UQ}_{ij} &\quad  C^{UU}_{ij}%}   \left. \right),
    \end{array}     \right),
\end{eqnarray}
where $\bs N$ is the noise covariance matrix and the rotation matrix $\bs R$ is given by:
\begin{equation}       %开始数学环境
 \label{QML_matrix_R}
\bm{R}(\alpha_{ij})=
  \left(
\begin{array}{ccc}  %该矩阵一共3列，每一列都居中放置
    1  & \quad 0                  &\quad  0\\ %第一行元素
    \\
    0  & \quad \cos(2\alpha_{ij})  &\quad  \sin(2\alpha_{ij})\\  %第二行元素
    \\
    0  & \quad -\sin(2\alpha_{ij}) &\quad  \cos(2\alpha_{ij})   %第二行元素
\end{array}     \right),
\end{equation}
which performs a rotation to a global frame where the reference directions are given by the meridians. $\bm{M}$ is the covariance matrix when the $Q$ and $U$ are defined with the reference direction being the great circle connecting the two points. So, it depends only on the angular separation between the two pixels. The explicit expression for this matrix is given by
\begin{eqnarray}       %开始数学环境
 \label{QML_matrix_M}
\bm{M}(\hat r_i \cdot \hat r_j )=
  \left(
\begin{array}{ccc}
    \langle T_i T_j \rangle  & \quad \langle T_i Q_j \rangle \  &\quad  \langle T_i U_j \rangle \\  %第一行元素
    \\
    \langle Q_i T_j\rangle  & \quad \langle Q_i Q_j \rangle  &\quad  \langle Q_i U_j \rangle \\  %第二行元素
    \\
    \langle U_i T_j \rangle  & \quad \langle U_i Q_j \rangle  &\quad  \langle U_i U_j \rangle  %第二行元素
  \end{array}     \right),
\end{eqnarray}
where
\begin{eqnarray}
 \label{QML_component_matrix_M}
\langle T_i T_j \rangle &=&  \sum_{\ell}\frac{2\ell+1}{4\pi}C^{TT}_{\ell}P_{\ell}(z),\nonumber\\[0.1cm]
\langle T_i Q_j \rangle &=& -\sum_{\ell} \frac{2\ell+1}{4\pi}C^{TE}_{\ell}F^{10}_\ell(z),\nonumber\\[0.1cm]
\langle T_i U_j \rangle &=& -\sum_{\ell} \frac{2\ell+1}{4\pi}C^{TB}_{\ell}F^{10}_\ell(z),\nonumber\\[0.1cm]
\langle Q_i Q_j \rangle &=&  \sum_{\ell}\frac{2\ell+1}{4\pi}[C^{EE}_{\ell}F^{12}_{\ell}(z)-C^{BB}_{\ell}F^{22}_{\ell}(z)],\nonumber\\[0.1cm]
\langle U_i U_j \rangle &=&  \sum_{\ell}\frac{2\ell+1}{4\pi}[C^{BB}_{\ell}F^{12}_{\ell}(z)-C^{EE}_{\ell}F^{22}_{\ell}(z)],\nonumber\\[0.1cm]
\langle Q_i U_j \rangle &=&  \sum_{\ell}\frac{2\ell+1}{4\pi}[F^{12}_{\ell}(z)+F^{22}_{\ell}(z)]C^{EB}_{\ell}.\nonumber
\end{eqnarray}
We have used $z=\hat r_i\cdot \hat r_j$ as the cosine of the angle between the $i^{\rm th}$ and $j^{\rm th}$ pixels. Note that $P_\ell$ denotes a Legendre polynomial, and the form of functions $F^{10}_\ell$, $F^{12}_\ell$ and $F^{22}_\ell$ are given in appendix \ref{apdx:imp_relations}.

Considering the matrix definitions given above, one can get minimum variance estimates of the CMB power spectra using equation \ref{QML_yrl}. From equation \ref{QML_matrix_C}, we find that the temperature fluctuation fields are coupled with the polarization fields. It has been pointed out by \citet{2001PhRvD..64f3001T} that this may be problematic for realistic noisy data, where systematic errors in the $\Delta{T}$ measurements could contaminate or bias the estimates of $E$- and $B$-mode power spectra, which have much lower magnitudes. To avoid this potential issue, we rewrite the covariance matrix in equation \ref{QML_matrix_C} as:
\begin{equation}
    \check{C}_{ij}=\langle \bm x_i \bm x^{t}_j \rangle=
   \left(
\begin{array}{ccc}
        C^{TT}  & 0            &\quad  0     \\   %第一行元素
    \\
    0       & \quad C^{QQ} &\quad  C^{QU}\\   %第二行元素
    \\
    0       & \quad C^{UQ} &\quad  C^{UU}  %第二行元素
    \end{array}     \right).
    \label{QML_new_matrix_C}
\end{equation}      
Note that this simplification is useful only when we are interested in the polarization auto-spectra and do not wish to compute the polarization temperature cross-spectra. This is done by simply dropping the temperature and polarization cross covariance terms in the full covariance matrix. Therefore, we can rewrite matrices $\bm E^r_l$   as:
\begin{equation}
   \check{\bm E}^{r\ell}=\frac{1}{2} \check{\bs C}^{-1} \frac{\partial \bs C}{\partial C^r_\ell} \check{C}^{-1}
   \label{QML_new_Erl},
\end{equation}	
following the definition in Eq. \ref{QML_Erl}. The expectation values of $y^r_l$ in equation \ref{QML_yrl} are
\begin{equation}	
	\langle y^r_\ell \rangle = \check{F}^{ur}_{\ell \ell'}C^{u}_{\ell'},
	\label{QML_brave_yrl}
\end{equation}
where we have used the Fisher matrix defined as:
\begin{equation}
    \check{F}^{ur}_{\ell \ell'}=\frac{1}{2}{\rm Tr}\left[\frac{\partial \bs C} {\partial C^u_{\ell'}} \check{\bs C}^{-1} \frac{\partial \bs C}{\partial C^{r}_{\ell}} \check{\bs C}^{-1}\right]
     \label{QML_Fisher_Matrix}.
\end{equation}
Therefore, $y^r_{\ell}$ can give the unbiased estimates of the actual power spectra $C^u_l$ with $u \in [T,E,B]$.

When we use the redefined covariance matrix of equation \ref{QML_new_matrix_C}, the power spectra estimates from relation \ref{QML_yrl} become sub-optimal. However, the penalty we pay is an increase in the error bars of the power spectra estimates. It can be seen that, by redefining the covariance matrix as in \ref{QML_new_matrix_C}, we can separate the scalar $T$ field from the polarization field. This allows us to only work on the $QU$ map for the $E$- and $B$-mode power spectra. Throughout this work, we will only work with the polarization part of the QML method.
If the matrix, $\bs F$, is invertible, one can define unbiased estimates of the true power spectra via
\begin{equation}
    \hat{C}^r_\ell=\check{\bs F}^{-1}\bs y^r
    \label{QML_true_PS}.
\end{equation}	
The covariance matrix of the estimates $y^r_\ell$ is then given by:
\begin{eqnarray}	
	\langle y^r_\ell y^u_{\ell'}\rangle - \langle y^r_\ell\rangle \langle y^u_{\ell'} \rangle \equiv F^{ru}_{\ell \ell'} = 2{\rm Tr}[\bs C \check{\bs E}^{r\ell} \bs C \check{\bs E}^{u\ell'}],
	\label{QML_covariance_matrix1}
\end{eqnarray}
being $\bs F^{ru}$ the Fisher matrix. Thus, the covariance matrix of the true power spectra estimates of equation  \ref{QML_true_PS} is given by
\begin{equation}	
	\langle \Delta \hat{C}_\ell \Delta \hat{C}_{\ell'} \rangle = \check{\bs F}^{-1} \bs F \check{\bs F}^{-1}.
	\label{QML_covariance_matrix2}
\end{equation}
In this work, the QML estimators have been implemented with the xQML\footnote{https://gitlab.in2p3.fr/xQML/xQML} python package \citep{Vanneste2018}. 

From these equations, we observe that the computational requirements for this standard QML estimator for polarization, in its minimal form, scales as $\mathcal{O}$($8N^3_{\rm pix, obs}$), where $N_{\rm pix, obs}$ is the number of observed pixels. This makes the QML method for polarization computationally prohibitive beyond the lowest resolution CMB maps.

% subsection qml_estimator (end)
% %*********************************************************************%

% %*********************************************************************%
% %*********************************************************************%

\subsection{QML-SZ estimator}
\label{sub:qml_sz_estimator}
In order to accelerate the computation speed of the QML estimators for CMB polarization power spectra, we introduce two new estimators. 
In these approaches, by applying the $E$-$B$ separation methods proposed in the literature, we first construct the pure $E$-type or $B$-type polarization maps from the observed $Q$ and $U$ maps, which are scalar (or pseudo-scalar) fields in two-dimensional sphere. Then, we can apply the QML estimator for scalar fields \citep{1997PhRvD..55.5895T}. In this article, we consider two different methods for $E$-$B$ separation, which are sufficiently fast, and have little information loss. 

In the first approach, we consider the $E$-$B$ separation method proposed in \citet{ebmixture9} and \citet{2007PhRvD..76d3001S}. To deal with the mixture in a partial sky polarization analysis, one can construct two scalar (pseudo-scalar) field quantities by using the spin-raising and spin-lowering operators \citep{Newman1966}, $\eth$ and $\bar \eth$, on $P_\pm$. Specifically, we define a new set of fields $\mathcal{E}$ and $\mathcal{B}$ as in \citep{2007PhRvD..76d3001S, 2010PhRvD..82b3001Z}: 
\begin{eqnarray}
\label{pseudo_E}
\mathcal{E}(\hat{n}) &=& -\frac{1}{2}[\bar{\eth}\bar{\eth}P_{+}(\hat{n}) + \eth \eth P_{-}(\hat{n})], \\
\label{pseudo_B}
\mathcal{B}(\hat{n}) &=& -\frac{1}{2i}[\bar{\eth}\bar{\eth}P_{+}(\hat{n}) - \eth \eth P_{-}(\hat{n})].
\end{eqnarray}
These define two mutually orthogonal scalar and pseudo-scalar fields, which are usually called the pure $E$- and pure $B$- fields in the literature. In this paper, we focus only on the $B$-modes since the $E$-mode power spectrum can be recovered satisfactorily with existing estimators. Expanding the $\mathcal{B}(\hat{n})$ component in spherical harmonics, we obtain:
\begin{equation}
\label{pseudo_B_harm}
\mathcal{B}(\hat{n}) \equiv \sum_{\ell m}\mathcal{B}_{\ell m}Y_{\ell m}(\hat{n}).
\end{equation}
The $\mathcal{B}_{\ell m}$ coefficient can then be computed from the pure $B$ field as:
\begin{equation}
 \label{Elm_2}
\mathcal{B}_{\ell m} = \int  \mathcal{B}(\hat{n})Y_{\ell m}^*(\hat{n})d\hat{n}.
\end{equation}

This new pure $B$-mode spherical harmonic coefficient is related to the coefficient $B_{\ell m}$ as \citep{zaldarriaga-b-mode}:
\begin{equation}
 \label{Elm_pseudo}
\mathcal{B}_{\ell m}=N_{\ell,2}B_{\ell m},
\end{equation}
and the power spectrum becomes
\begin{equation}
\label{pseudo_cl}
C_{\ell}^{\mathcal{BB}}\equiv\langle \mathcal{B}_{\ell m} \mathcal{B}_{\ell m}^* \rangle=N_{\ell ,2}^2 C_{\ell}^{BB}. %\frac{1}{2\ell+1}\sum_m
\end{equation}
with $N_{\ell ,s} = \sqrt{(\ell + s)! / (\ell - s)!}$.
 
For an incomplete sky observation defined by the window function $W(\hat{n})$, the partial-sky harmonic coefficients of the pure $E$- and $B$-fields are defined as \citep{efstathiou2004},
\begin{eqnarray}
\label{pseudo_EBlm}
\tilde{\mathcal E}_{\ell m} &=& \int  W(\hat{n}) \mathcal{E}(\hat{n})Y_{\ell m}^*(\hat{n})d\hat{n},\\
\label{pseudo_EBlm2}
\tilde{\mathcal B}_{\ell m} &=& \int  W(\hat{n}) \mathcal{B}(\hat{n})Y_{\ell m}^*(\hat{n})d\hat{n}.
\end{eqnarray}
Following the SZ method detailed in \citet{ebmixture9}, we write the partial-sky pure-field harmonic coefficients as:
%From hereafter we will refer to this approach as the SZ method
\begin{eqnarray}
\tilde{\mathcal E}_{\ell m} &=& -\frac{1}{2}\int d\hat{n} \bigg\{P_{+}(\hat n)\left[\bar\eth \bar\eth\left(W(\hat{n})Y_{\ell
m}(\hat{n})\right)\right]^\ast   \nonumber \\
& & +P_-(\hat n)\left[\eth\eth\left(W(\hat{n})Y_{\ell m}(\hat{n})\right)\right]^\ast \bigg\}, \label{pureee}\\
\tilde{\mathcal B}_{\ell m} &=& -\frac{1}{2i}\int d\hat{n}\bigg\{P_+(\hat n)\left[\bar\eth\bar\eth\left(W(\hat{n})Y_{\ell
m}(\hat{n})\right)\right]^\ast  \nonumber  \\
& & -P_-(\hat n)\left[\eth\eth\left(W(\hat{n})Y_{\ell m}(\hat{n})\right)\right]^\ast \bigg\} \label{purebb}.
\end{eqnarray}
These expressions can be expanded and simplified further for implementation. Full expressions can be found in appendix \ref{apdx:imp_relations}. Once the coefficients $\tilde{\mathcal E}_{\ell m} $ and $\tilde{\mathcal B}_{\ell m}$ are derived, the scalar fields in our observation window $W(\hat{n})\mathcal{E}(\hat{n})$ and $W(\hat{n})\mathcal{B}(\hat{n})$ can be directly obtained by inverting the relations in (\ref{pseudo_EBlm}) and (\ref{pseudo_EBlm2}).

Similar to equation \ref{eq:EB_partsky}, the partial-sky pure harmonic coefficients $\tilde{\mathcal E}_{\ell m}$ and $\tilde{\mathcal B}_{\ell m}$ are related to the full-sky harmonic coefficients $E_{\ell m}$ and $B_{\ell m}$ as follows,
\begin{eqnarray}
\tilde{\mathcal E}_{\ell m} &=&\sum_{\ell'm'}[\mathcal{K}^{EE}_{\ell m,\ell'm'}a^E_{\ell'm'}+i\mathcal{K}^{EB}_{\ell m,\ell'm'}a^B_{\ell'm'}], \label{Elm}\\
% \label{Blm}
\tilde{\mathcal B}_{\ell m} &=&\sum_{\ell'm'}[-i\mathcal{K}^{BE}_{\ell m,\ell'm'}a^E_{l'm'}+\mathcal{K}^{BB}_{\ell m,\ell'm'}a^B_{\ell 'm'}], \label{Blm}
\end{eqnarray}
where $\mathcal{K}^{ru}_{\ell' m' \ell m}$ are the pure field mixing kernels, which in general can be all different, non-vanishing, and non-diagonal in both $\ell$ and $m$. Just as in the case of equation \ref{eq:EB_partsky}, the $\mathcal{K}^{EB}_{\ell' m' \ell m}$ and $\mathcal{K}^{BE}_{\ell' m' \ell m}$ are the mixing terms between the two polarization modes. However, it can be shown that for pure-$E$ and pure-$B$ construction, the mixing terms are orders of magnitude smaller than the standard case of equation \ref{eq:EB_partsky}. This indicates that the pure fields are nearly orthogonal with very small mixing between the two polarization modes.

%This indicates that the pure fields are mutually orthogonal with much reduced mixing between them.

In the previous subsection, the redefined covariance matrix in equation \ref{QML_new_matrix_C} indicates that QML estimator for the scalar temperature field can be separated from the polarization part. By using the pure-$E$/$B$ fields, we can construct the two scalar (pseudo-scalar) polarization modes. Thus, it is conceivable to use the scalar QML method to estimate the power spectrum of the decoupled scalar (pseudo-scalar) polarization fields. Let us briefly outline the scalar QML implementation, which we use to estimate the power spectrum of the pure-$B$ field.

Let $x^\mathcal{S}_i$ denote the $i^{th}$ pixel value in the scalar map $x^\mathcal{S}_i = \mathcal{S}_i + n^\mathcal{S}_i$, where $n^\mathcal{S}_i$ is the noise in the individual pixel. The covariance matrix $\bs C_\mathcal{S}$ of input data $x^\mathcal{S}_i$ can be written as:
\begin{equation}       %开始数学环境
\label{QML-SZ_matrix_C}
C_{s,ij}=\langle x^\mathcal{S}_i (x^\mathcal{S}_j)^t \rangle =  \sum_{\ell}\frac{2 \ell + 1}{4 \pi} C_{\ell}^{\mathcal{SS}} P_{\ell}(z)+N_{\mathcal{S},ij} ,
\end{equation}
where $C_{\ell}^{\mathcal{SS}}$ means the power spectrum corresponding to the signal of scalar field map $\mathcal{S}(\hat{n})$ and $\bs N_\mathcal{S}$ is the noise variance matrix. 

According to the QML approach, we can construct the optimal estimator by making use of $C_{S,ij}$ and $x^\mathcal{S}_i$, 
\begin{equation} 	
 \label{QML_SZ_yl}
	y^\mathcal{S}_{\ell} =x^\mathcal{S}_i x^\mathcal{S}_j E^{\ell}_{S,ij} - b^\mathcal{S}_\ell.
\end{equation}
The matrices $\bs E^{\ell}_\mathcal{S}$ have a similar form to equation \ref{QML_Erl}
\begin{equation}	
 \label{QML_SZ_Erl}
	\bs E^{\ell}_\mathcal{S} = \frac{1}{2} \bs C_\mathcal{S}^{-1} \frac{\partial{\bs C_\mathcal{S}}}{\partial{C_{\mathcal{S},\ell}}} \bs C_\mathcal{S}^{-1}.
\end{equation}
Similarly, the Fisher matrix $F_{S,\ell \ell'}$ expression becomes:
\begin{equation} 	
\label{QML_SZ_Fllp}
	F_{S,\ell \ell'} = \frac{1}{2}Tr \left[\frac{\partial \bs C_\mathcal{S}}{\partial C_{\mathcal{S},\ell}}  \bs C^{-1}_\mathcal{S}\frac{\partial \bs C_\mathcal{S}}{\partial C_{\mathcal{S},\ell'}}  \bs C_\mathcal{S}^{-1} \right].
\end{equation}

The QML estimator for scalar power spectrum $\hat C_{\ell}^{\mathcal{SS}}$ is given by
\begin{equation} 	
 \label{QML_SZ_true_PS}
	\hat{C}_{\ell}^{\mathcal{SS}}=\left(\tilde{\bs F}_\mathcal{S}\right)^{-1}\bs y^\mathcal{S}.
\end{equation}
We apply this scalar QML method to estimate the power spectrum of the decoupled pure-$B$ polarization field, which is denoted as QML-SZ estimator in this work.

In the SZ approach to separate the $E$ and $B$ modes, we must use a proper sky apodization, instead of the binary window function, to avoid numerical divergences in the calculation of the window function derivatives. A Gaussian smoothing kernel has been shown to induce very small leakage in the final $B$-map \citep{Wang, 2011A&A...531A..32K}. So, this is our apodization choice for obtaining the pure-$B$ map for QML-SZ method. For the $i^{\rm th}$ pixel in the region allowed by the binary mask, the apodized window is defined as:
% \begin{equation}

\begin{numcases} {W_i=} 
         \frac{1}{2} +\frac{1}{2}{\rm erf}\left(\frac{\delta_i-\frac{\delta_c}{2}}{\sqrt{2}\sigma}\right), & $\delta_i < \delta_c$  \nonumber \\
         1, & $\delta_i > \delta_c$ \label{QML_SZ_Gaussian_window_function}
\end{numcases}
% \end{equation}
where $\delta_i$ is the shortest distance between the $i^{\rm th}$ observed pixel from the boundary of the allowed region,  $\sigma={\rm FWHM}/\sqrt{8\ln 2}$ with FWHM denoting the full width at half maximum of the Gaussian kernel, and $\delta_c$ is the apodization length which acts as an additional adjustable parameter.

Now, we summarize the construction of the QML-SZ estimator as follows. For the given observed $Q$ and $U$ polarization maps, we construct a partial-sky pure $B$-type map $\mathcal{B}(\hat{n})$, following the SZ method with Gaussian apodization of the binary window function. Then, we use $\mathcal{B}(\hat{n})$ map and $C_{\ell}^{\mathcal{BB}}$ as input parameters to replace $x^\mathcal{S}_i$ and $C_{\ell}^{\mathcal{SS}}$ in the scalar QML method to estimate the $B$-mode power spectrum $C_\ell^{BB}$. In comparison with traditional QML estimator, our goal with the QML-SZ method is to simplify the calculation, without compromising significantly on the accuracy or error bars. We implement the QML-SZ estimator with the modified xQML python package.
% subsection qml_sz_estimatior (end)

\subsection{QML-TC estimator} % (fold)
\label{sub:qml_template_cleaning_estimator}
In this subsection, we describe a similar procedure by constructing a $B$-mode map, which is leakage free, and then use scalar QML method to estimate the power spectrum. We use the $E$-mode recycling method proposed in \citet{2019arXiv190400451L, 2019PhRvD.100b3538L} to obtain a leakage free $B$-mode map. This procedure essentially uses the $E$-mode signal in the CMB map to estimate the leakage due to the incompleteness of the sky. This template for the leakage is then used to clean the $B$-mode map. The mathematical principle of this idea is the following:

The true $E$-to-$B$ leakage in the pixel domain is given by the following equation:
\begin{eqnarray}\label{equ:real leakage}
\bm{L}(\bm{n})_{\mathrm{true}} = \int G_{EB}(\bm{ n},\bm{ n}') \bm{P}(\bm{n}')\, d \bm {n}',
\end{eqnarray}
where $G_{EB}(\bm{ n},\bm{ n}')$ is the convolution kernel of the $E$-to-$B$ leakage given in eq. (3.2) of \citet{2019arXiv190400451L}, and $\bm{P}(\bm{n}')$ includes both the $Q$ and $U$ Stokes parameters. Note that this convolution kernel is fixed and does not change with different realizations of the CMB sky.

The above integral should be calculated over the entire sphere. This explains why the true $E$-to-$B$ leakage cannot be precisely estimated in the case of partial sky coverage. However, the above equation also tells us that if there is no additional information of the mission's sky region, then the estimation of the $E$-to-$B$ leakage is given by
\begin{equation}
\label{equ:available leakage}
\bm{L}(\bm{n})_{\mathrm{blind}} = \int G_{EB}(\bm{ n},\bm{ n}') \bm{P}(\bm{n}')\bm{M}(\bm{n}')\, d \bm {n}',
\end{equation}
where $\bm{M}(\bm{n}')$ is the sky mask. The above equation can be further decomposed into the combination of two integrals:
\begin{equation}\label{equ:available leakage1}
\bm{L}(\bm{n})_{\mathrm{blind}} = \int G_B(\bm{ n},\bm{ n}')\bm{M}(\bm{n'}) \,d \bm {n}' 
  \int G_E(\bm{n}',\bm{n}'')\bm{M}(\bm{n''})\bm{P}(\bm{n''})\,d \bm {n}'', 
\end{equation}
where $G_E(\bm{ n},\bm{ n}')$ and $G_B(\bm{ n},\bm{ n}')$ are the pixel domain convolution kernels of the $E$- and $B$-mode signals, respectively. Mathematically, the integrals with these two kernels are nothing but standard forward-backward spherical harmonic transforms of the $E$- and $B$-modes.

Therefore, the algorithm for the template cleaning method is:
\begin{enumerate} 
    \item Starting with the data vector $\bs x$ from equation \ref{define_x} and the binary window function for the observed sky $W$, we obtain the spherical harmonic coefficients $a^r_{\ell m}$ with $r \in [T, E, B]$.
    
    \item We reconstruct a $QU$ map with the $a^E_{\ell m}$s only. In the CMB case, the power in the $B$-mode is much smaller than that in the $E$-mode. So, compared to the $E$-mode signal, the leakage from $B$-to-$E$ is negligible. Therefore, this $QU$ map represents the actual $E$-mode-only CMB polarization in the observed sky patch.
    
    \item Next, we mask the $E$-mode-only $QU$ maps again with the binary window function and obtain harmonic coefficients $\bar a^r_{\ell m}$, with $r \in [E, B]$. Since these $QU$ maps were constructed from only $E$-mode information, any $B$ modes generated by the harmonic transformation is produced by the $E$-to-$B$ leakage. So we can construct a scalar $B$-mode leakage template by using the $\bar a^B_{\ell m}$ obtained this way.
    
    \item We use the original $a^B_{\ell m}$ to obtain a scalar $B$-mode map, which is contaminated by $E$-to-$B$ leakage.
    With the contaminated $B$-mode map,
    we can obtain a linear fit of the leakage template, which is subtracted from the contaminated $B$-mode map in pixel space to obtain the leakage cleaned $B$-map. 
\end{enumerate}

This cleaned $B$ map is essentially the isolated $B$-mode pseudo-scalar field, and  we can apply the scalar QML estimator to reconstruct its power spectrum. However, the template cleaning method is not perfect so we will have some residual leakage that gets left in the cleaned $B$ maps. Therefore, the cleaned $B$ map can be treated as $B_i+x^R_i+n^B_i$, where $B_i$ is the $i$-th pixel value of the cleaned $B$ modes, $x^R_i$ is the residual leakage, and $n^B_i$ is the noise. This residual leakage term will be treated similar to the noise contribution. In practice, we compute the residual leakage covariance matrix from simulations. Then the scalar covariance matrix shown in equation \ref{QML-SZ_matrix_C} becomes:
\begin{equation}       %开始数学环境
\label{QML_TC_matrix_C}
C_{B,ij}=\langle x^B_i (x^{B}_j)^t \rangle = \sum_{\ell}\frac{2 \ell + 1}{4 \pi} C_{\ell}^{BB} P_{\ell}(z)+N_{ij}+R_{ij},
\end{equation}
where $R_{ij}$ is the covariance matrix of residual leakage.

The bias term $b^B_\ell$ is now given by $Tr[\boldsymbol{E}_\ell(\boldsymbol{N} + \boldsymbol{R})]$.
We find that, for $\ell > 5$, the impact of the residual leakage can be removed by masking pixels near the mask boundary in the cleaned $B$ map, and its impact on the final results can be ignored in most cases. In this paper, we mask all pixels $3.5^\circ$ from the mask boundary. The results of the QML methods discussed in here are also not very sensitive to the choice of the fiducial $C_\ell^{BB}$ or $C_\ell^{\mathcal{BB}}$ chosen in computing the signal covariance matrix. For example a different choice of the tensor-to-scalar ratio does not make any difference in the final results. 

% \textcolor{red}{When we regard the residual as a kind of noise, the true $B$-mode map becomes the combination of signal element and noise element. Obviously, the power spectrum of the true $B$-mode map is $C^{BB}_{\ell}$, using this together with the scalar covariance matrix shown in equation \ref{QML-SZ_matrix_C}, we obtain that:
% \begin{equation}       %开始数学环境
% \label{QML_TC_matrix_C}
% C_{B,ij}=\langle x^B_i (x^{B}_j)^t \rangle = \sum_{\ell}\frac{2 \ell + 1}{4 \pi} C_{\ell}^{BB} P_{\ell}(z)+N_{t,ij}.
% \end{equation}
% }
%Obviously, the power spectrum of the template cleaned map is the $B$-mode power spectrum $C^{BB}_{\ell}$, instead of $C^{\mathcal{B}\mathcal{B}}_{\ell}$. Using this together with the scalar covariance matrix shown in equation \ref{QML-SZ_matrix_C}, we obtain that

%\begin{equation}       %开始数学环境
%\label{QML_TC_matrix_C}
%C_{B,ij}=\langle x^B_i (x^{B}_j)^t \rangle = \sum_{\ell}\frac{2 \ell + 1}{4 \pi} C_{\ell}^{BB} P_{\ell}(z).
%\end{equation}

Using the template cleaned $B$-mode map as input and the covariance matrix above, we use the scalar QML (equations \ref{QML_SZ_Erl} to \ref{QML_SZ_true_PS}) to obtain the QML-TC estimator for the $B$-mode power spectrum. Similar to the QML-SZ estimator, the QML-TC estimator is a much simpler implementation that should have reasonably good performance at the largest scales. In our calculation, the QML-TC estimator is also implemented with the xQML python package.
% %*********************************************************************%

% subsection qml_template_cleaning_estimator (end)

% %**********************************************************************
% subsection PCL estimator
% %**********************************************************************
\subsection{Pure $B$-mode PCL estimator}
\label{sub:PCL_estimator}
The most common method that is adopted in the CMB data analysis is the so-called pseudo-$C_\ell$ (PCL) estimators, which are both computationally inexpensive and near-optimal at high multipoles. 
%This method was developed for temperature by \citet{Hivon2002} and extended for polarization by \citet{Hansen2003,2009PhRvD..79l3515G}. 
For polarization analysis with PCL estimators, the preferred method is to work with pure-$E$ and pure-$B$ fields. In case of an incomplete sky, equations \ref{Elm} and \ref{Blm} give the relation between the partial sky pure $E$/$B$ mode harmonics and the full sky $E$ and $B$ modes. They can be used to obtain a relation between the partial sky pure $E$/$B$-power spectra and the actual $E$/$B$- mode power spectra:
\begin{eqnarray}
    \tilde{C}_\ell^{\mathcal{EE}} &=& \sum_{\ell'} \left[ \mathcal{M}_{\ell \ell'}^{EE}C^{EE}_{\ell'} + \mathcal{M}_{\ell \ell'}^{EB}C^{BB}_{\ell'}\right], \label{eq:pureEClmix}\\
    \tilde{C}_\ell^{\mathcal{BB}} &=& \sum_{\ell'} \left[ \mathcal{M}_{\ell \ell'}^{BE}C^{EE}_{\ell'} + \mathcal{M}_{\ell \ell'}^{BB}C^{BB}_{\ell'}\right]. \label{eq:pureBClmix}
\end{eqnarray}
In these expressions $\mathcal{M}^{ru}_{\ell \ell'}$ is the mixing matrix that relates the two sets of power spectra. The mixing matrices are defined as:
\begin{equation}
    \mathcal{M}^{ru}_{\ell \ell'} = \frac{1}{2\ell + 1} \sum_{m m'} |\mathcal{K}^{ru}_{\ell m \ell' m'}|^2,
\end{equation}
with $r,u \in [E,B]$. The exact expressions for the mixing matrices can be found in \citet{2009PhRvD..79l3515G}. We can solve the set of equations \ref{eq:pureEClmix} and \ref{eq:pureBClmix} for the $E$- and $B$-mode power spectra. These are called the pseudo full sky power spectra estimates. In this work, we will compare the performance of the three QML methods with the PCL method in the multipole range of interest. 

When we perform spherical harmonic transformation with binary window functions it will lead to severe leakage and mode mixing. Therefore we have apodized our observation window with a `C2' (cosine) apodization function for the PCL estimates in this work. The weight in the $i^{\rm th}$ pixel is given as \citep{Alonso2019}:

    \begin{numcases} {W_i =}
    \frac{1}{2}\left[1 - \cos (\pi \delta^r_i)\right] & $\delta^r_i < 1$ \nonumber\\
    1 & ${\rm otherwise}$,
    \end{numcases}

where $\delta^r_i = \sqrt{(1-\cos \delta_i)/(1-\cos \delta_c)}$. We compute all PCL estimator results in this work with the C2 apodization. The PCL estimator for this work has been implemented with the python package of NaMaster \footnote{https://github.com/LSSTDESC/NaMaster} \citep{Alonso2019}.
% %*********************************************************************%

% %*********************************************************************%

\begin{figure}[t]
\centering
\includegraphics[width=0.48\textwidth]{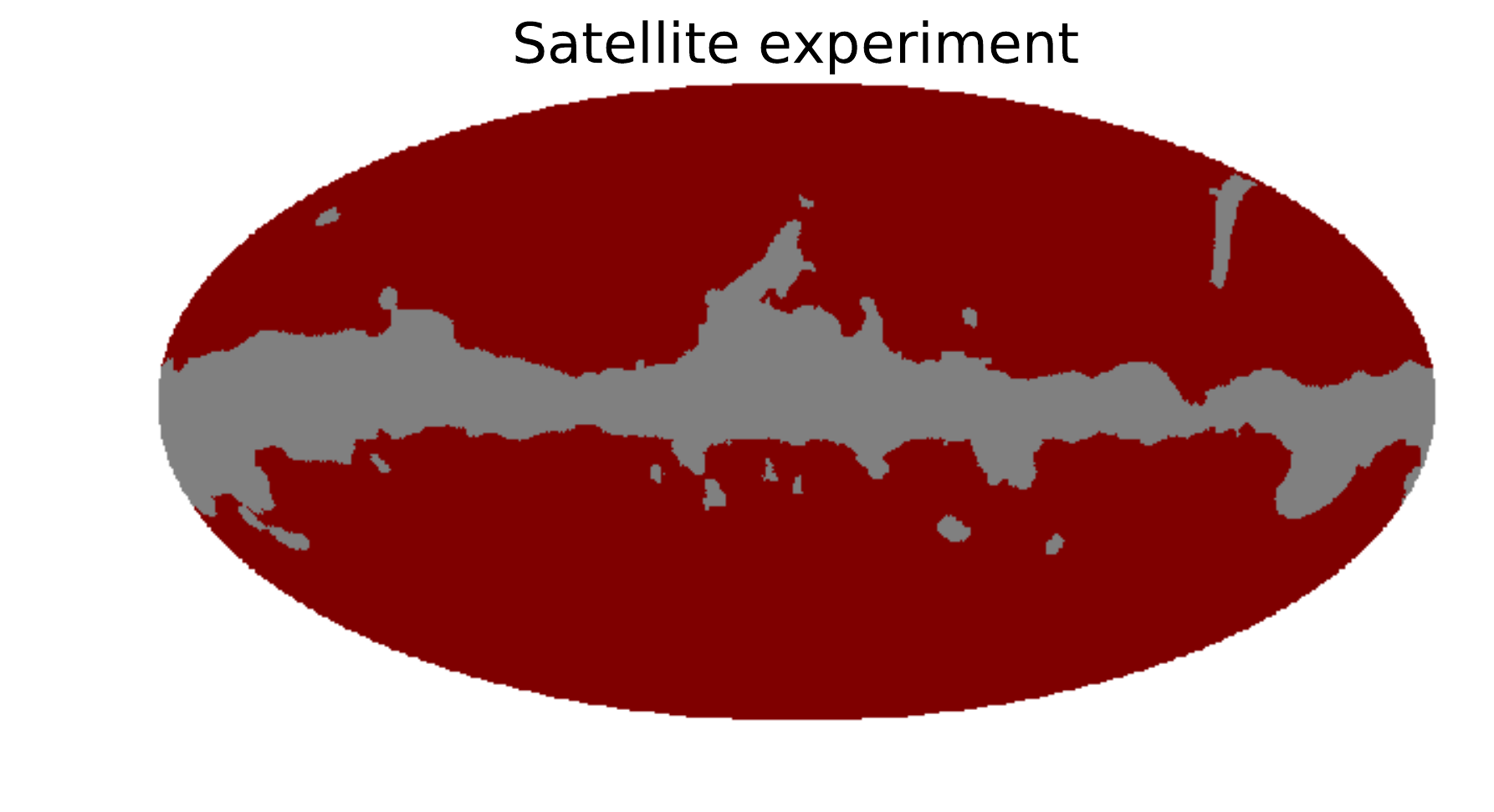}
\includegraphics[width=0.38\textwidth]{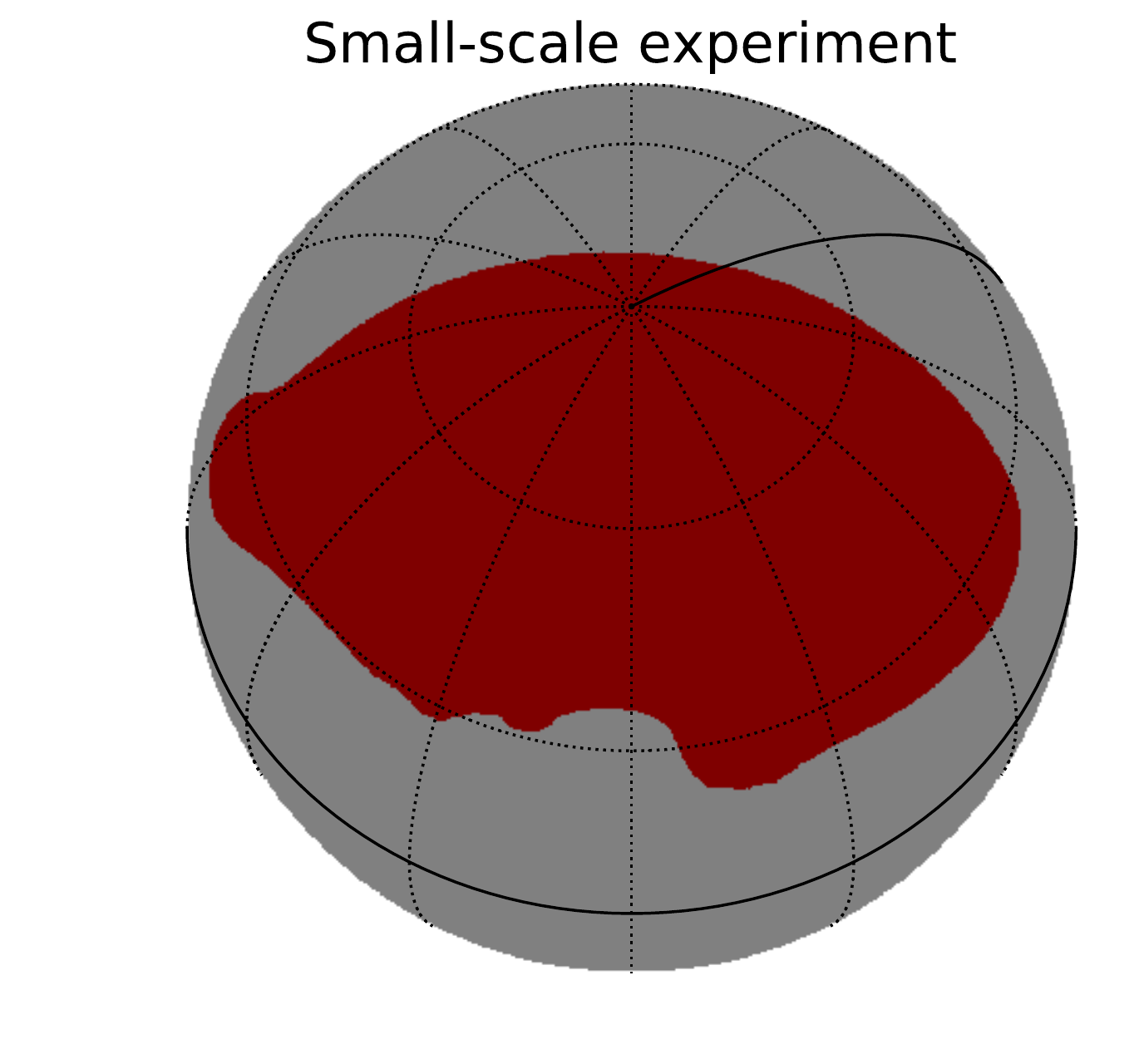}
\caption{Binary masks showing the observed sky patch for a space-based experiment (top) and for the ground-based experiment (bottom) as considered in this work. The red area is the observed area and the gray area is the mask area. The plots are in Galactic coordinate system. The sky fractions are $78.8\%$ and $15.1\%$, respectively.} 
\label{sky_coverage} 
\end{figure}

% %*********************************************************************%
\section{SIMULATION SETUP AND IDEALIZED TESTS} % (fold)
\label{sec:tests_and_applications}

In this work, we consider two cases of future CMB polarization experiments: a space-based experiment and a ground-based polarization experiment. In both cases, due to astrophysical foreground and/or survey limitations, only an incomplete sky patch can be used for scientific analysis. Thus, for each experimental scenario the $E$-to-$B$ leakage is going to be a challenge. For the space-based experiment case, we consider the 2018 Planck common polarization mask, which masks the galactic foregrounds and the point sources resolved in Planck maps. Using HEALPix \texttt{process\_mask} subroutine we fill-in all the point source smaller than $5^\circ$ which are masked, as well as the extended source masking at high galactic latitudes ($|b|>45^\circ$).  We assume this is the observed sky patch with $f_{\rm sky} \sim 78.8\%$, which is representative of a space-based polarization experiment. In addition, we consider a $f_{\rm sky} \sim 15.1 \%$ sky patch in the northern hemisphere, based on the AliCPT-1 experiment \citep{Hong2017, 2021arXiv210109608S}, for the observed sky of the ground-based experiment simulations. We show the binary mask for these two sky patches in figure \ref{sky_coverage}. 
%*********************************************************************%

\begin{figure*}
\centering
\includegraphics[width=\textwidth]{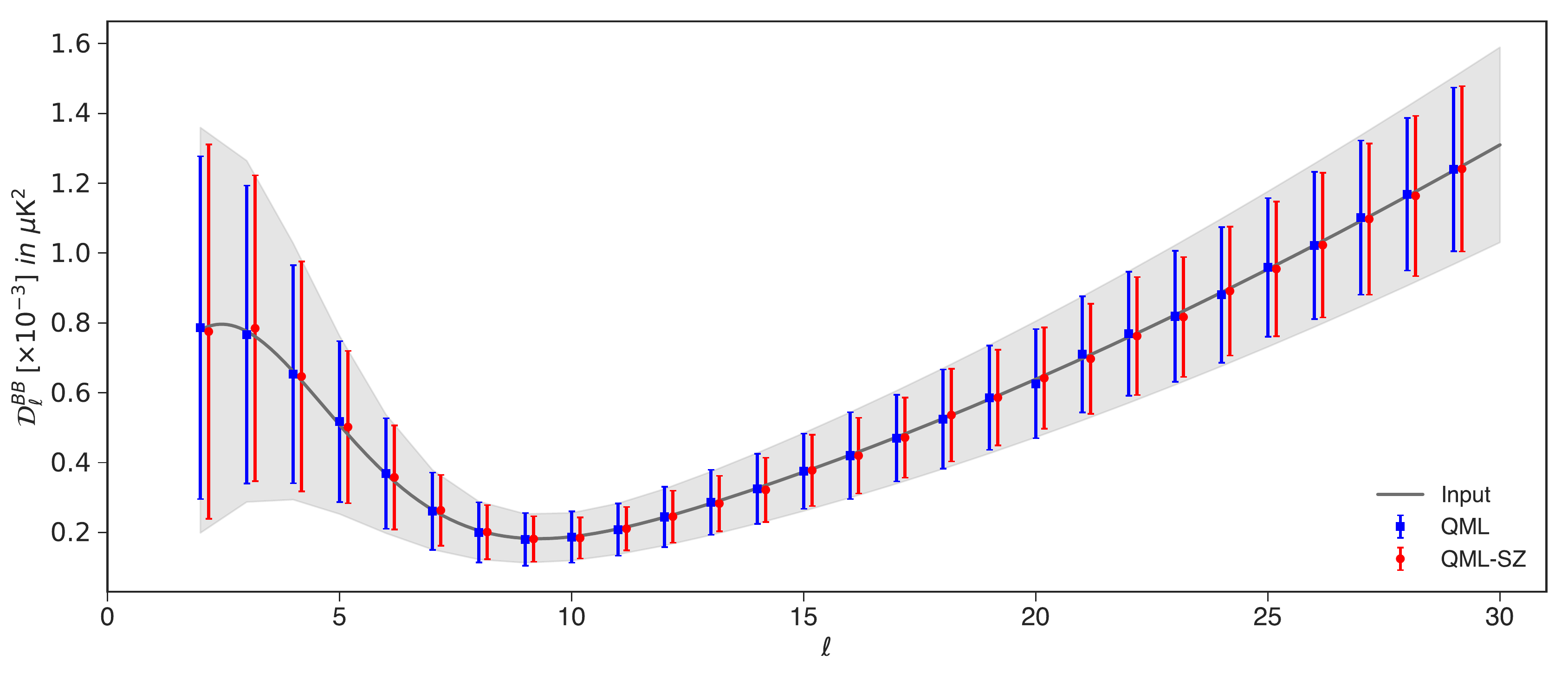}
\caption{Plot of $B$-mode power spectrum estimates for testing the effects of $f_{\rm sky}$ for the space-based experiment. The input $B$-mode power spectrum is represented by the black curve, the blue square markers indicate the classic QML method estimators and the red circle markers show results from QML-SZ scalar method. The gray band shows the error bars calculated with the analytical formula in Eq.(\ref{eq:cosmic_var}).} 
\label{fig:mask_effect_72}
\end{figure*}

%*********************************************************************%

In our calculation, the CMB map simulations are produced using the \texttt{synfast} subroutine of HEALPix\footnote{http://healpix.sourceforge.net} with the input CMB power spectra generated by CAMB\footnote{http://camb.info} \citep{Lewis:1999}. In this article, we consider the 2018 Planck cosmological parameters \citep{Planck2018VI}, including weak lensing contributions, and setting the tensor-to-scalar ratio $r$ to 0.05. For the realistic examples shown in section \ref{sec:realistic_examples}, we additionally compute the results for $r=0$. The choices for the tensor-to-scalar ratio denote the current upper and lower limits on the value of $r$ \citep{2015PhRvL.114j1301B}.

Before considering the realistic simulations to test the performance of the three estimators, we first perform some idealized tests to identify the impact on the results of various factors of our simulation setup. These analysis independently take into account the effects due to masking, downgrading procedure, and different white noise levels. For optimizing our simulation pipeline considering these idealized tests, we use full sky maps. Therefore,  we do not produce scalar maps of pure-$B$ or template cleaned $B$-mode map from masked $TQU$ maps, in order to prevent any additional complications that might arise from any residual $E$-to-$B$ leakage in the scalar maps. In summary, we simulate a full $TQU$ CMB map in order to test the optimal simulation settings for the standard QML method. We simulate a scalar pure-$B$ map, $\mathcal{B}(\hat{n})$, with $C_\ell^{\mathcal{BB}}$ as input to optimize the QML-SZ estimator and finally, for QML-TC estimator, we simulate a scalar $B$-mode map with input $C^{BB}_\ell$.

%In these idealized tests, that we use for further optimization of our simulation pipeline, we do not attempt to produce scalar maps of pure-$B$ or template cleaned $B$-mode map from masked $TQU$ maps. This is done to prevent any additional complications that might arise from any residual $E$-to-$B$ leakage in the scalar maps. So, for testing the optimal simulation settings for classic QML method, we simulate a full $TQU$ CMB map. For optimization QML-SZ estimator simulations, we simulate a scalar pure-$B$ map, $\mathcal{B}(\hat{n})$, with $C_\ell^{\mathcal{BB}}$ as input. Finally, for QML-TC estimator, we simulate a scalar $B$-mode map with input $C^{BB}_\ell$.

The computational requirement for a QML method scales with the size of the data vector $\bs x$. For the full standard QML method with temperature and polarization, the size of data vector $N_d$ is $3 \times N_{\rm pix, obs}$, where $N_{\rm pix, obs}$ is the number of pixels in the observed sky patch. However, as described in section \ref{sub:qml_estimator}, we will work with only the polarization part of the QML estimator. This reduces the size of the data vector to $2 \times N_{\rm pix, obs}$ by dropping the $T$ fields. For both the scalar methods, the data vector size is $N_d = N_{\rm pix, obs}$. The size of the computational requirement is set by the size of the covariance matrix, which is $N_d \times N_d$. The inversion of the covariance matrix is therefore an $O(N_d^3)$ operation. Thus, the scalar QML method has significant advantage regarding memory and computation time required in the estimation. This is also a limiting factor for the HEALPix map resolution we can work with for a given experiment. For the space-based experiment, since the observed sky fraction is quite large, we work at \texttt{NSIDE}=16, with $\ell_{\rm max}=2\times$\texttt{NSIDE}. But for a small sky patch of the ground-based experiment, we can choose a higher resolution of \texttt{NSIDE}=32, with $\ell_{\rm max}=3\times$\texttt{NSIDE}.

Unfortunately, the template cleaning method has residuals that impact the power spectrum estimation at the lowest multipoles. Therefore, in the satellite case, we will not consider the QML-TC estimator, since, in this particular case, we expect accurate estimation even in lowest multipoles. We discuss this issue in more detail in appendix \ref{apdx:QML-TC_method}. 

%This impacts the power estimation for satellite experiments where we would typically want accurate estimation of even the lowest few multipoles. Therefore we will not consider the QML-TC method for the satellite experiment situation in this article. We discuss this issue in more detail in appendix \ref{apdx:QML-TC_method}.

\subsection{Effect of $f_{\rm sky}$} % (fold)
\label{sub:effect_of_mask}

In an incomplete sky, besides the mixtures between the $E$ and $B$ modes, we have to deal with the well-known mode mixing problem. In this subsection, we will test the QML estimator, as well as the QML-SZ and QML-TC estimators introduced previously. As discussed, we simulate the pure-$B$ and $B$-mode maps directly from $C_\ell^{\mathcal{B}\mathcal{B}}$ and $C_\ell^{BB}$, so there is no $E$-to-$B$ leakage residual contribution. We simulate at \texttt{NSIDE} of 16 and 32 for satellite and ground experiments, respectively, withthe noise level set to 0.1 $\mu$K-arcmin, which is close to the noise-free case. Therefore, we can only study the effect of mode mixture caused by the partial-sky surveys.

%On an incomplete sky, there exist the leakages of power spectra between different multipoles. This is the usual mode mixing problem, which is present even for a scalar field like temperature fluctuations. In case we have polarization $QU$ fields, there are additional mixtures between the $E$ and $B$ modes. In this subsection, by simulation, we will test the QML estimator, as well as the QML-SZ and QML-TC estimators introduced in this article. As discussed we simulate the pure-$B$ and $B$-mode maps with \texttt{NSIDE} of 16 and 32 for satellite and ground experiments respectively, directly from the pure-$B$ and $B$-mode power spectrum as a scalar map. We set the noise level to 0.1 $\mu$K-arcmin, which is close to the noise-free case. This means that there is no $E$-to-$B$ leakage residual contribution. Therefore, we can only study the effect of mode mixture caused by the partial-sky surveys.

%*********************************************************************%
\begin{figure}[htbp]
\centering
\includegraphics[width=0.45\textwidth]{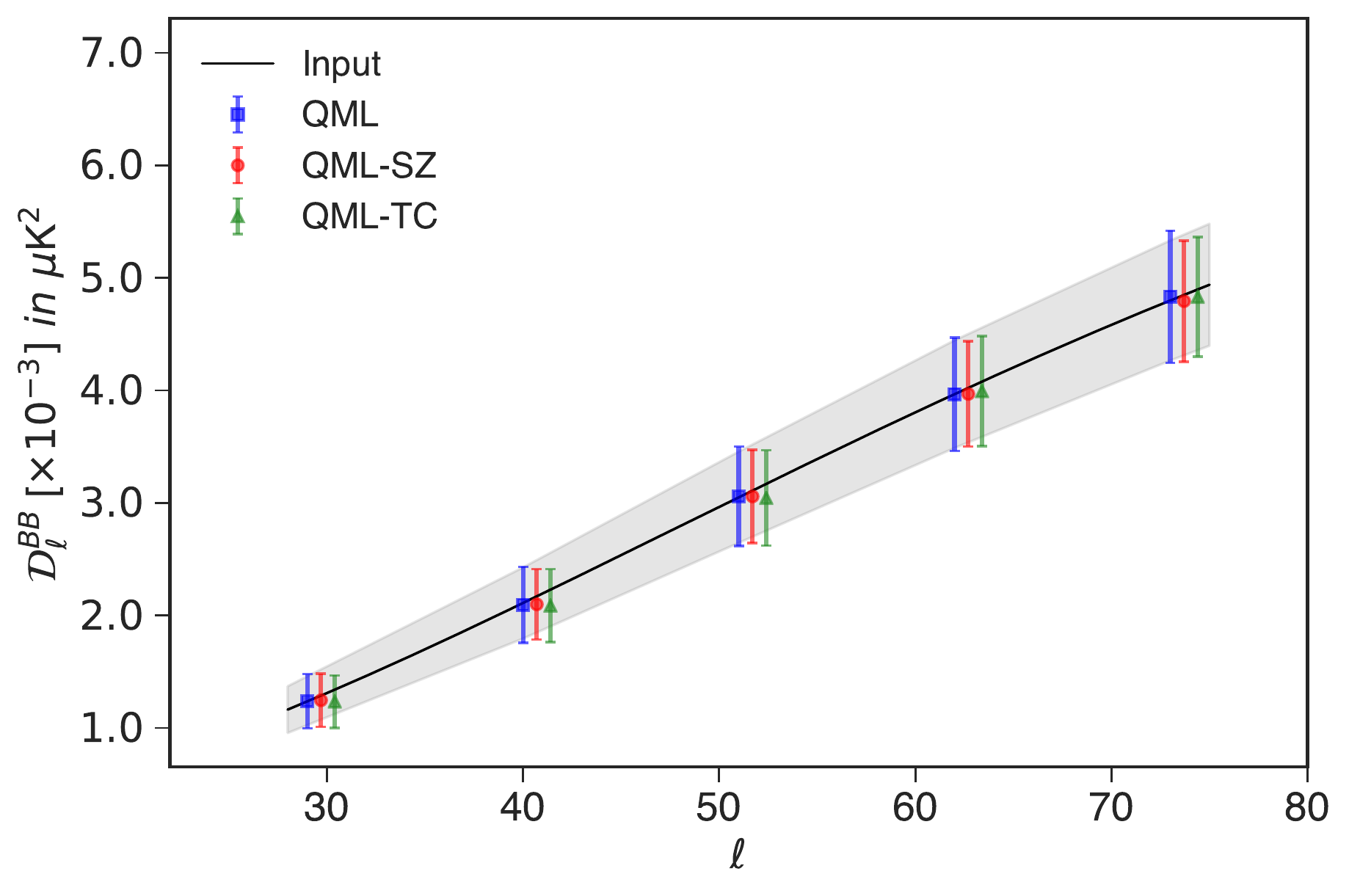}
\caption{Plot of $B$-mode power spectrum estimates for testing the effects of $f_{\rm sky}$ for the ground-based experiment. The input $B$-mode power spectrum is represented by the black curve, the blue square markers indicate the classic QML method estimators, red circle markers show results from QML-SZ scalar method, while green triangle markers show QML-TC estimators. The gray band shows the error bars calculated with the analytical formula in Eq.(\ref{eq:cosmic_var}).} 
\label{fig:mask_effect_12} 
\end{figure}{}
%*********************************************************************%

%*********************************************************************%

\begin{figure*}[ht]
\centering
\includegraphics[width=\textwidth]{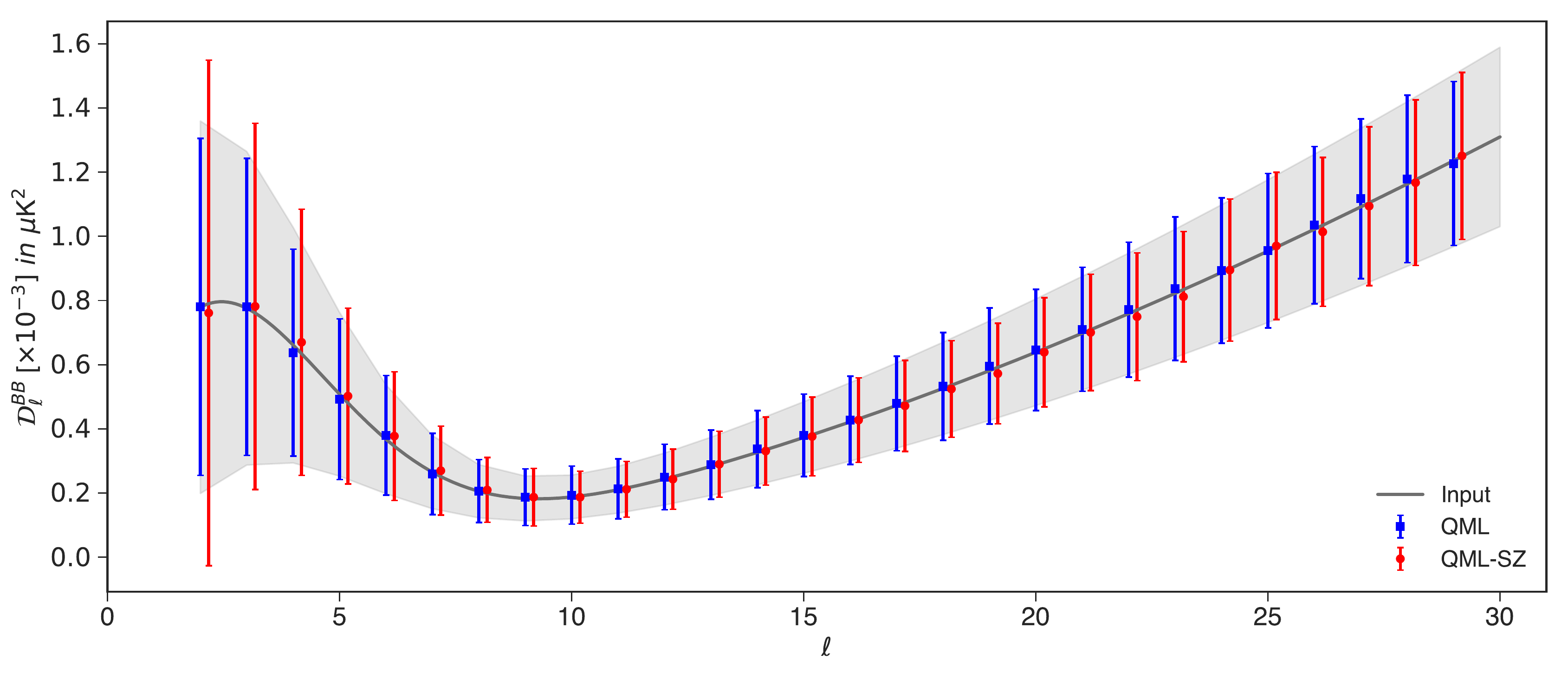}
\caption{Plot of power spectrum estimates for testing the impacts of \texttt{ud\_grade} for the space-based experiment. All maps are produced at \texttt{NSIDE}=512 with $\ell_{\rm max}$=32, and then downgraded to \texttt{NSIDE}=16. The input $B$-mode power spectrum is represented by the black curve, the blue square markers indicate the classic QML method estimates and the red circle markers show results from QML-SZ scalar method. The gray band shows the optimal error limit from cosmic variance and noise variance.}
\label{fig:downgrade_impact_72}
\end{figure*}
%*********************************************************************%

In figures \ref{fig:mask_effect_72} and \ref{fig:mask_effect_12}, we show the power spectra estimates plotted for space-based and ground-based experiments, respectively. Note that we have binned the power spectra with the following definition,
\begin{equation}
    \mathcal{D}^{BB}_\ell = \frac{1}{\Delta \ell}\sum_{\ell'=\ell-\Delta \ell/2}^{\ell'=\ell+\Delta \ell/2} \frac{\ell'(\ell' + 1)}{2 \pi} C^{BB}_{\ell'},
    \label{eq:binnedCl}
\end{equation}
where $\Delta \ell$ is the bin width. For the space-based experiment, due the large sky coverage, it is reasonable to estimate the power spectra at each multipole, hence we set $\Delta \ell =1$. For ground-based experiments,  we bin the power spectrum into bands with $\Delta \ell = 11$ to mitigate the greater mode mixing problem due to a smaller observation patch. Throughout this work, our power spectra estimates for any estimator is a mean of 1000 random simulations, and the errors are computed as the standard deviation of the samples. For comparison, we also compare our results with the simple analytical estimate of error bars, which is given by,

%For ground-based experiments, smaller observation patch causes greater mode mixing and results in strong correlations between nearby harmonics. Therefore we bin the power spectrum into bands with $\Delta \ell = 11$ for all ground-based experiment cases in this paper. Throughout this work, our power spectra estimates for any estimator is a mean of 1000 random simulations, and the errors are computed as the standard deviation of the samples. For comparison, we also compare our results with the simple analytical estimate of error bars, which is given by,

\begin{equation}
    \Delta \mathcal{D}^{BB}_{\ell, {\rm optimal}} \cong \sqrt{\frac{2}{(2\ell + 1) \Delta \ell f_{\rm sky}}} \left[ \mathcal{D}^{BB}_\ell +\mathcal{N}^{BB}_\ell \right],
    \label{eq:cosmic_var}
\end{equation}
where $\mathcal{N}^{BB}_\ell$ is the noise power spectrum binned using relation \ref{eq:binnedCl}.  

As anticipated, the results in figures \ref{fig:mask_effect_72} and \ref{fig:mask_effect_12} show that all three methods can obtain unbiased estimates for the $B$-mode power spectrum. Then, we will investigate the impact of $f_{\rm sky}$ on the error bars. In the case with Planck mask, we find the RMS errors of these QML methods are even smaller than the analytical results in Eq.\ref{eq:cosmic_var}, since the large-scale information in the masked region can be partly recovered by QML analysis, which is consistent with the results in \citet{efstathiou2004, 2006MNRAS.370..343E}. For smaller scales, the errors of the three methods are optimal, and all methods perform equally. Our results show that the mode mixing due to $f_{\rm sky}$ is sufficiently well corrected in all the presented methods, and its impact on the final results is negligible for either of the two sky patches considered in this work.

%%---------- I think this paragraph is not necessary
%%Even for this simplified study, the effect of a sky mask adds various complications to the analysis, including the pixelization scheme, map resolution and mask geometry. Here, we have chosen the `complicated' and realistic mask, instead of simple geometric masks, to test effect of realistic incomplete skies. Fortunately, our results show that the mode mixing due to $f_{\rm sky}$ is sufficiently well corrected in all the QML methods, and its impact on the final results is negligible for either of the two sky patches considered in this work.
%\vfill\null
 
% subsection effect_of_mask (end)

%*********************************************************************%
\begin{figure}[t]
\centering
\includegraphics[width=0.45\textwidth]{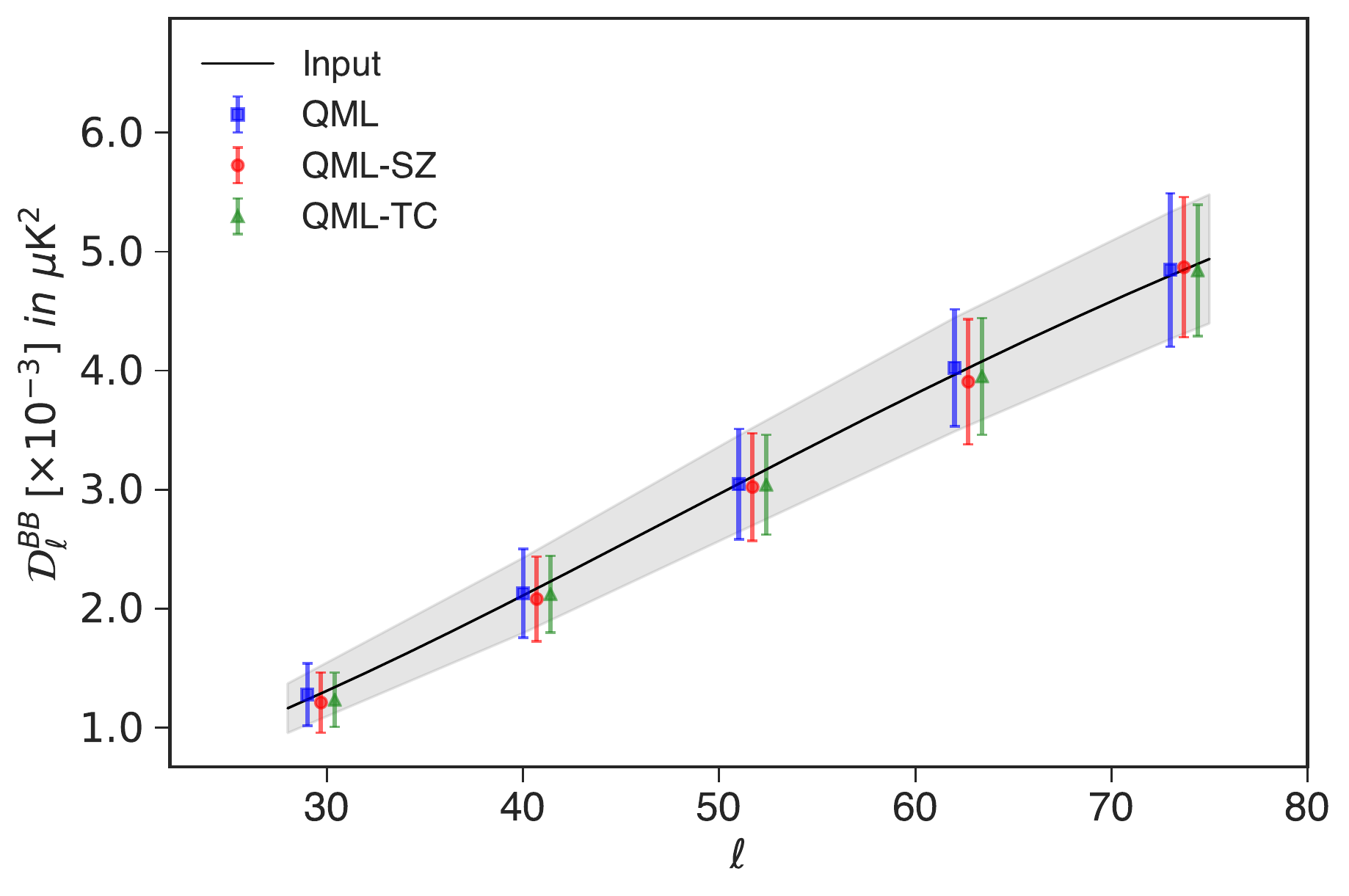}
\caption{Plot of power spectrum estimates for testing the impacts of \texttt{ud\_grade} for the ground-based experiment. All maps are produced at \texttt{NSIDE}=512 with $\ell_{\rm max}$=96, and then downgraded to \texttt{NSIDE}=32. The input $B$-mode power spectrum is represented by the black curve, the blue square markers indicate the classic QML method estimates, red circle markers show results from QML-SZ scalar method, while green triangle markers show QML-TC estimates. The gray band shows the optimal error limit from cosmic variance and noise variance.}
\label{fig:downgrade_impact_12} 
\end{figure}
%*********************************************************************%

\subsection{Impact of downgrading map} % (fold)
\label{sub:effect_of_down_grade}

%*********************************************************************%
\begin{figure}[t]
\centering
\includegraphics[width=0.45\textwidth]{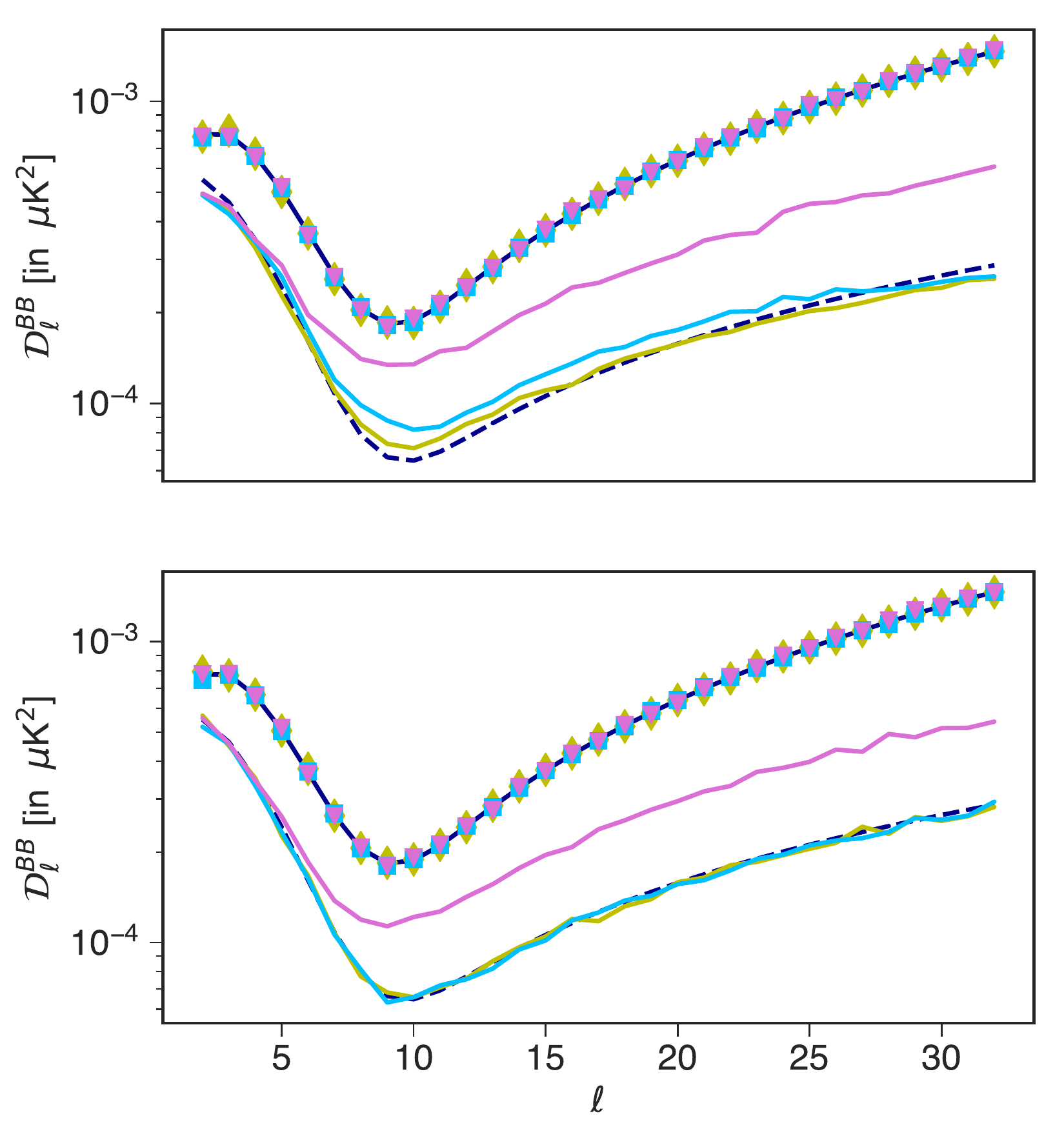}
\caption{Plots for power spectrum reconstruction with the different QML methods with three different noise levels for the space-based experiment. The upper rows show QML results and the lower rows show QML-SZ method results. The solid black line indicates the input theoretical $B$-mode power spectra. The dashed black line shows the cosmic variance limit which can be obtained from equation \ref{eq:cosmic_var} by setting the noise to zero. We show the mean power spectra estimates with markers, and standard deviation errors with solid colored lines. The noise levels in the plot: 10 $\mu$K-arcmin (pink, inverted triangle), 1 $\mu$K-arcmin (cyan, square), and 0.1 $\mu$K-arcmin (yellow, diamond).}
\label{fig:noise_levels_72}
\end{figure}
%*********************************************************************%

Actual CMB polarization maps are usually at much higher resolution than the resolution imposed by computational limitations in calculating the QML estimator. Any analysis using QML estimators will then require a smoothing process to reduce the influence of higher multipoles, followed by a downgrading process to reduce the \texttt{NSIDE} of the map. The map degrading process for this work will be done with the \texttt{ud\_grade} subroutine from HEALPix. In this subsection, we test the impact of \texttt{ud\_grade} alone (without considering the effect of smoothing). 

%Actual CMB polarization experiments are usually at much higher resolution than the resolution limits on QML estimator from the computational requirements. So, any analysis of experimental observations with QML estimators will require a smoothing process to reduce the influence of higher multipoles, followed by a downgrading process to reduce the \texttt{NSIDE} of the map to the target resolution of the analysis. The map degrading process for this work will be done with the \texttt{ud\_grade} facility from HEALPix. In this subsection, we test the impact of \texttt{ud\_grade} alone without the effect of smoothing. 

For this test, we simulate CMB maps at \texttt{NSIDE}=512, with $\ell_{\rm max}$ set to 32 and 96 for space-based and ground-based cases, respectively. %These represent the $2\times$\texttt{NSIDE} and $3\times$\texttt{NSIDE} $\ell_{\rm max}$ for the target \texttt{NSIDE} for our QML analysis. 
This is done to mimic the cases with smoothing that cuts off the modes above the set $\ell_{\rm max}$ values. Thus, it helps to isolate the effect of the downgrading procedure only. The full sky maps at \texttt{NSIDE}=512 is masked with appropriate binary masks and then downgraded to the targeted \texttt{NSIDE} by \texttt{ud\_grade}. We also downgrade the masks with \texttt{ud\_grade}, and set any pixels with values $<0.99$ to zero. We multiply this downgraded mask by the downgraded maps, which act as the input maps for the QML estimators.

%*********************************************************************%
\begin{figure}[t]
\centering
\includegraphics[width=0.45\textwidth]{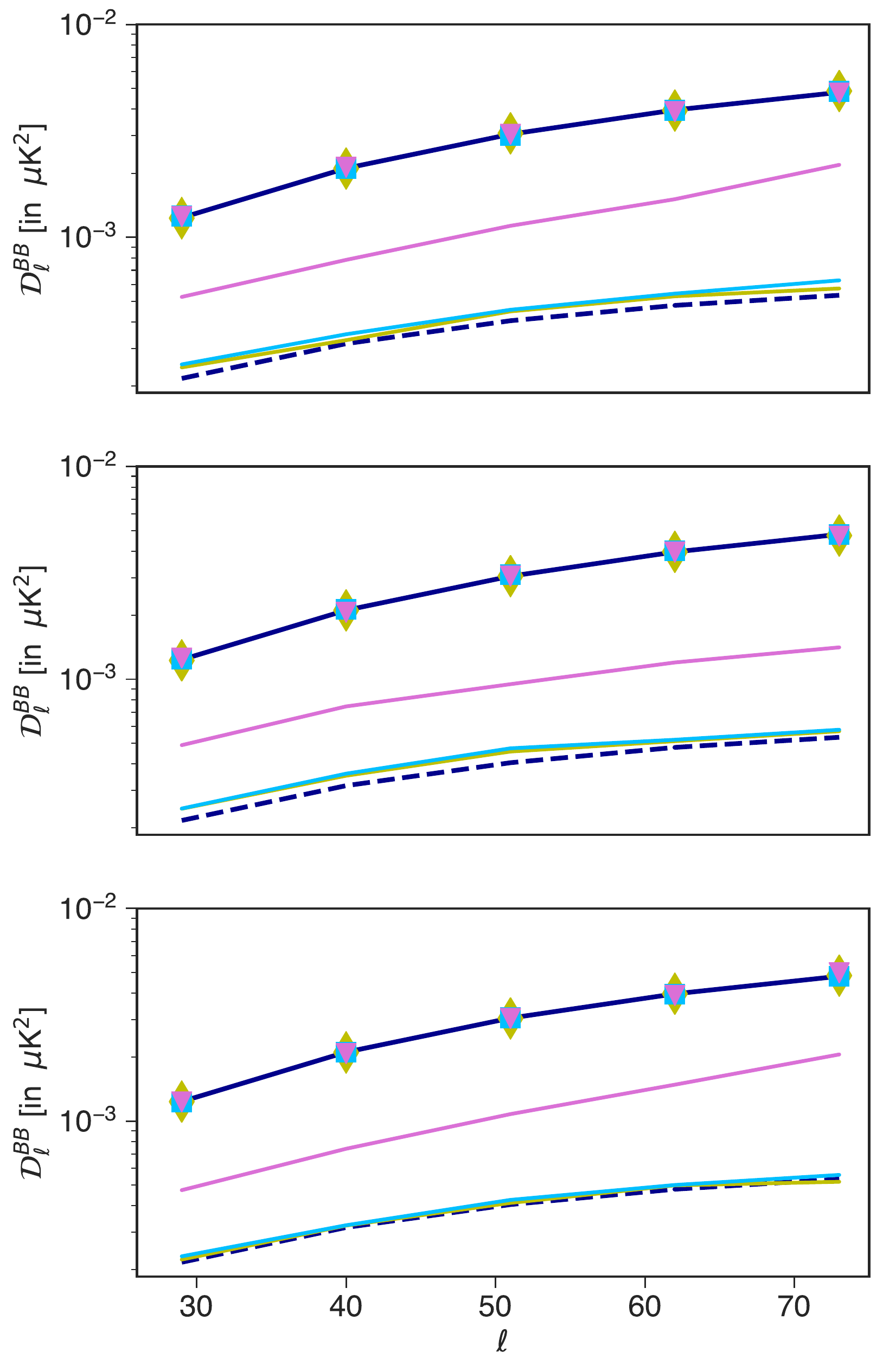}
\caption{Plots for power spectrum reconstruction with the different QML methods with three different noise levels for the ground-based experiment. The upper rows show QML results, the middle rows show QML-SZ method results, and the lower rows show QML-TC method results. The solid black line indicates the input theoretical $B$-mode power spectra. The dashed black line shows the cosmic variance limit which can be obtained from equation \ref{eq:cosmic_var} by setting the noise to zero. We show the mean power spectra estimates with markers, and standard deviation errors with solid colored lines. The noise levels in the plot: 10 $\mu$K-arcmin (pink, inverted triangle), 1 $\mu$K-arcmin (cyan, square), and 0.1 $\mu$K-arcmin (yellow, diamond).}
\label{fig:noise_levels_12}
\end{figure}
%*********************************************************************%

%*********************************************************************%
%*********************************************************************%
\begin{figure*}[th]
\centering
\includegraphics[width=\textwidth]{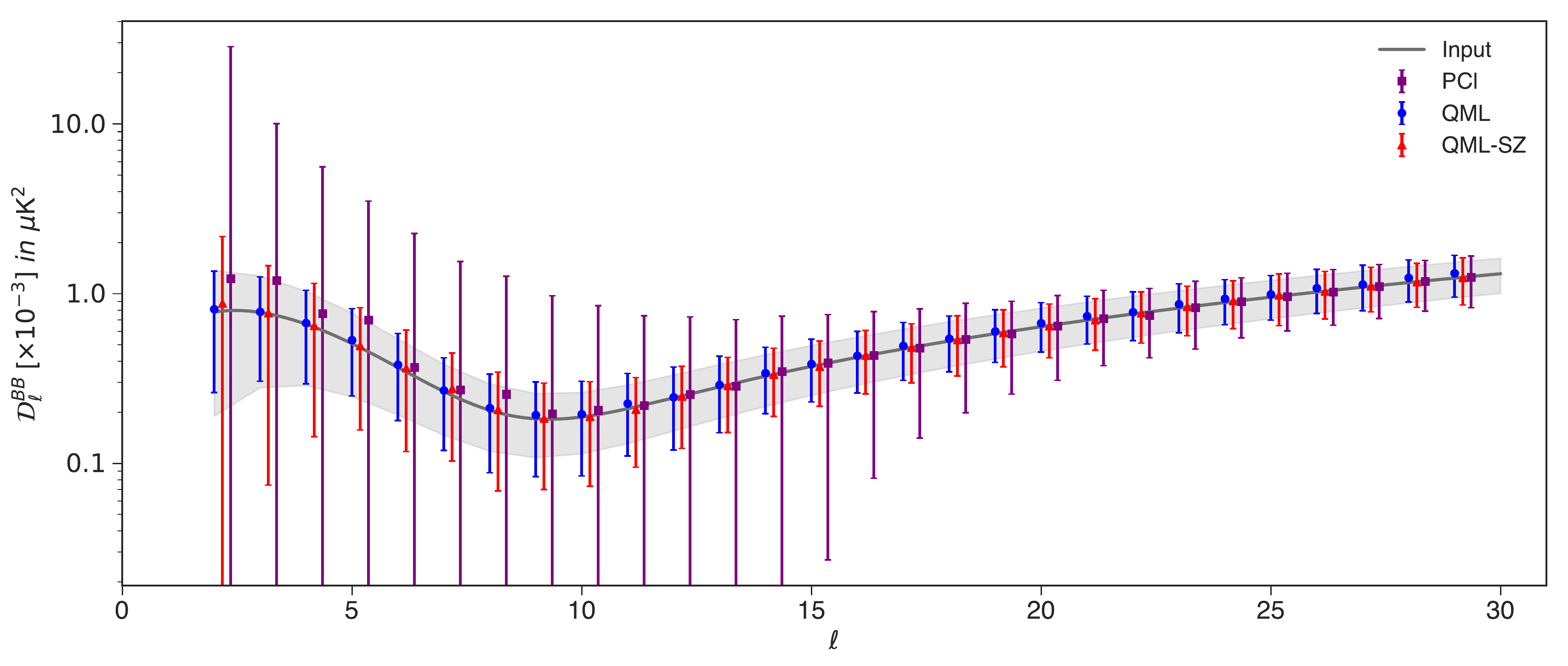}
\includegraphics[width=\textwidth]{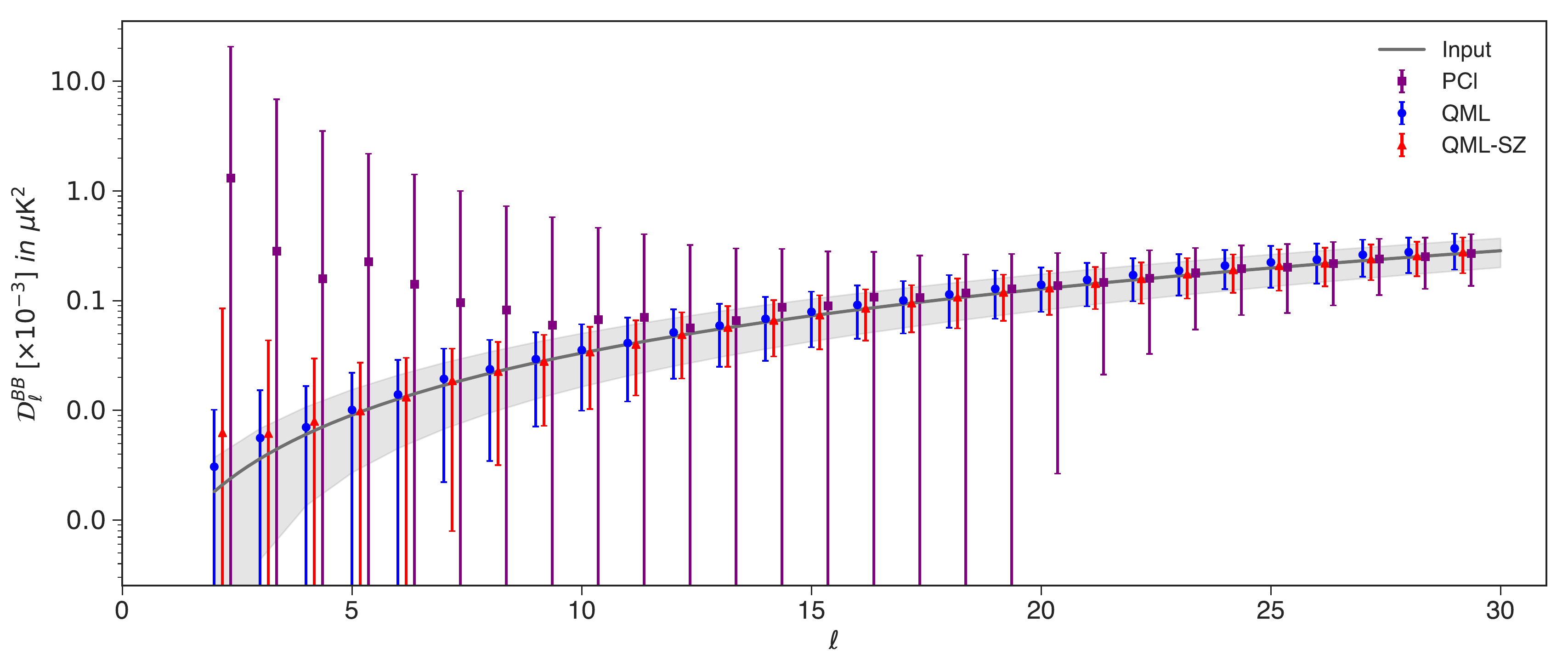}
\caption{Plot of the results for $B$-mode power spectrum estimates for the realistic space-based CMB experiment, the upper panel and the lower panel are $r=0.05$ and $r=0$ respectively. The observed sky with 3 $\mu$K-arcmin noise, is simulated at \texttt{NSIDE}=512 with $\ell_{\rm max}$=1024. The input $B$-mode power spectrum is shown with the black solid curve. The classic QML method results are shown with blue, square markers, QML-SZ method results with red, circular markers. These results are computed at \texttt{NSIDE}=16 with $\ell_{\rm max}$=47. We also show PCL estimator results, obtained with NaMaster, with $\delta_c = 6^\circ$ C2 apodization, with purple, inverted triangle markers. The gray region denotes the analytical approximation of the error bounds. The data points are the mean of 1000 estimates, and the error bars are given by the standard deviation of the estimates.}
\label{fig:realex_72_homo}
\end{figure*}
%*********************************************************************%

%************************************************************
\begin{figure*}[ht]
\centering
\includegraphics[width=0.48\textwidth]{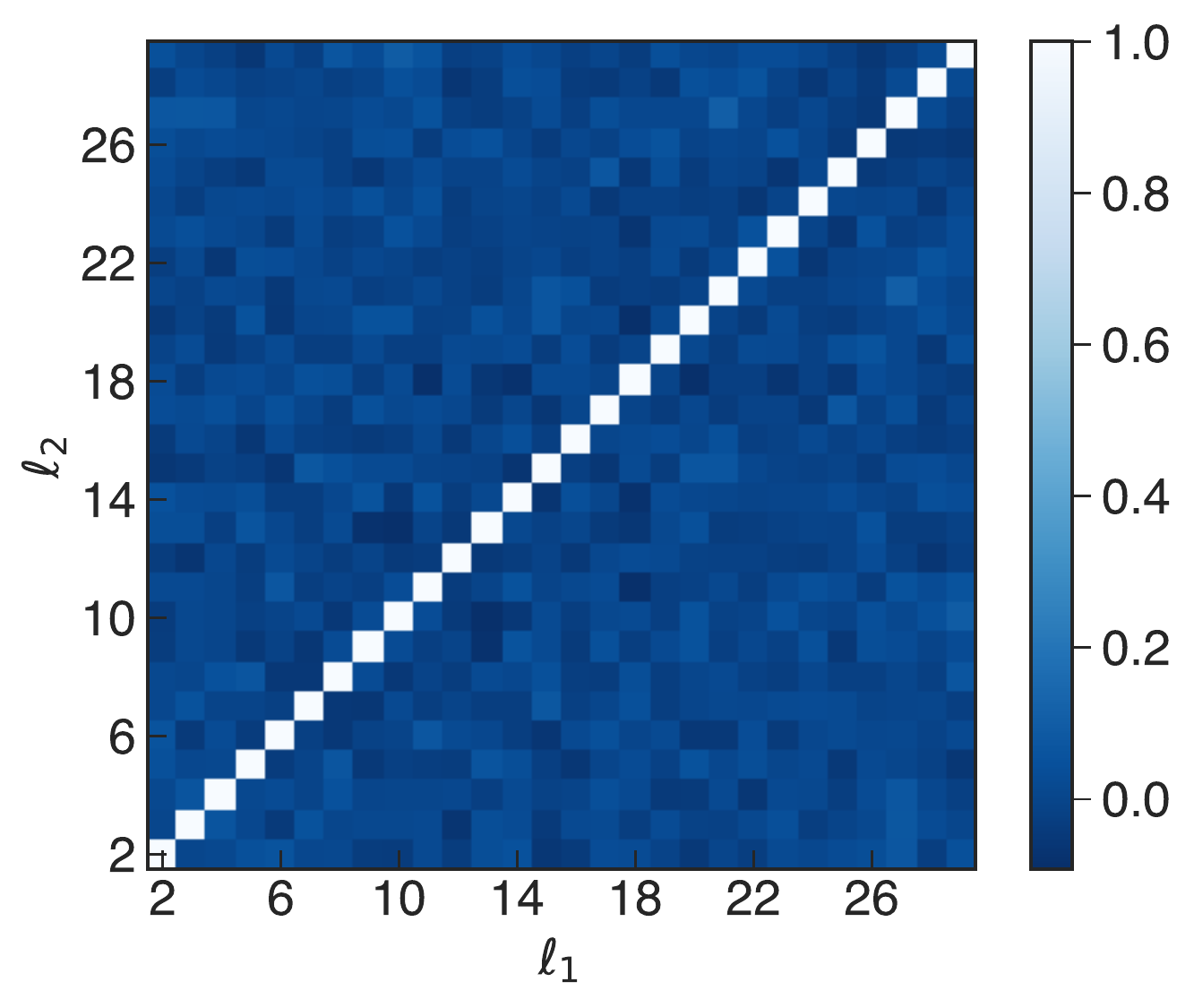}
\includegraphics[width=0.48\textwidth]{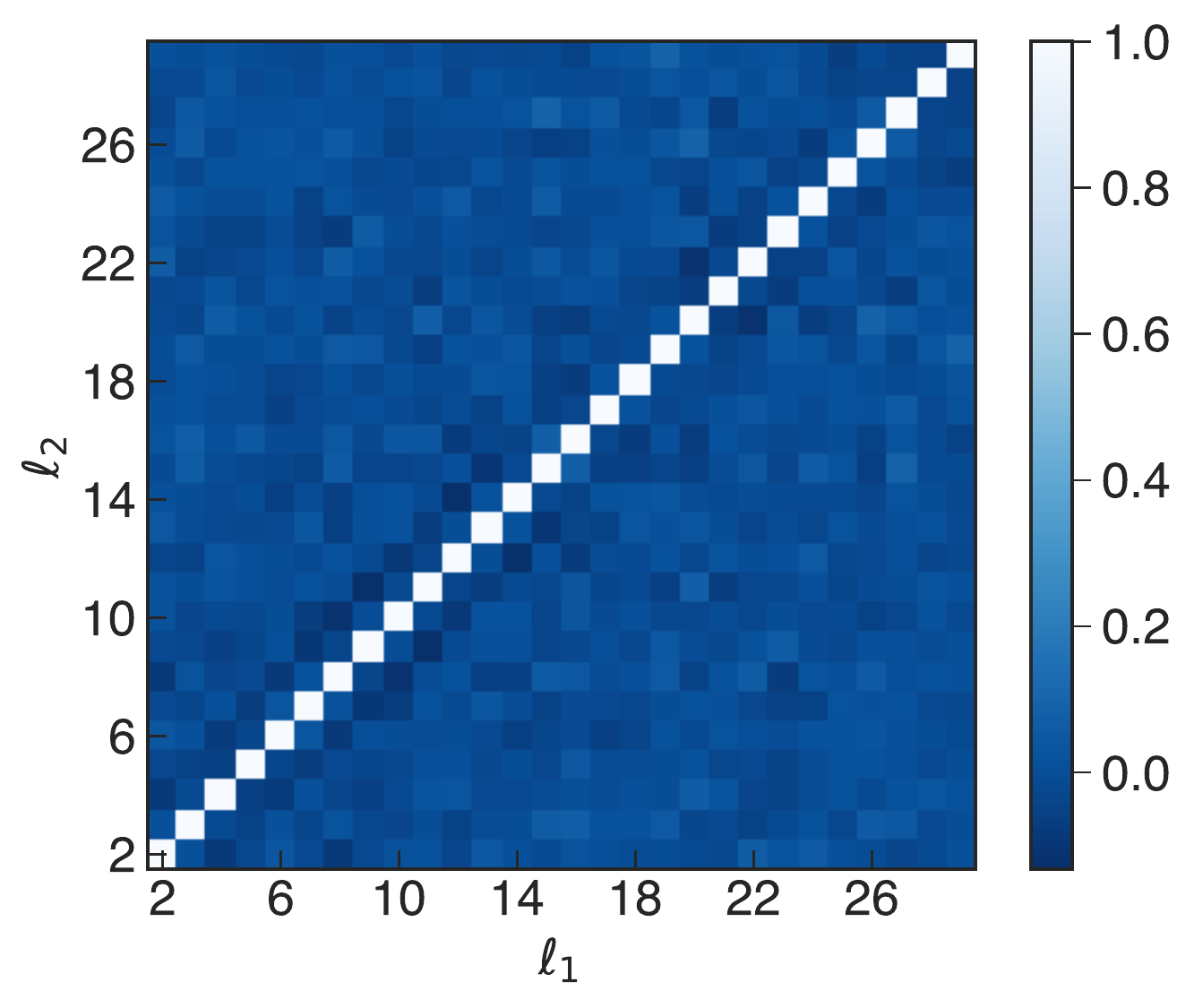}
\includegraphics[width=0.48\textwidth]{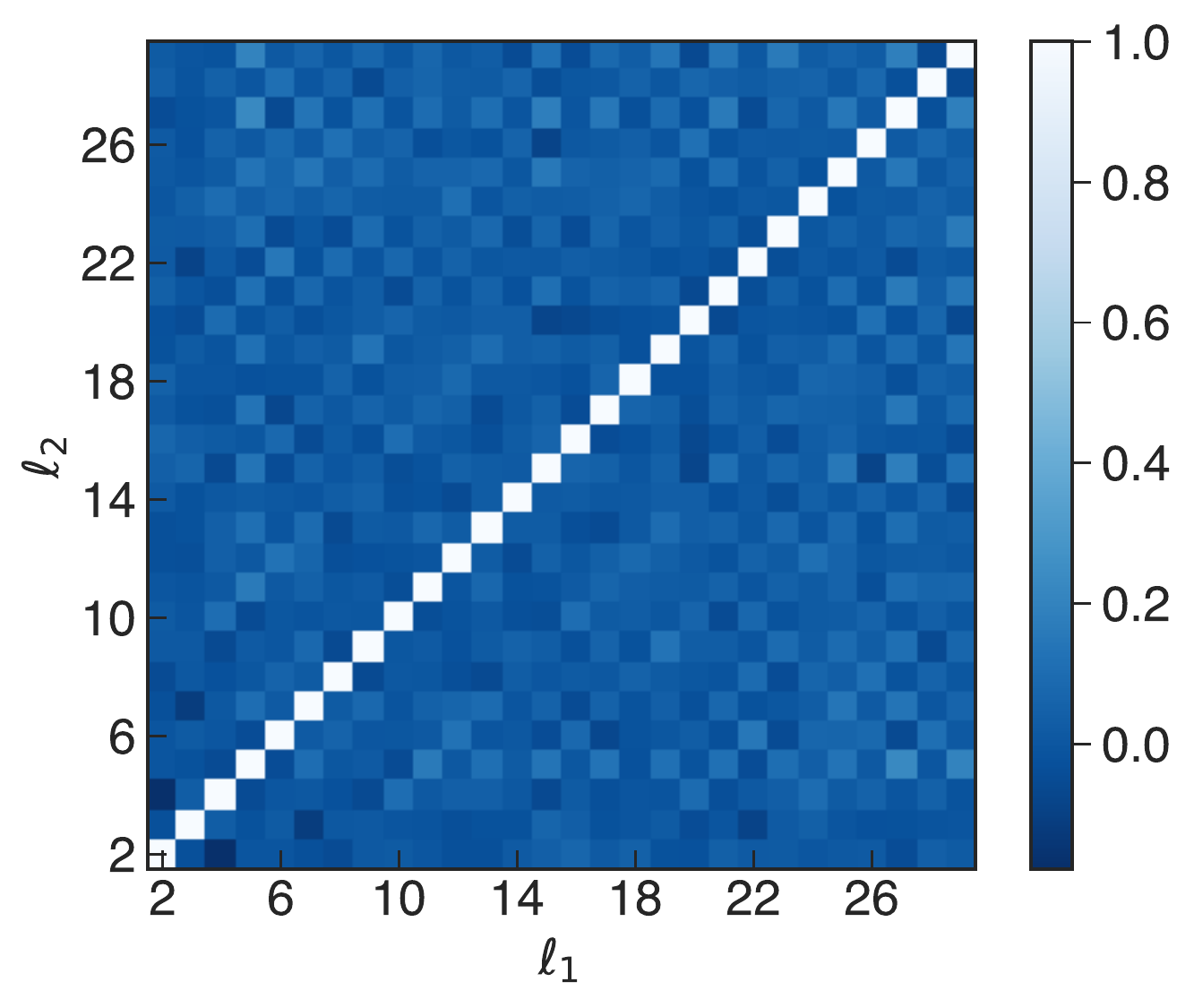}
\includegraphics[width=0.48\textwidth]{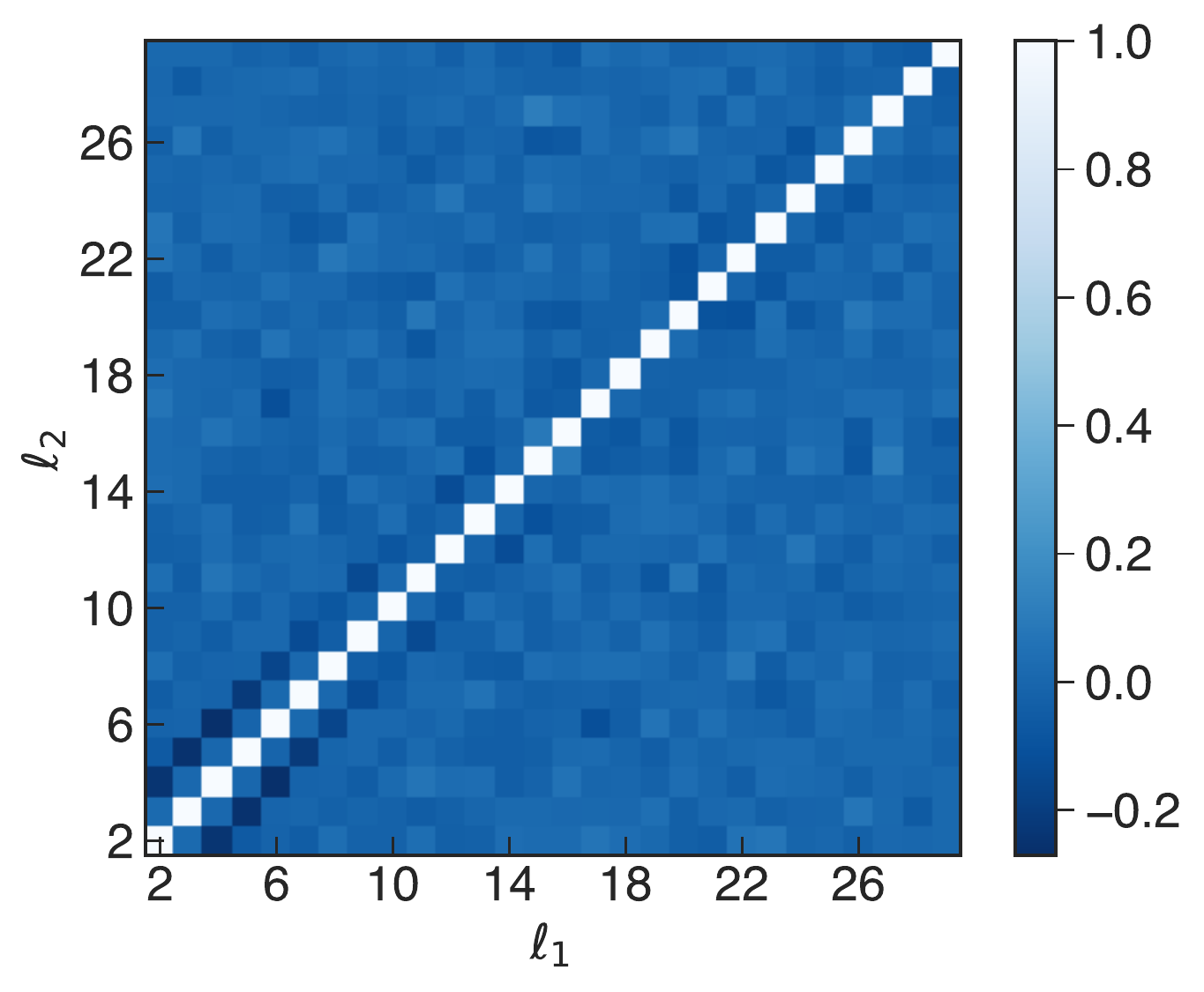}
\caption{Normalized covariance matrices $\mathcal{C}_{\ell, \ell'} = {\rm cov}(\hat{C}_\ell, \hat{C}_{\ell'}) /\sqrt{{\rm var}(\hat{C}_\ell){\rm var}(\hat{C}_{\ell'})}$ of the three QML methods for the space-based experiment. The matrices are obtained from estimates of 1000  simulations for classic QML estimator (left) and for QML-SZ estimator (right). The upper diagrams show the covariance matrices of $r=0.05$ case, the lower diagrams show equivalent plots of $r=0$ case. }
\label{fig:norm_cov_72_homo}
\end{figure*}
%************************************************************

%*********************************************************************%
\begin{figure}[ht]
\centering
\includegraphics[width=0.50\textwidth]{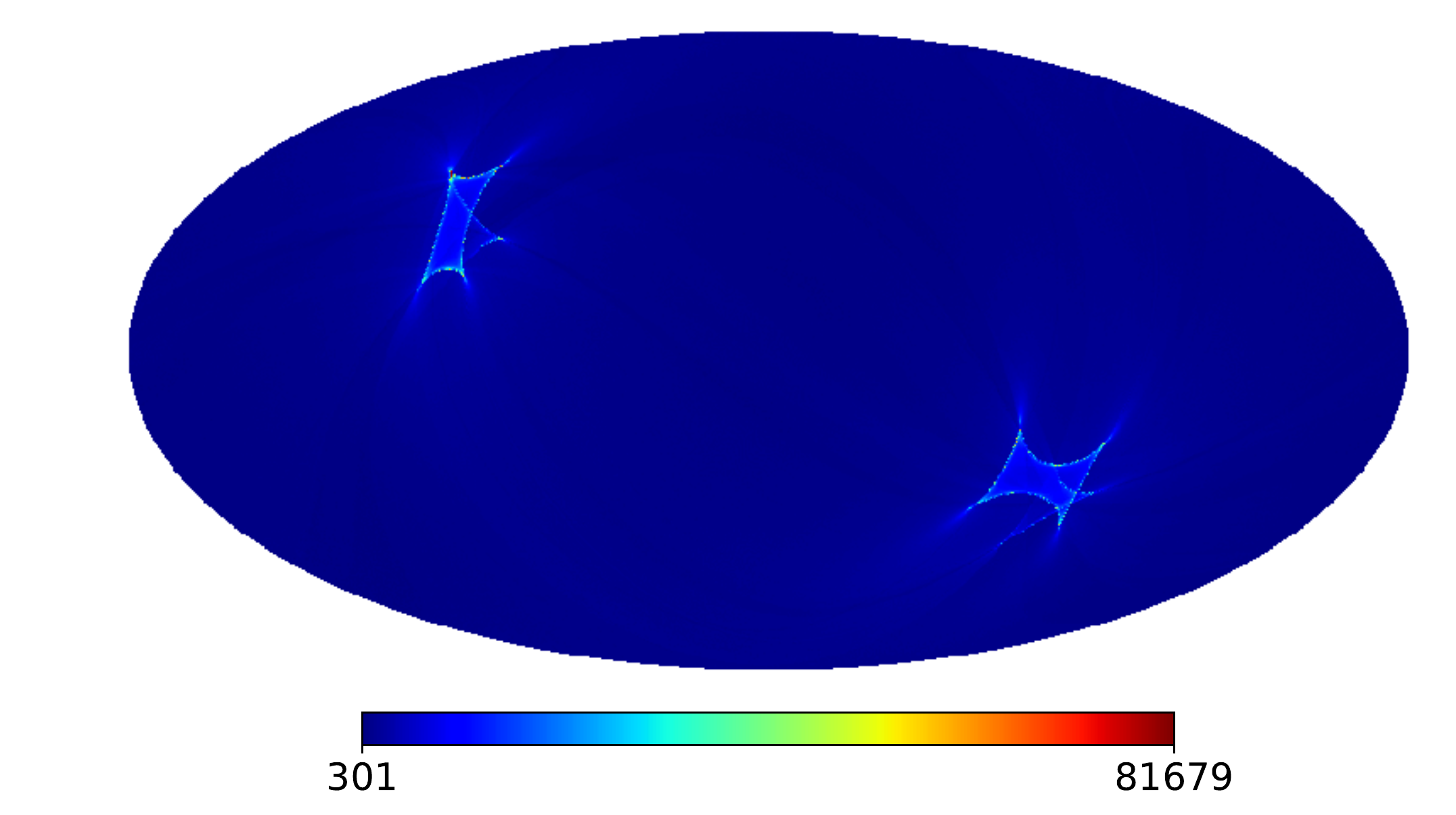}
 \caption{A hit map in Galactic coordinates of Planck 100GHz channel, obtained for Nside = 2048. Pixel values quantify the number of observations of the pixel. Areas near the ecliptic poles are observed several times more frequently than regions of the sky near the ecliptic plane. The lowest values are about 301 while the highest are about $8.2\times 10^4$.}
\label{fig:noise_map_bigfsky} 
\end{figure}
%**********************************************************************%

%*********************************************************************%
The results for the impact of \texttt{ud\_grade} are shown in figures \ref{fig:downgrade_impact_72} and \ref{fig:downgrade_impact_12} for the space-based and ground-based experiment cases, respectively. For both sky patches, we find that downgrading the map has negligible impact on the mean estimates. Thus, this process does not generate any additional biases to the QML estimators. For the space-based experiment case, the error bars for $\ell \le 5$ show a small increase for the QML estimator while showing significant increase for the QML-SZ estimator. This behaviour can be explained by the power spectrum of QML-SZ method, which is given by $N^2_{\ell,2}C_{\ell}^{BB}$. By downgrading the map, the power from high multipoles leaks to the lower multipoles and increases the uncertainty on the large angular scales. We also find that errors for the standard QML estimator slightly increase throughout the multipole range, though the errors are still near optimal. For the ground-based experiment, we find negligible impact of the downgrading process on the error bars.

% effect_of_down_grade (end)

\subsection{Impact of noise level} % (fold)
\label{sub:effect_of_noise}

In every CMB experiment, we have to suitably mitigate the impact of the noise. The noise level depends on various factors, such as, instrument sensitivity, survey duration, survey strategy, etc. As mentioned before, the RMS noise levels can vary due to various factors, but for this work we test three cases: 10 $\mu$K-arcmin, 1 $\mu$K-arcmin, and 0.1 $\mu$K-arcmin, the latter one acting as the noise free limit of our estimator performance.

For this study, we simulated the different noise maps at the targeted  $\texttt{NSIDE}=16$ for the large sky fraction patch and $\texttt{NSIDE}= 32$ for the small sky patch. The RMS noise levels are suitably converted to get the noise variance per pixel at one of those \texttt{NSIDE} values. This is used to generate Gaussian white noise. The CMB maps are simulated at the targeted \texttt{NSIDE} with $\ell_{\rm max}$ of 32 and 96. The two maps are then co-added and masked to produce the input maps for the QML estimators.

In figures \ref{fig:noise_levels_72} and \ref{fig:noise_levels_12}, we show the results for the power spectra estimated with the QML methods for the large and small sky fraction patches. In these figures, we compare the errors with the cosmic variance limit that can be obtained from equation \ref{eq:cosmic_var} by setting $\mathcal{N}^{BB}_\ell$ to zero. We can see that the variation in the noise level does not affect the mean of the power spectra estimates. This is expected as we have the $b^r_\ell$ term in equation \ref{QML_yrl} to debias for the noise. We also find that the noise level will impact the error bars, but at a low noise level it almost behaves as the cosmic variance limited case. Considering the same noise level, for every tested case, all the QML estimators behave very similar to each other.

% subsection impact of noise levels (end)

% %*********************************************************************%

\section{REALISTIC EXAMPLES} % (fold)
\label{sec:realistic_examples}
In the previous section, we have performed some idealized tests for three QML estimators. In this section, we consider more realistic simulations, which show a precise implementation of the pipeline to obtain the power spectrum estimates from CMB observations with all three estimators. Finally, we will obtain a detailed comparison of the computational requirements for these three methods. Here, we are comparing the performance of our estimators in four situations: satellite and ground-based experiments, both with homogeneous noise or inhomogeneous noise.

%Here, we are comparing the estimator performances in the following situations: satellite experiment with homogeneous noise and inhomogenous noise, ground-based experiment with homogeneous noise and inhomogeneous noise. 

To simulate the CMB sky, we use the 2018 Planck cosmological parameters as given by \citet{Planck2018VI} for the $E$-mode input signal. For the $B$-mode input signal, we include lensing and primordial $B$-modes with both $r = 0.05$ and $r = 0$, which represent the upper and lower limits on $r$. 

Here, we will outline the common simulation setup for our realistic examples. We simulate full sky CMB realizations at $\texttt{NSIDE}=512$ with $\ell_{\rm max}=1024$ for both $r=0.05$ and $r=0$ by using the \texttt{synfast} subroutine of HEALPix.  To simulate the noise map, we consider two different cases. For the homogeneous noise case, the RMS white noise level for our realistic examples are set to 3 $\mu$K-arcmin. This equates to a white noise level of 0.44 $\mu$K-pixel at $\texttt{NSIDE}=512$. We simulate Gaussian white noise at $\texttt{NSIDE}=512$ on full sky. For the inhomogeneous noise case, we use different ways to generate noise maps for the space-based and ground-based experiments (see subsections \ref{sub:satellite_example_inhomo} and \ref{sub:ground_example_inhomo} for details). The signal and noise maps are finally coadded to generate our `observed' CMB map.

%For inhomogeneous noise case,  the space-based experiment and ground-based experiment use different way to generate noise maps and the details can be found in subsection \ref{sub:satellite_example_inhomo} and subsection \ref{sub:ground_example_inhomo}. Then the signal and noise maps are coadded to generate the 'observed' CMB map.
%we simulate inhomogeneous noise map using the noise level per pixel shown in Fig.\ref{fig:noise_map} at \texttt{NSIDE}=512. In Fig.\ref{fig:noise_map} the value of a pixel is the noise standard variance at that pixel in unit $\mu$K. 

In the satellite-based experiment case, we use the HEALPix \texttt{smoothing} subroutine to smooth the `observed' CMB map with FWHM=$8^{\circ}$ to suppress higher multipoles. While, for the ground-based experiment, the coadded maps are multiplied by a particular binary mask (see Fig.\ref{sky_coverage}) to keep only the fraction of sky observed in the considered experiment. This produces our simulated CMB observations from the two experimental setups considered.

%************************************************************

In the next step, we first need to prepare our scalar pure-$B$ map, $\mathcal{B}(\hat n)$, at $\texttt{NSIDE}=512$ by using a Gaussian apodized mask with $\sigma = 10^{-6}$, $\delta_c =1^{\circ}$ for Planck mask and $\sigma = 10^{-4}$, $\delta_c =0.5^{\circ}$ for AliCPT mask in Eq. (\ref{QML_SZ_Gaussian_window_function}). We use the expression \ref{eq:pureB_implement} for this computation. Similarly, we use the template cleaning algorithm detailed in section \ref{sub:qml_template_cleaning_estimator} to produce a leakage template cleaned scalar $B$-mode map. Thus, we derive $TQU$ maps for the QML method, a scalar $\mathcal{B}$-map for the QML-SZ method, and a template cleaned $B$-map for the QML-TC. As we have discussed in section \ref{sub:PCL_estimator}, we intend to compare the results of these QML methods with the PCL estimators. For PCL estimators, we smooth the masked $TQU$ map with a Gaussian smoothing with FWHM$=20'$ and directly analyze these maps without the downgrading process.

%In the next step, we first need to prepare our scalar maps at \texttt{NSIDE}=512. To prepare our scalar pure-$B$ map, $\mathcal{B}(\hat n)$, we use a Gaussian apodized mask with $\sigma = 10^{-6}$, $\delta_c =1^{\circ}$ for Planck mask and $\sigma = 10^{-4}$, $\delta_c =0.5^{\circ}$ for AliCPT mask in Eq. (\ref{QML_SZ_Gaussian_window_function}). We use the expression \ref{eq:pureB_implement} for this computation.

\subsection{Space-based experiment: homogeneous noise} % (fold)
\label{sub:satellite_example}
%*********************************************************************%
\begin{figure*}[th]
\centering
\includegraphics[width=\textwidth]{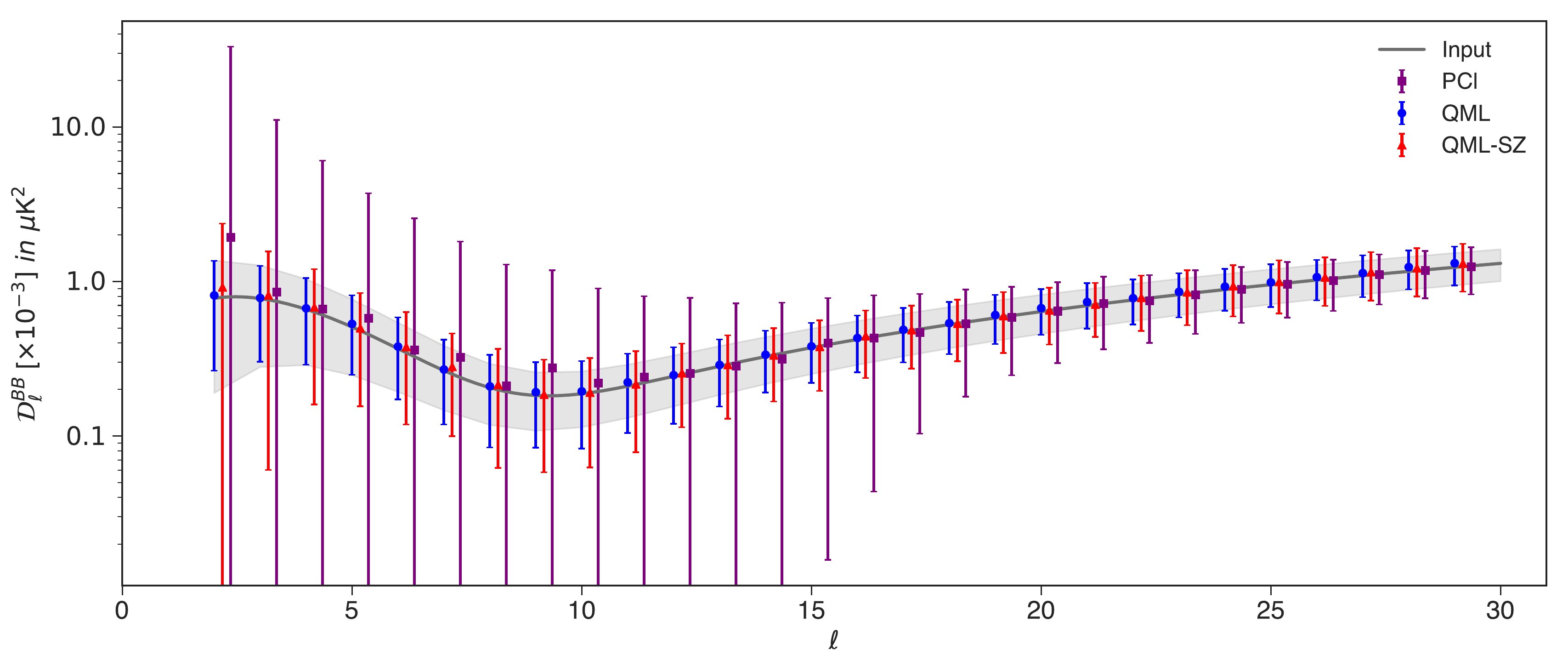}
\includegraphics[width=\textwidth]{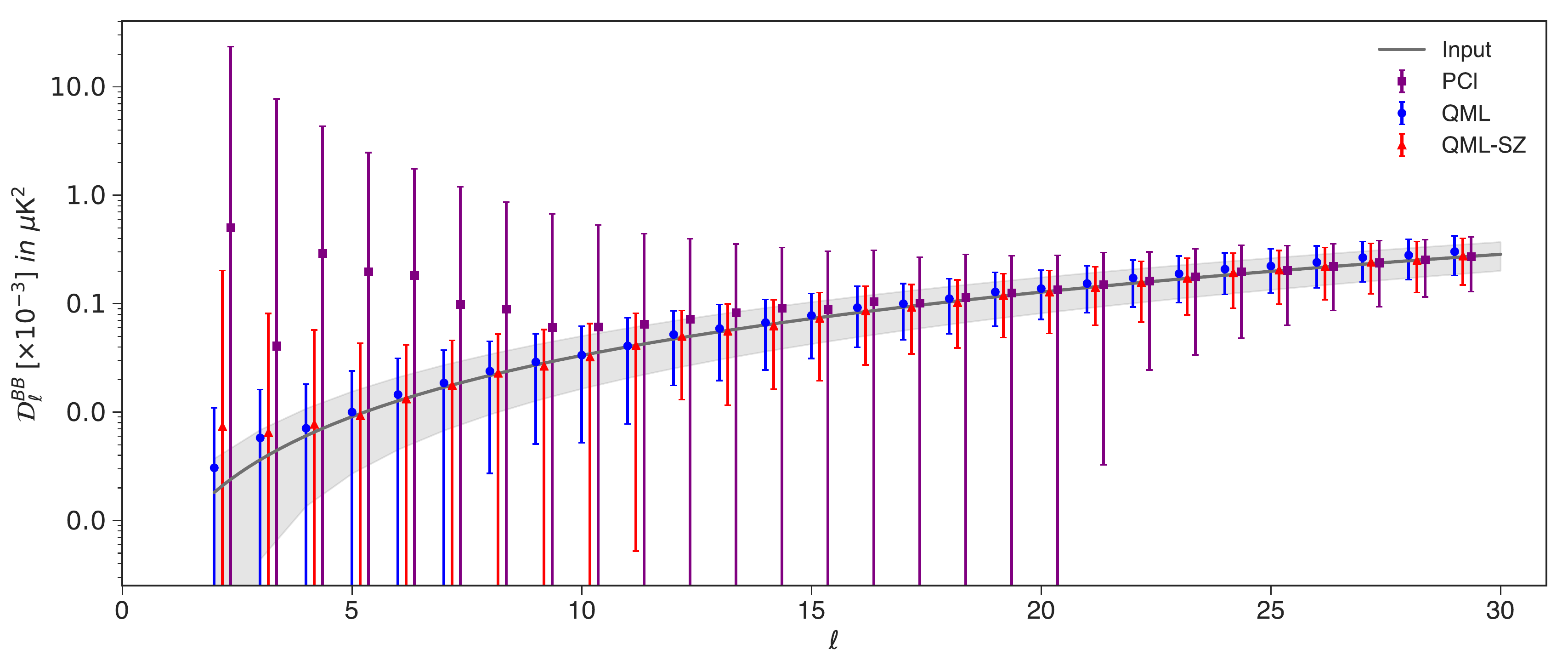}
\caption{Plot of the results for $B$-mode power spectrum estimates for the realistic space-based CMB experiment, the upper panel and the lower panel are $r=0.05$ and $r=0$ respectively. The observed sky with inhomogeneous noise, is simulated at \texttt{NSIDE}=512 with $\ell_{\rm max}$=1024. The input $B$-mode power spectrum is shown with the black solid curve. The classic QML method results are shown with blue, square markers, QML-SZ method results with red, circular markers. These results are computed at \texttt{NSIDE}=16 with $\ell_{\rm max}$=47. We also show PCL estimator results, obtained with NaMaster, with $\delta_c = 6^\circ$ C2 apodization, with purple, inverted triangle markers. The gray region denotes the analytical approximation of the error bounds. The data points are the mean of 1000 estimates, and the error bars are given by the standard deviation of the estimates.}
\label{fig:realex_72_inhomo}
\end{figure*}
%*********************************************************************%

%************************************************************
\begin{figure*}[ht]
\centering
\includegraphics[width=0.48\textwidth]{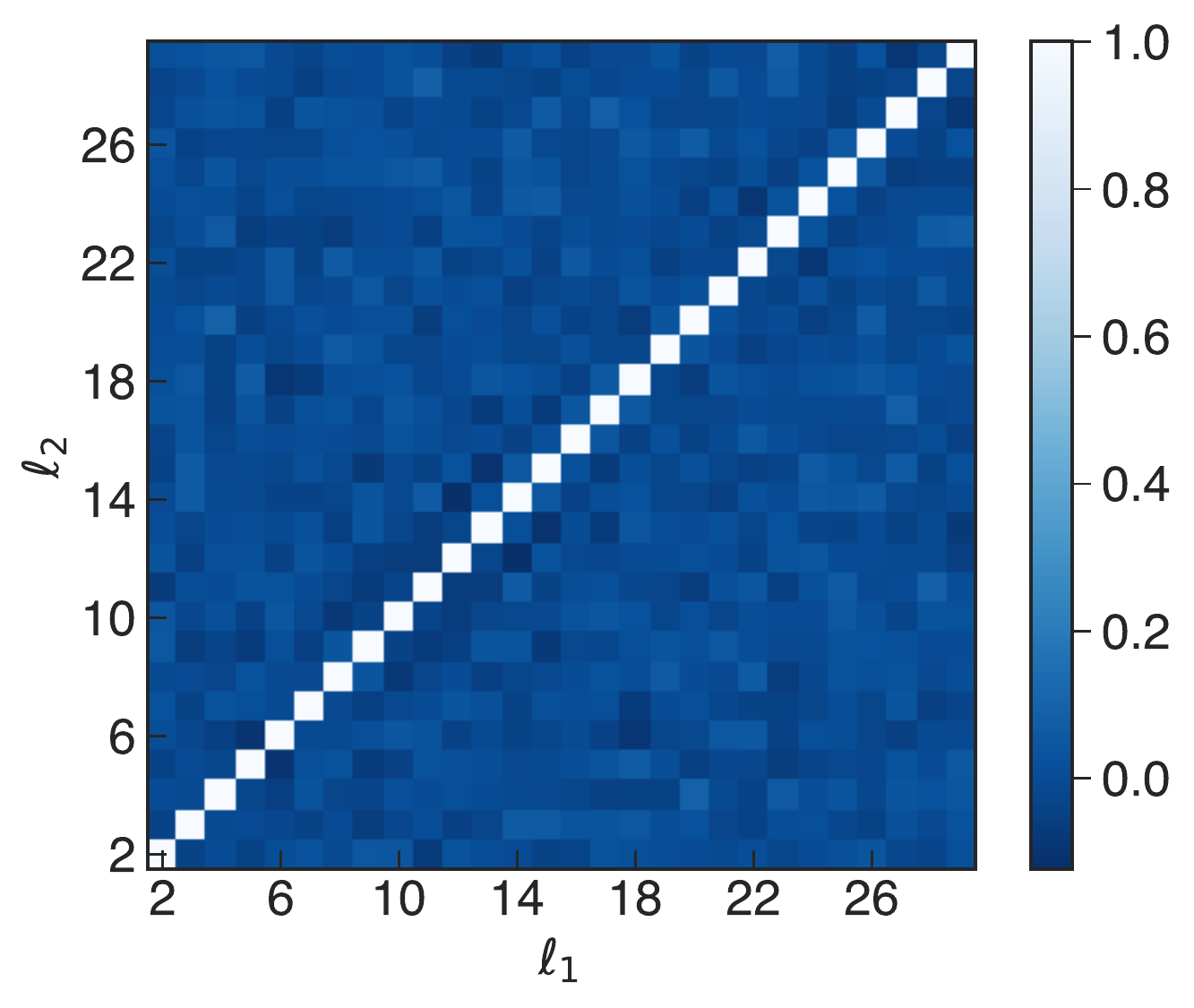}
\includegraphics[width=0.48\textwidth]{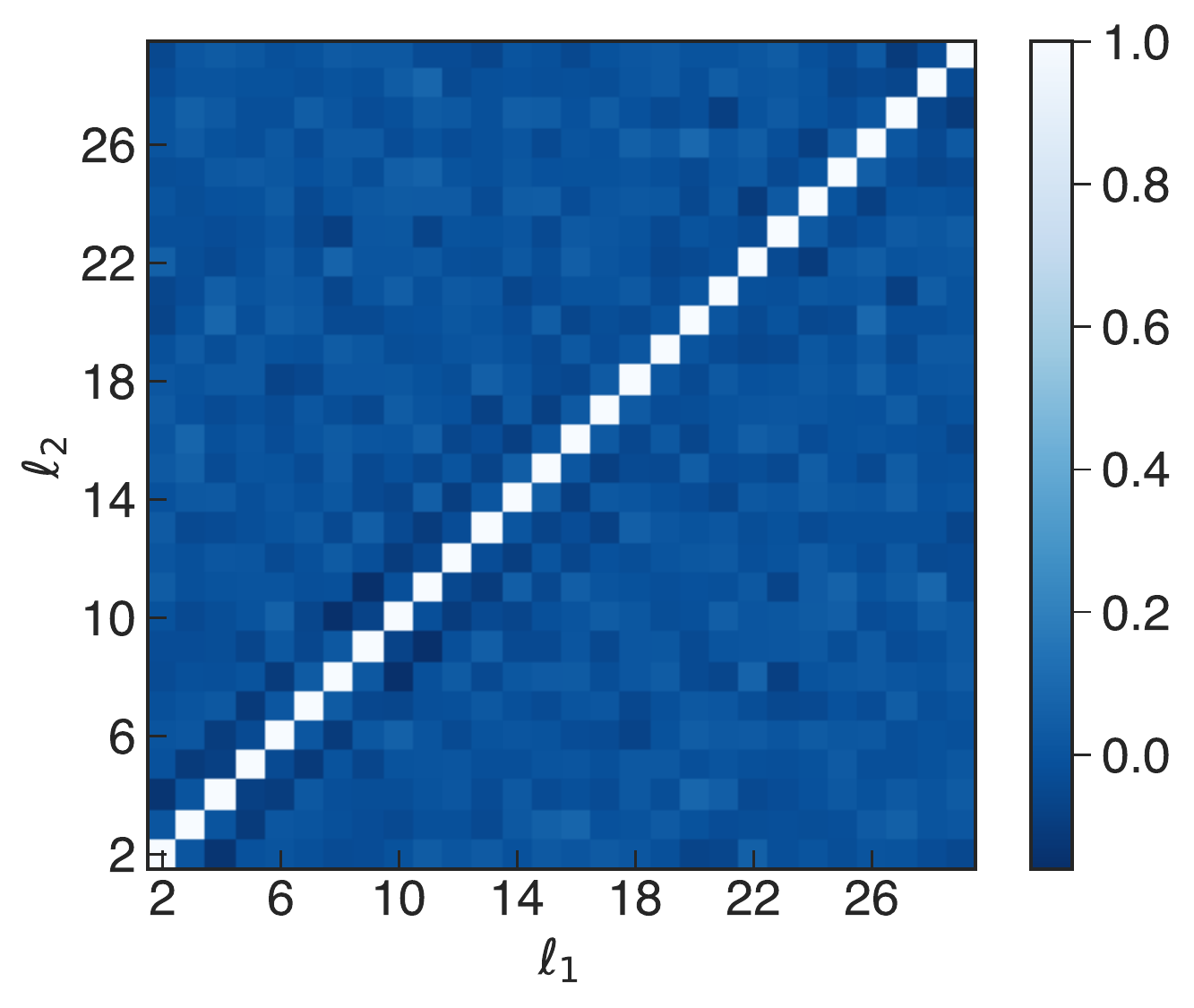}
\includegraphics[width=0.48\textwidth]{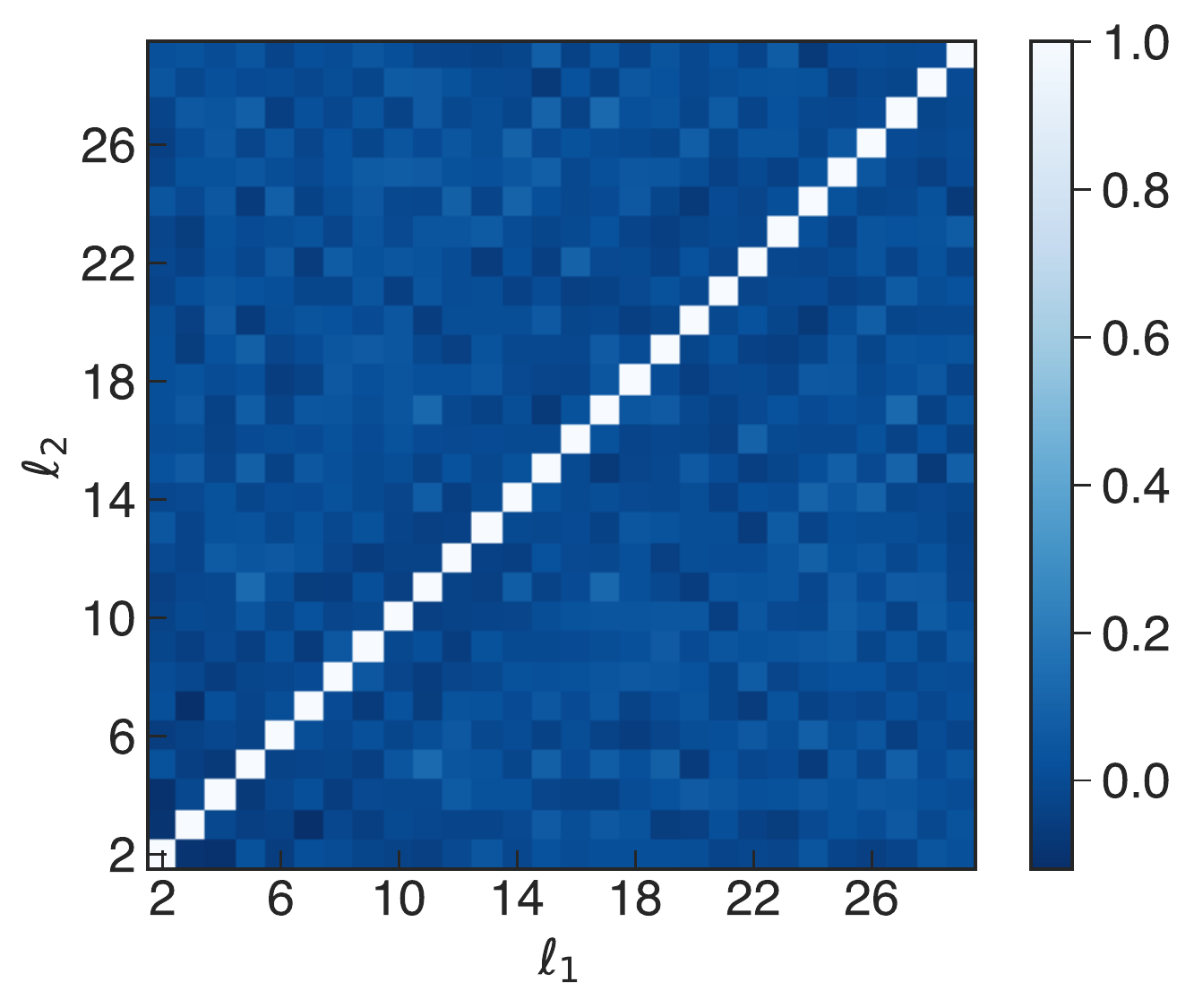}
\includegraphics[width=0.48\textwidth]{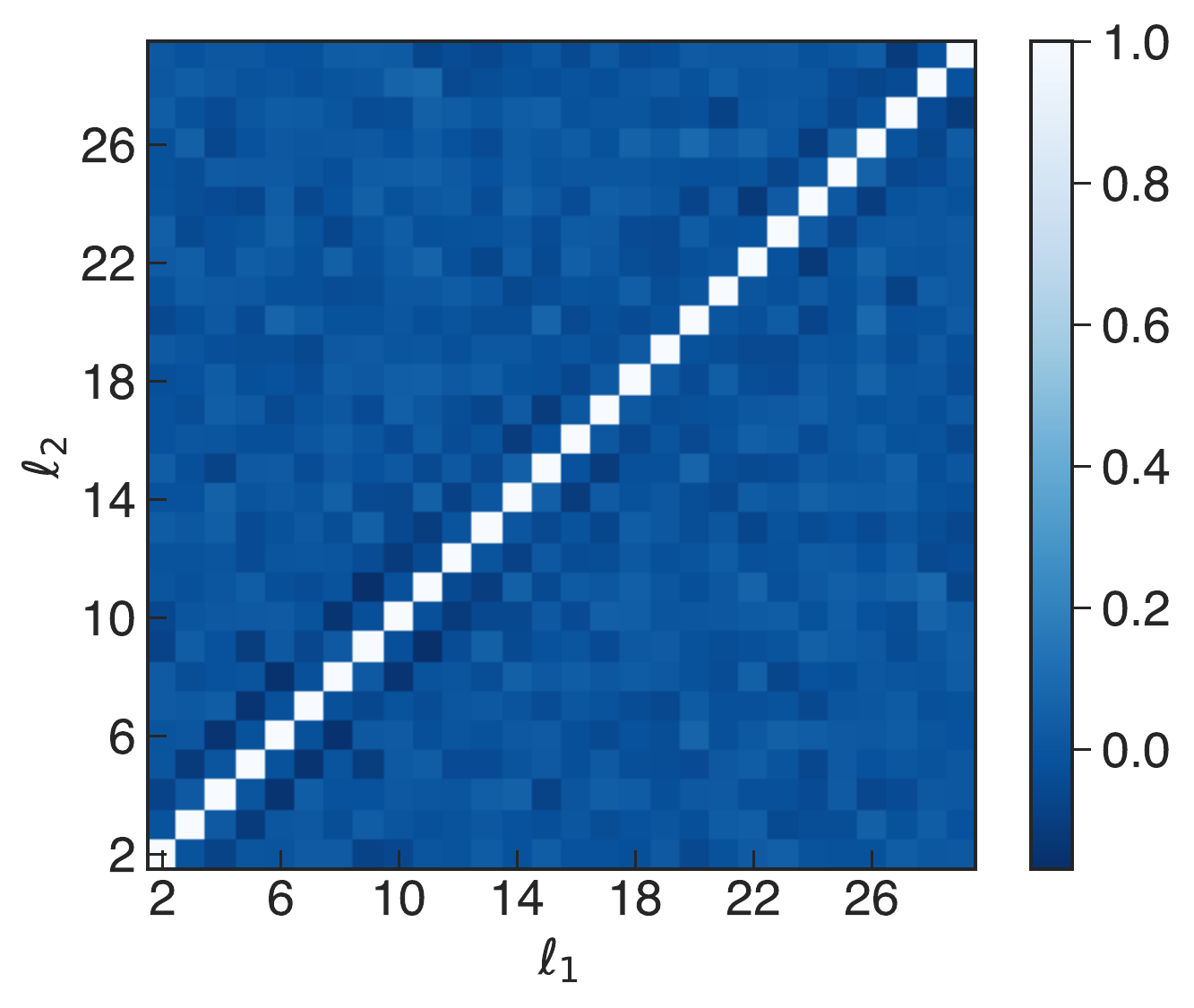}
\caption{Normalized covariance matrices $\mathcal{C}_{\ell, \ell'} = {\rm cov}(\hat{C}_\ell, \hat{C}_{\ell'}) /\sqrt{{\rm var}(\hat{C}_\ell){\rm var}(\hat{C}_{\ell'})}$ of the three QML methods for the space-based experiment with inhomogeneous noise. The matrices are obtained from estimates of 1000  simulations for classic QML estimator (left) and for QML-SZ estimator (right). The upper diagrams show the covariance matrices of $r=0.05$ case, the lower diagrams show equivalent plots of $r=0$ case. }
\label{fig:norm_cov_72_inhomo}
\end{figure*}
%************************************************************
The space-based experiments have a major advantage of being able to observe the full sky. However, the Galactic plane must be masked out due to the strong polarized foreground contribution from our galaxy. Even though part of the sky must be removed to avoid the astrophysical Galactic contamination, satellite experiments are our best bet for observing the largest angular modes. For the satellite case, we then set our target to $\texttt{NSIDE}=16$ and $\ell_{\rm max}=32$ for all three QML estimators.  The $QU$ and scalar maps at $\texttt{NSIDE}=512$ are downgraded to the targeted $\texttt{NSIDE}=16$, using \texttt{ud\_grade} Healpix subroutine, as well as the masks, setting any pixel with values $<0.99$ to zero. Note that in this section, for QML-SZ estimator, we downgrade the apodized mask instead of the binary mask as SZ method uses an apodized mask. We multiply this downgraded mask to the downgraded maps. These are our input maps for QML estimators.  We will compare these results with the PCL estimator results. 

%as shown in the upper panel of Fig.\ref{mask_effect_72}--> check figure reference

%However, uncertainties arising from foreground removal mean the patches of the sky obscured by galactic polarized foreground needs to be masked at the time of analysis. The satellite CMB experiment mask used in this work is representative of such sky coverage. Despite masking for the galactic foregrounds, satellite experiments are our best bet for observing the largest angular modes. For this example, we set our target \texttt{NSIDE}=16 and $\ell_{\rm max}$=32 for all three QML estimators.  The $QU$ and scalar maps at \texttt{NSIDE}=512 are downgraded to the target \texttt{NSIDE}=16, using \texttt{ud\_grade} facility. We also downgrade the masks with \texttt{ud\_grade} facility and set any pixels with values $<0.99$ to zero. Note that in this section, for QML-SZ estimator, we downgrade the apodized mask instead of the binary mask as SZ method uses apodized mask. We multiply this downgraded mask to the downgraded maps. These are our input maps for QML estimators.  We will compare these results with the PCL estimator results. 

%************************************************************

The results for this first case are shown in Fig.\ref{fig:realex_72_homo} for both $r=0.05$ and $r=0$. We plot the mean of the power spectra estimates from 1000 simulations with error bars given by the standard deviation of these samples. In addition, we plot the results of pure $B$-mode PCL estimator for the same case. The PCL results are obtained with C2 apodization with $\delta_c$ of $6^\circ$. We find that all the QML methods outperform the PCL results in the entire multipole range. We notice that while the standard QML method has nearly-optimal error bars throughout the entire multipole range, the QML-SZ method has sub-optimal error bars for the lowest multipoles because we downgrade the input map, as discussed in section \ref{sub:effect_of_down_grade}.

On the other hand, all QML estimators are tested with the same binary mask as defined in section \ref{sub:effect_of_down_grade}. However, for QML-SZ estimator,  we use the Gaussian apodized window to replace the binary mask, which causes the effective $f_{\rm sky}$ of the QML-SZ estimator to be smaller than the standard QML one and enlarge the uncertainties of the QML-SZ estimator for the entire multipole range. While the performance of the QML-SZ estimator is not as good as the standard QML method, it is still a fast and reliable solution for power spectrum estimation, except for the lowest few multipoles.

%The QML methods give the unbiased estimates for the entire multipole range, which is hard for PCL estimator at low multipoles due to the much larger error bars.  

%As discussed in section \ref{sub:effect_of_down_grade}, this effect is caused by downgrading the map with \texttt{ud\_grade}.

%, which means all tests in section \ref{sub:effect_of_down_grade} have same $f_{\rm sky}$

%, the upper panel and the lower panel are $r=0.05$ and $r=0$, respectively. 

%Therefore, the downgrading of the $\mathcal{B}$-map is the primary reason for suboptimal error bars for the QML-SZ method. In fact, if we ignore the result at $\ell=2$ of QML-SZ method, it is still a reliable and fast way to reconstruct the $B$-mode power spectra at large scale situation. 

Due to the partial-sky analysis, the coupling between different multipole is inevitable. In order to quantify it, we calculate the normalized covariance matrices defined as
\begin{equation}
    \mathcal{C}_{\ell \ell'} = \frac{{\rm cov}\left(\hat{C}^{BB}_\ell, \hat{C}^{BB}_{\ell'}\right)}{\sqrt{{\rm var}(\hat{C}^{BB}_\ell){\rm var}(\hat{C}^{BB}_{\ell'})}},
    \label{eq:cov_mat}
\end{equation}
and present the results in Fig.\ref{fig:norm_cov_72_homo}. For the space-based experiment considering homogeneous noise, the power spectra estimates only weakly couple among different multipoles for every QML method tested here, with the covariance matrices being approximately diagonal.
% \vfill\null

\subsection{Space-based experiment: inhomogeneous noise} % (fold)
\label{sub:satellite_example_inhomo}

In this subsection, we will study the performance of the QML methods for a satellite experiment with inhomogeneous noise. We generate inhomogeneous noise maps using the hitmap for Planck HFI 100 GHz channel (shown in Fig.\ref{fig:noise_map_bigfsky}) following the prescription given in \citet{2013MNRAS.429.2104D}. We set the white noise level of these maps ($\sigma_{\rm isotropic noise}$ of \citet{2013MNRAS.429.2104D}) to 5 $\mu$K-arcmin. All the calculation steps and parameters are consistent with the last subsection. We will still compare the results for the QML methodology with the ones for PCL estimator.

The results for this case are shown in Fig.\ref{fig:realex_72_inhomo}, where the upper and lower panels represent $r=0.05$ and $r=0$, respectively. Comparing the results for the different methods with inhomogeneous noise, we can find that the different QMLs still outperforms the PCL in our entire multipole range. The standard QML method still have nearly-optimal error bars throughout the entire multipole range, while the QML-SZ method has slightly larger error bars for the lowest multipoles. 
% {\color{red}We note that the results showed for the inhomogeneous noise case use an isotropic noise level $\sigma_\text{isotropic noise}$ of 5 $\mu$K-arcmin which is higher than the noise level we considered in the homogeneous noise case. So with increased isotropic noise level and with inhomogeneous noise model, both the QML methods has greater advantage over the PCL method.}

%In addition to the different noise model used in this subsection, the white noise level for the inhomogeneous noise case is higher than that for homogeneous noise case. Comparing the results for different noise models, we find that as the noise level increases, the QML method has a bigger advantage for the $r=0$ case.

%All QML methods give unbiased estimates for the entire multipole range, which is not possible with the PCL estimator at low multipoles

We also show the normalized covariance matrices for the power spectra estimators with the QML methods for this case in Fig.\ref{fig:norm_cov_72_inhomo}. Similarly, we find that the covariance matrices are approximately diagonal, which indicates that the cross-correlations between different modes are weak for the presented QML methods.  

% \vfill\null
\subsection{Ground-based experiment: homogeneous noise} % (fold)
\label{sub:ground_example_homo}

Ground-based CMB experiments cannot account for full sky observations, since they are limited by their geographical location in terms of the total sky area available for survey. On the other hand, they have longer mission plans, during which they undergo instrumental upgrades allowing for higher sensitivity. For ground-based survey, $f_{\rm sky}$ is usually quite small. For most cases, we have $f_{\rm sky}\lesssim 10\%$, which allows us to choose larger \texttt{NSIDE}, in comparison with the ones chosen for the space-based experiment case.
%as $N_d$ is much smaller than the space-based experiment case. 
%*********************************************************************%
\begin{figure}[th]
\centering
\includegraphics[width=0.45\textwidth]{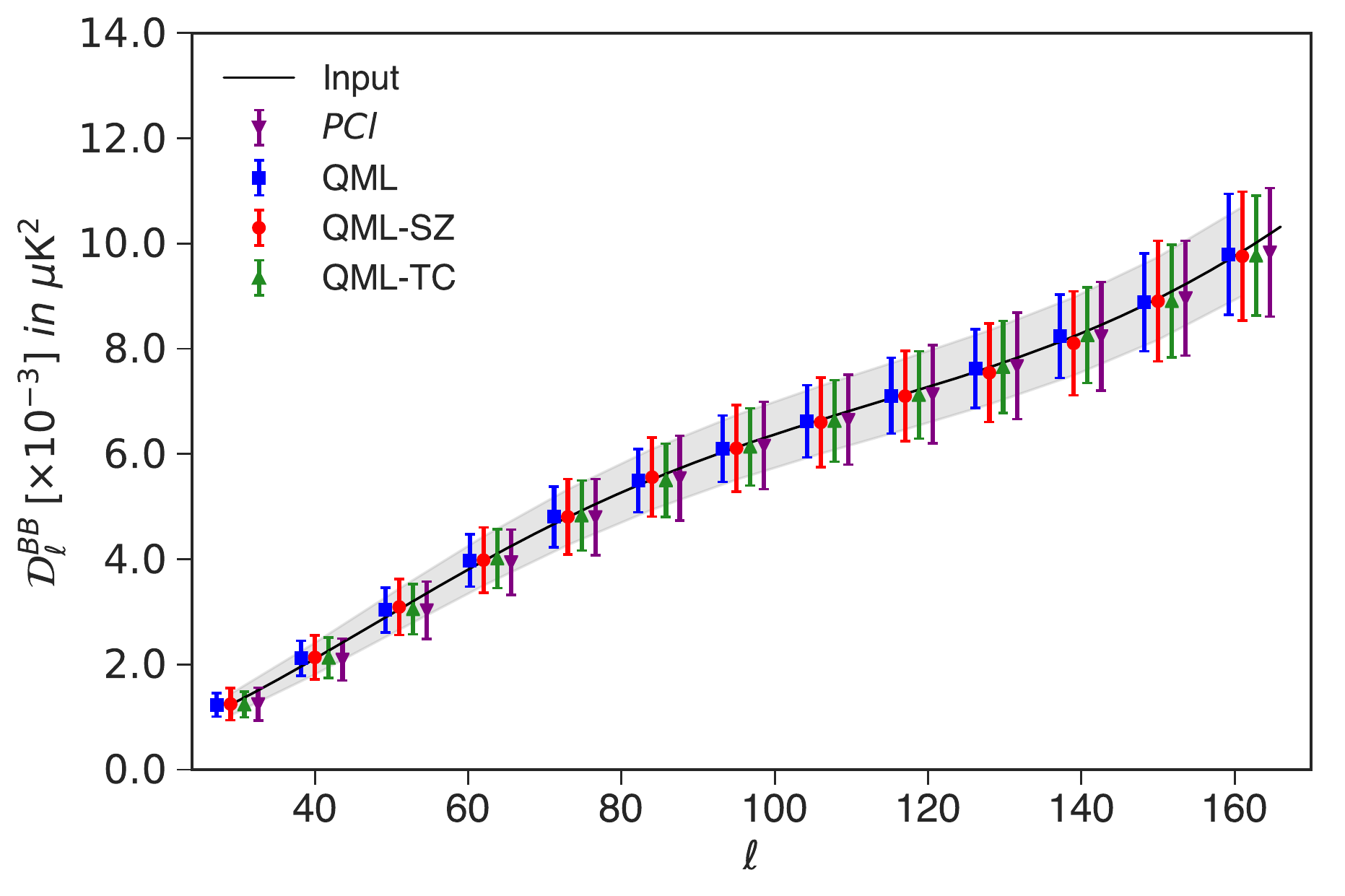}
\includegraphics[width=0.45\textwidth]{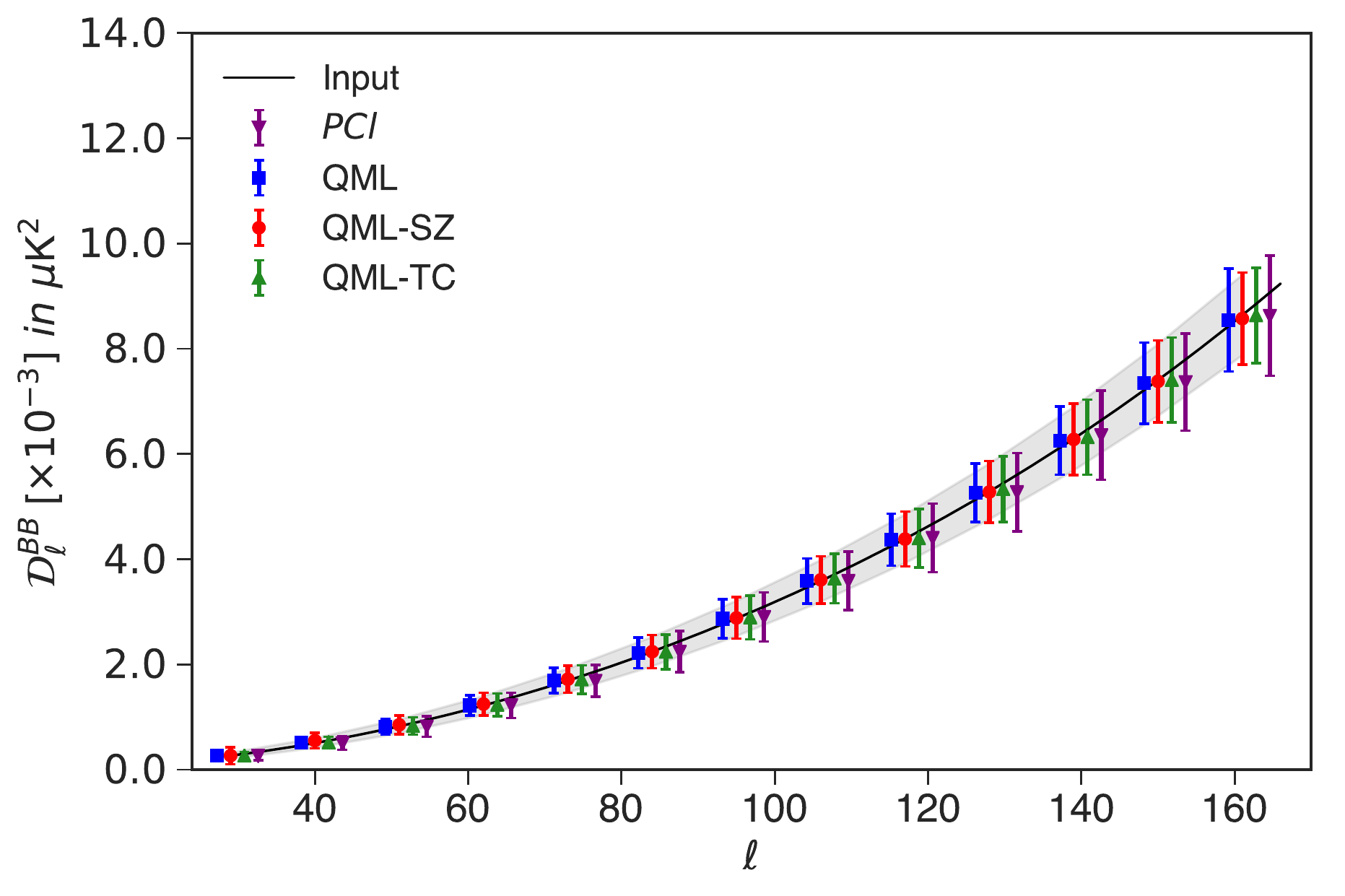}
\caption{Plot of the results of $B$-mode power spectrum estimates for realistic ground-based CMB experiment with homogeneous noise, the upper panel and the lower panel are $r=0.05$ and $r=0$ respectively. The observed sky with 3 $\mu$K-arcmin noise, is simulated at \texttt{NSIDE}=512 with $\ell_{\rm max}$=1024. The input $B$-mode power spectrum is shown with the solid, black curve. The classic QML method results are computed at \texttt{NSIDE}=64 with $\ell_{\rm max}$=192 (blue, square markers), QML-SZ method results are computed at \texttt{NSIDE}=64 with $\ell_{\rm max}$=192 (red, circle markers), and QML-TC method results are also computed at \texttt{NSIDE}=64 with $\ell_{\rm max}$=192 (green, triangle markers). We also show PCL estimator results, obtained with NaMaster, using $\delta_c = 6^\circ$ for $r=0.05$, and $\delta_c = 10^\circ$ for $r=0$,  (purple, inverted triangle markers). The gray region denotes the optimal error bounds. The data points are mean of 1000 estimates and the error bar is given by the standard deviation of the estimators.} 
\label{fig:realex_10} 
\end{figure}
%**********************************************************************%
%************************************************************
\begin{figure*}[ht]
\centering
\includegraphics[width=0.33\textwidth]{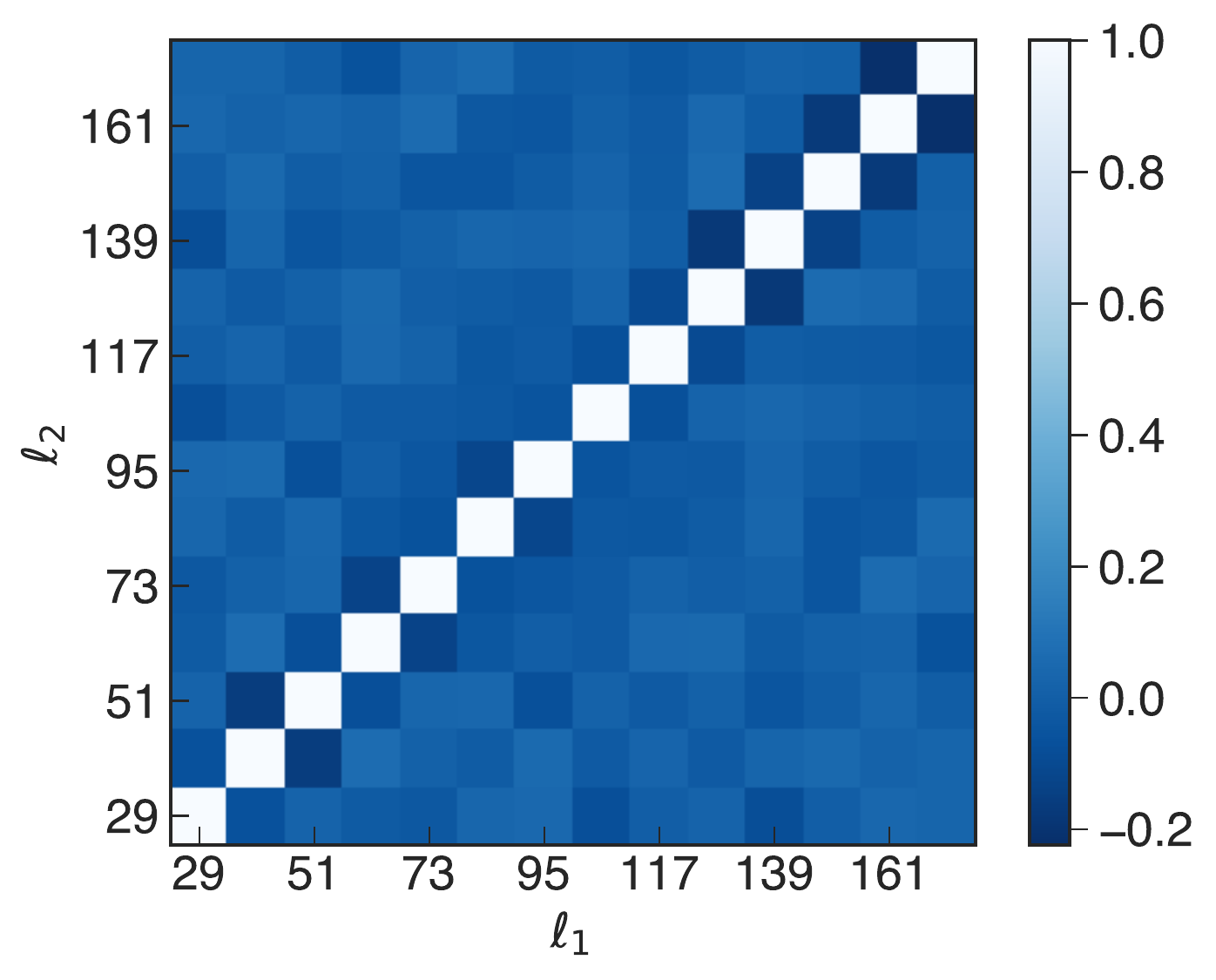}
\includegraphics[width=0.33\textwidth]{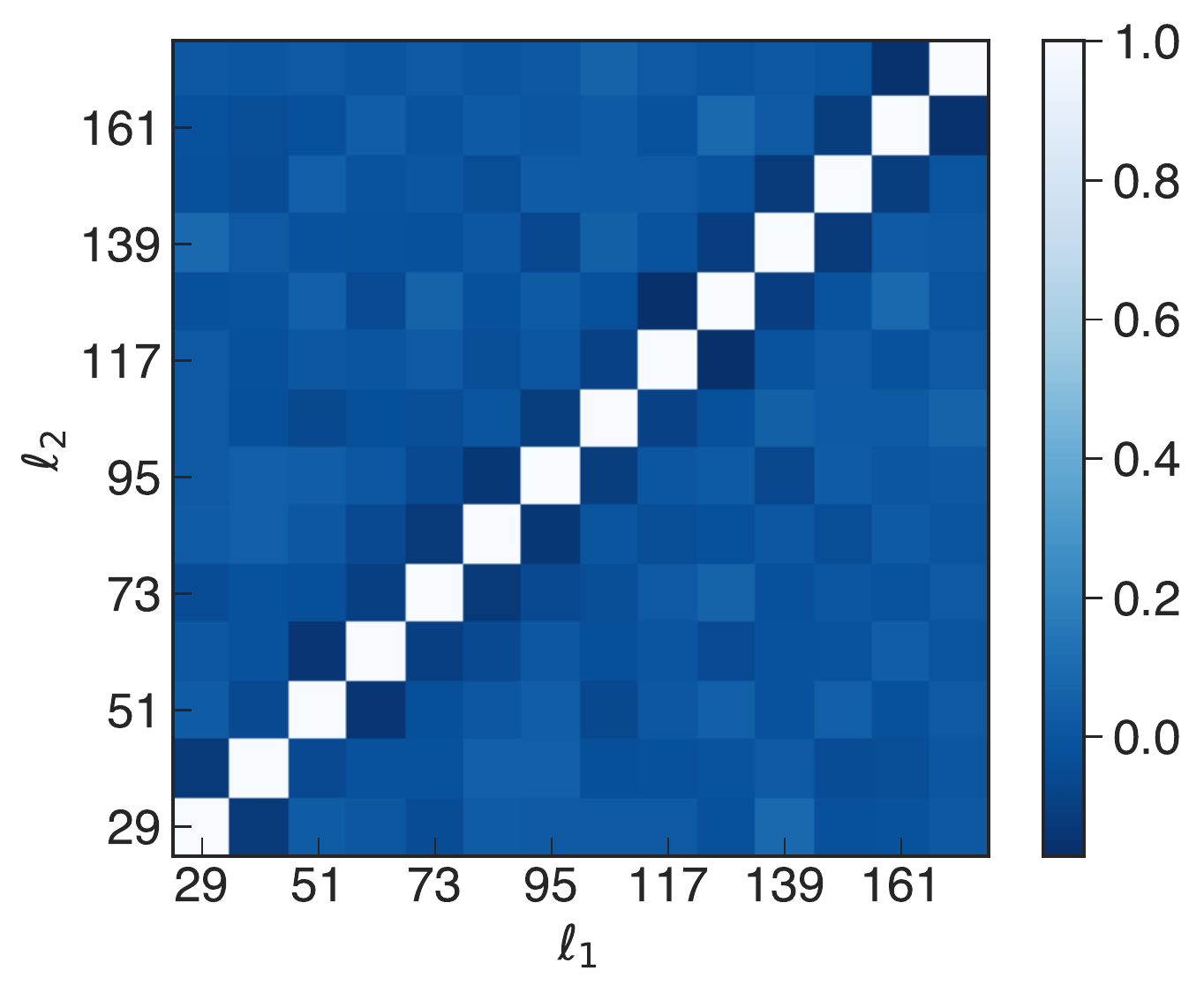}
\includegraphics[width=0.33\textwidth]{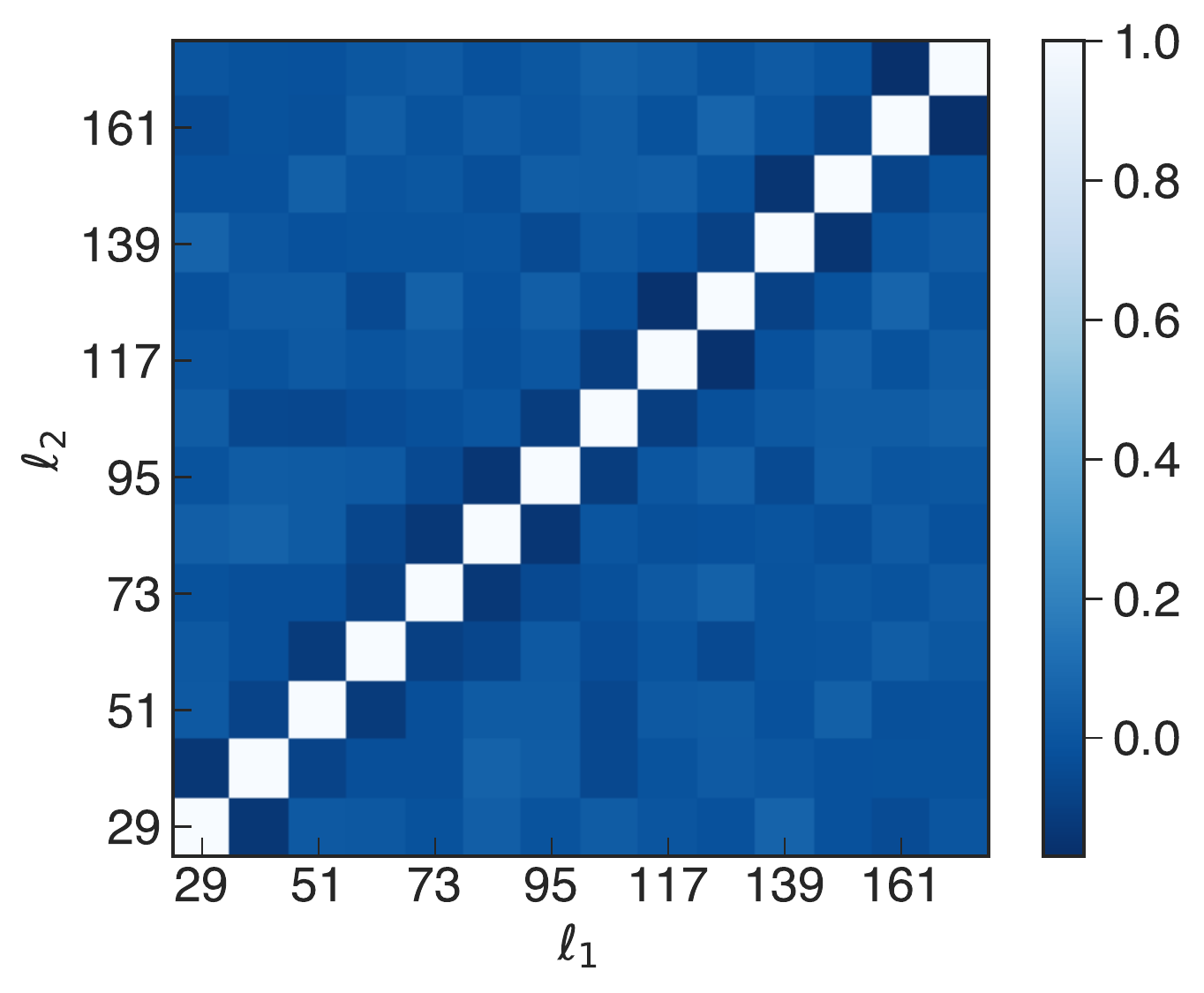}
\includegraphics[width=0.33\textwidth]{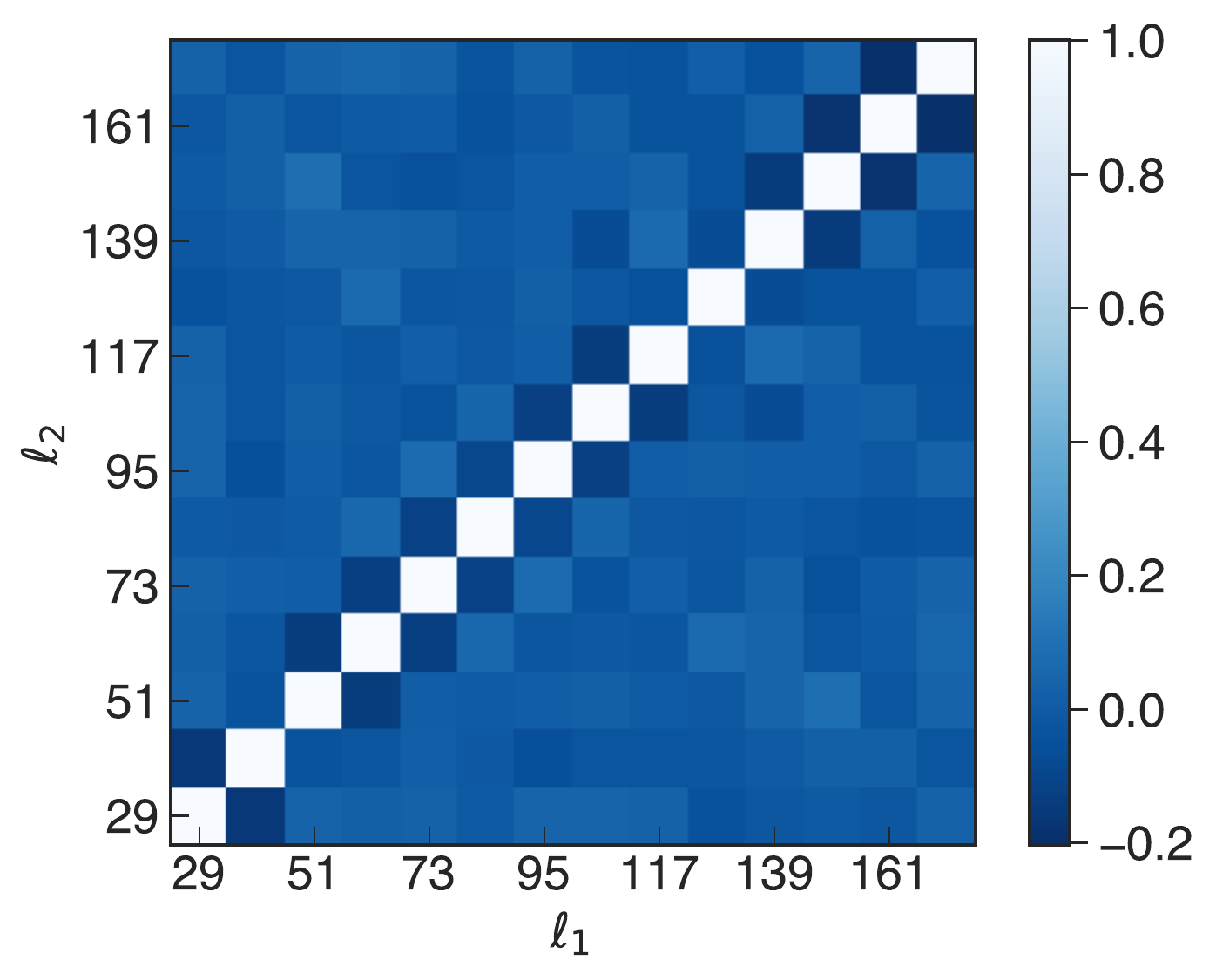}
\includegraphics[width=0.33\textwidth]{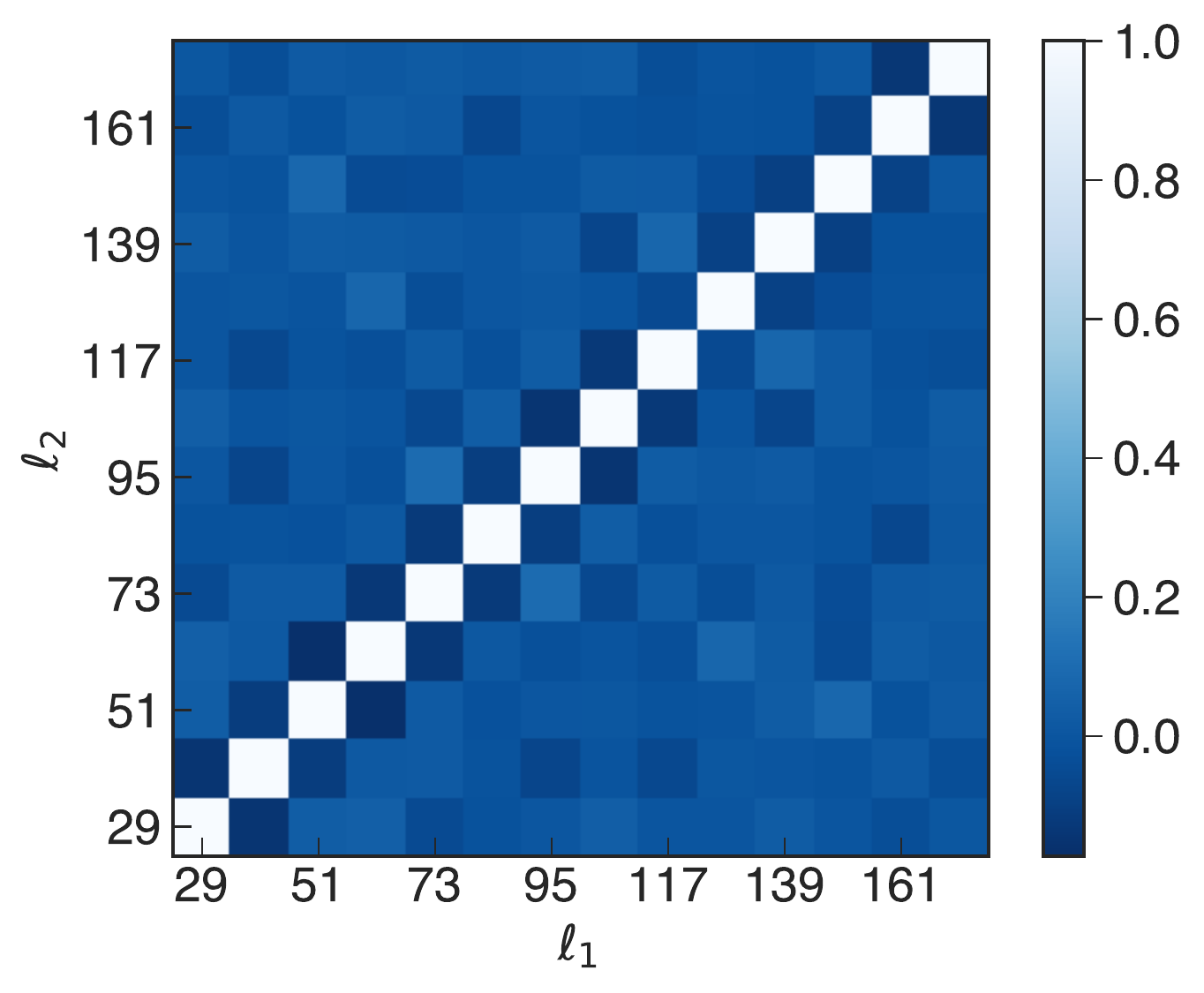}
\includegraphics[width=0.33\textwidth]{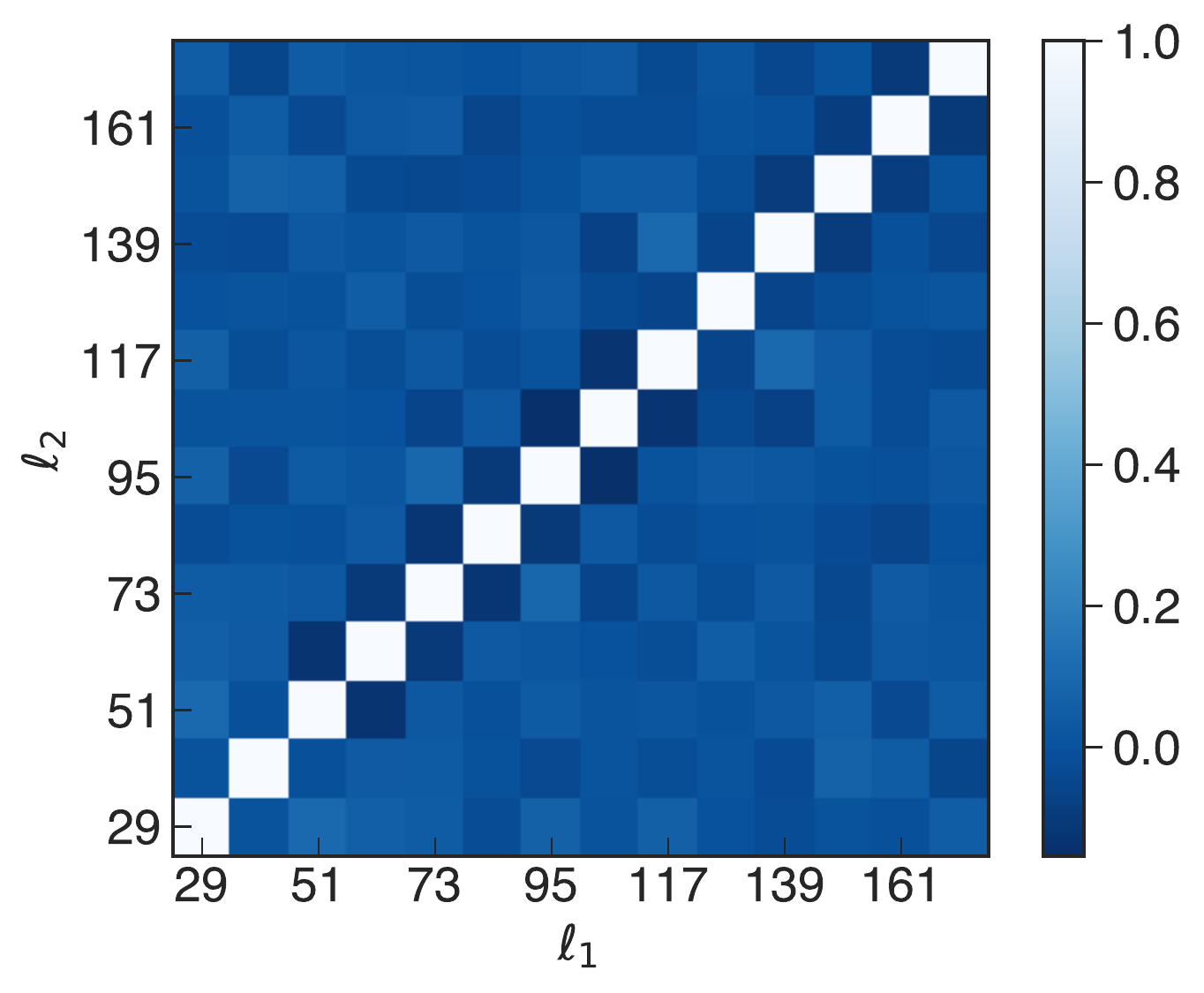}
\caption{Binned normalized covariance matrices $\mathcal{C}_{\ell, \ell'} = {\rm cov}(\hat{C}_\ell, \hat{C}_{\ell'}) /\sqrt{{\rm var}(\hat{C}_\ell){\rm var}(\hat{C}_{\ell'})}$ of the three QML methods for the ground-based experiment, with homogeneous noise. The upper diagrams show the covariance matrix for r=0.05 case and the lower diagrams show the covariance matrix for r=0 case. The matrices are obtained from estimates of 1000  simulations for classic QML estimators (left), QML-SZ estimators (center), and QML-TC estimators (right).}
\label{fig:norm_cov_12}
\end{figure*}
%************************************************************

For the analysis in this subsection, we downgrade the high resolution maps to $\texttt{NSIDE}=64$ ($\ell_{\rm max}=192$) for all three QML methods.  %As discussed in section \ref{sec:tests_and_applications}, the scalar QML has $N_d=N_{\rm pix, obs}$. This allows us to choose even higher \texttt{NSIDE} for these methods. For the analysis in this subsection, we downgrade the observed maps to $\texttt{NSIDE}=32$ ($\ell_{\rm max}=96$) for the standard QML, and $\texttt{NSIDE}=64$ ($\ell_{\rm max}=192$) for the scalar QML methods.
As in our previous examples, we also compare these QML results with pure $B$-mode PCL estimator, obtained with C2 apodization. The apodization length $\delta_c$ is set to $6^\circ$ and $10^\circ$ for $r=0.05$ and $r=0$, respectively.

The process we use for suppressing the higher multipoles in the simulated maps at high resolution is the major difference for a ground-based experiment. Instead of smoothing the map, we start by obtaining the spherical harmonic coefficients of the maps at $\texttt{NSIDE}=512$. We set all harmonic coefficients to zero above $\ell_{\rm max}$ values stated above. We use the $a_{\ell m}$'s with this cut-off to reconstruct the map at $\texttt{NSIDE}=512$, but removing the information above $\ell_{\rm max}$. This methodology is applied to the $QU$ maps at $\texttt{NSIDE}=512$ with $\ell_{\rm max}=192$, and to the scalar $\mathcal{B}$ and $B$-mode maps at $\texttt{NSIDE}=512$ with $\ell_{\rm max}=192$ too. Then we downgrade the map to the targeted \texttt{NSIDE} of 64. 

In this subsection, we consider the homogeneous noise with 3 $\mu$K-arcmin noise level to study the performance of QML methods on a small sky patch. The results for the power spectrum estimator are shown in Fig.\ref{fig:realex_10}, from which we find that all the methods discussed here give unbiased estimates of the $B$-mode band powers for both $r=0$ and $r=0.05$ cases. However, when we focus on the error bars, $r=0$ and $r=0.05$ cases have obvious differences. For the result of $r=0.05$, all the methods (QMLs and PCL) have near-optimal error bars in the entire multipole range.

However, for the $r=0$ case, the CMB signal of $B$-mode for $\ell<60$ is so weak that the error bars are too small, and we cannot tell which method performs better. However, for $\ell>60$, we find that all QML methods have smaller error bars than those in PCL method. Which means that for the case with small $r$, QML methods perform better in reconstructing the $B$-mode power spectrum.In Fig.\ref{fig:norm_cov_12}, we have shown the normalized covariance matrices for the power spectra estimates with the three QML methods for this case. For the ground-based experiment, we also find our covariance matrices to be approximately diagonal, showing that the band power leakages have been suitably removed.

\subsection{Ground-based experiment: inhomogeneous noise} % (fold)
\begin{figure}[th]
\centering
\includegraphics[width=0.48\textwidth]{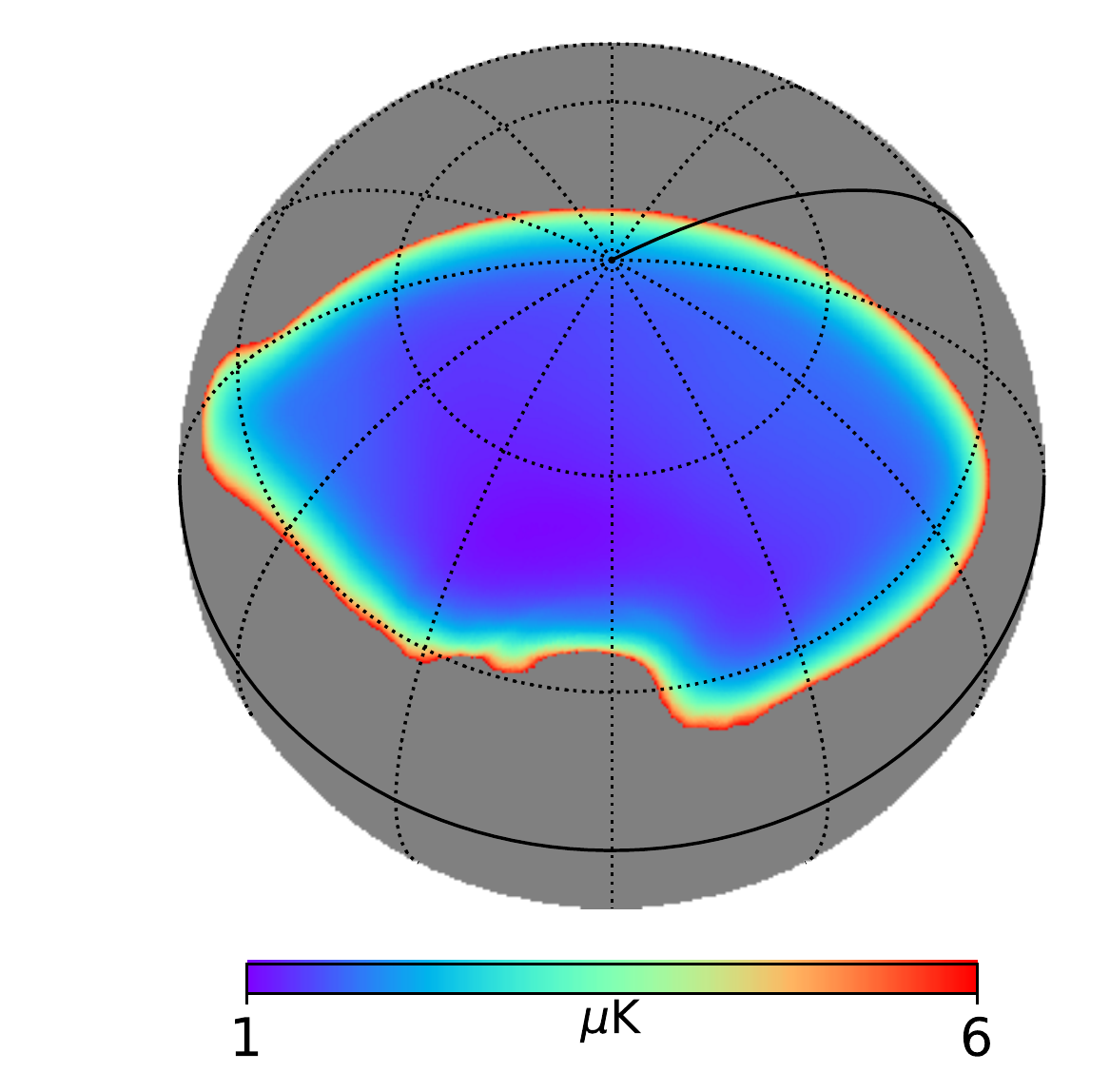}
 \caption{Map for the standard deviation of noise of per pixel (in unit of $\mu {\rm K}$), for AliCPT 95 GHz channel for one possible scan strategy}
\label{fig:noise_map} 
\end{figure}

\label{sub:ground_example_inhomo}
In this subsection, we will study the performance of QML methods with the noise profile of the AliCPT-1 experiment as a realistic example. A map of the noise standard deviation per pixel for AliCPT-1 is shown in Fig.\ref{fig:noise_map}. The major difference between the inhomogeneous noise case and homogeneous noise case is the way we deal with the noise. QML methods require a precise knowledge of the pixel noise matrix $\bs N$ to compute the bias term $b_l$ in Eq \ref{QML_yrl}. For homogeneous case, noise covariance matrix is a diagonal matrix with all diagonal elements equal, and we can calculate $\bs N$ with the noise power spectrum directly. %We can use the $\langle T_i T_j \rangle$ term of Eq. \ref{QML_matrix_M} to calculate the diagonal elements of noise matrix, just need to replace $C_\ell^{TT}$ by $N_\ell$. 
However, in the inhomogeneous case, it is difficult to characterize the noise covariance matrix by an analytical formula. Here, we estimate the noise covariance matrix from numerical simulations. We generate 1000 noise samples, and use the SZ method (or the TC method) to produce scalar noise maps of QML-SZ method (or QML-TC method). Then, we downgrade the scalar noise maps, as well as the original noise maps, to the targeted resolution. The covariance matrices of these noise-only simulations are the $\bs N$ matrices that we use in this case.
%*********************************************************************%
%**********************************************************************%
%*********************************************************************%
\begin{figure}[th]
\centering
\includegraphics[width=0.45\textwidth]{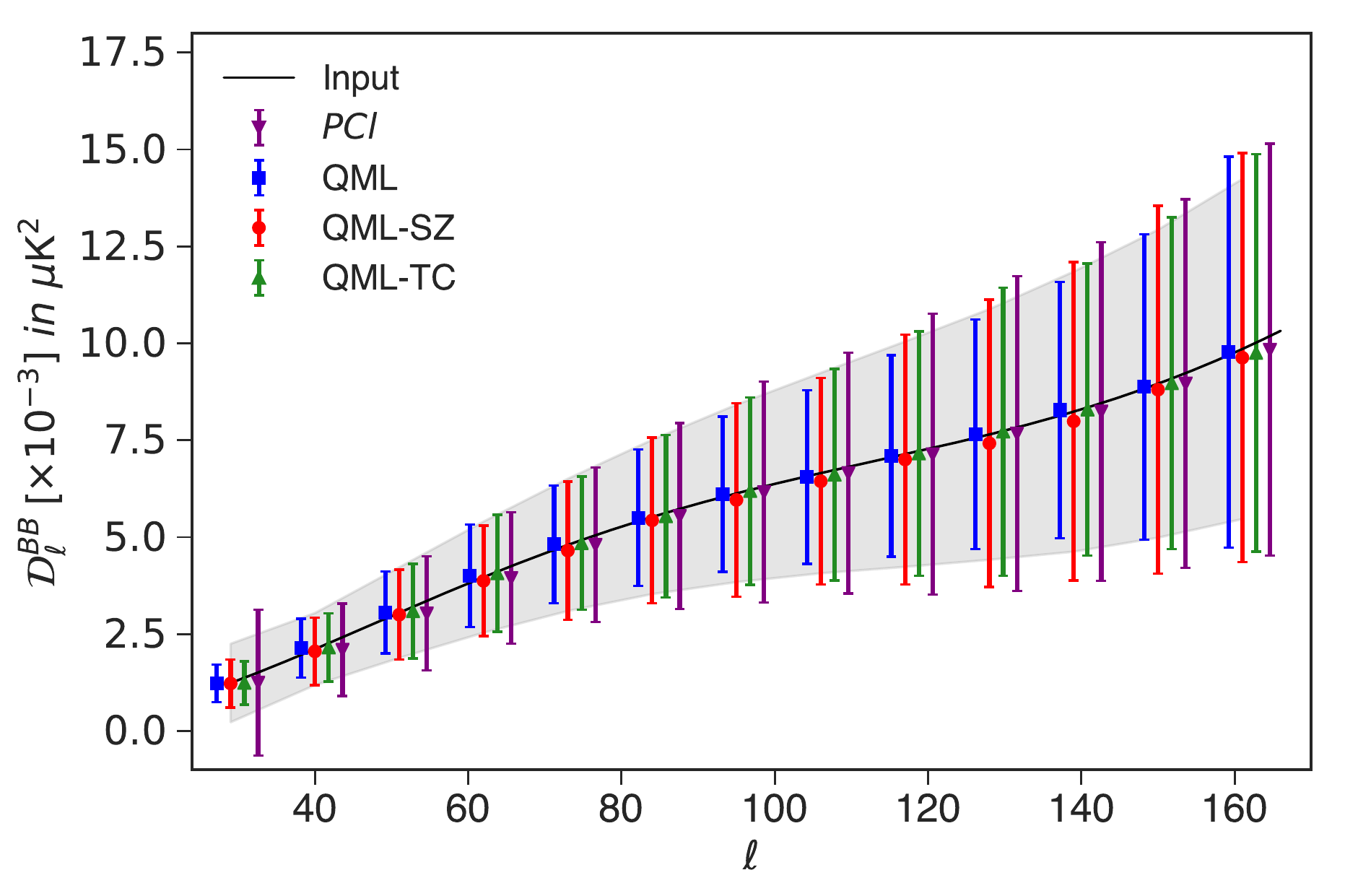}
\includegraphics[width=0.45\textwidth]{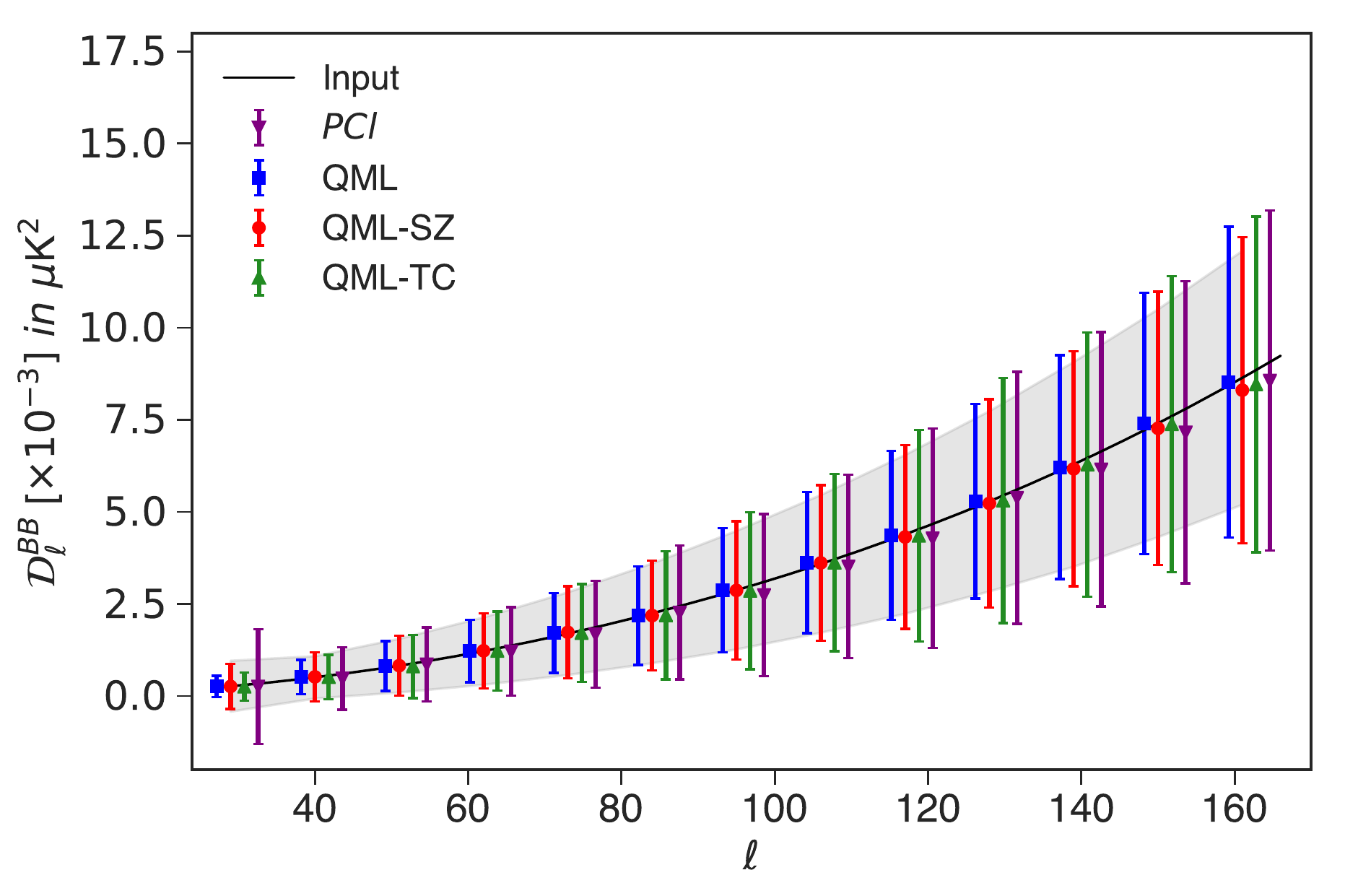}
\caption{Plot of the results of $B$-mode power spectrum estimates for realistic ground-based CMB experiment with inhomogeneous noise, the upper panel and the lower panel are $r=0.05$ and $r=0$, respectively. The observed sky with inhomogeneous noise, is simulated at \texttt{NSIDE}=512 with $\ell_{\rm max}$=1024. The input $B$-mode power spectrum is shown with the black solid  curve. The classic QML method results are computed at \texttt{NSIDE}=64 with $\ell_{\rm max}$=192 (blue square markers), QML-SZ method results are computed at \texttt{NSIDE}=64 with $\ell_{\rm max}$=192 (red, circle markers), and QML-TC method results are also computed at \texttt{NSIDE}=64 with $\ell_{\rm max}$=192 (green triangle markers). We also show PCL estimator results, obtained with NaMaster, using $\delta_c = 10^\circ$ for both $r$=0.05 and $r$=0, (purple inverted triangle markers). The gray region denotes the analytical error bounds. The data points are mean of 1000 estimates and the error bars are given by the standard deviation of the samples.} 
\label{fig:realex_12} 
\end{figure}
%**********************************************************************%
%************************************************************
\begin{figure*}[ht]
\centering
\includegraphics[width=0.33\textwidth]{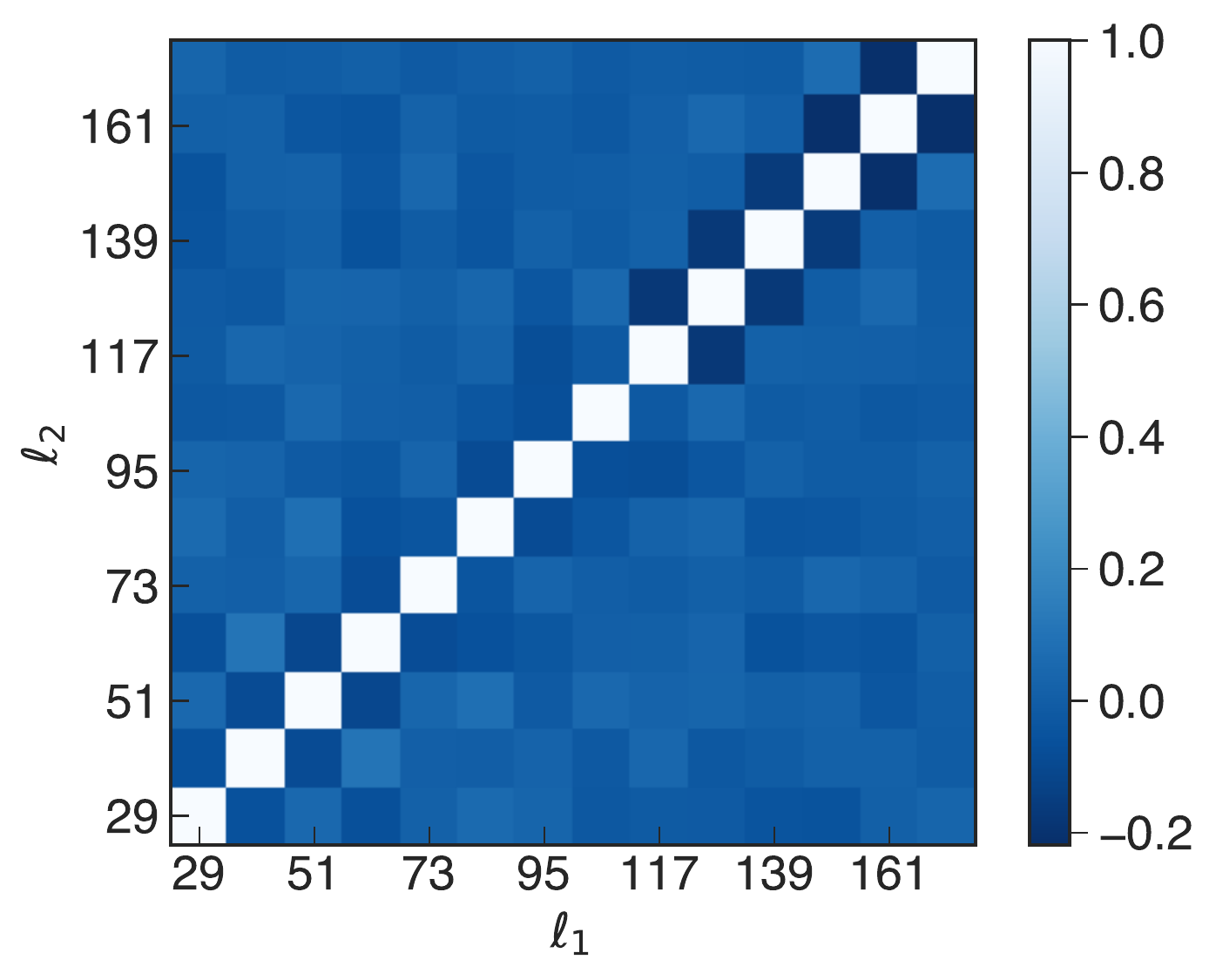}
\includegraphics[width=0.33\textwidth]{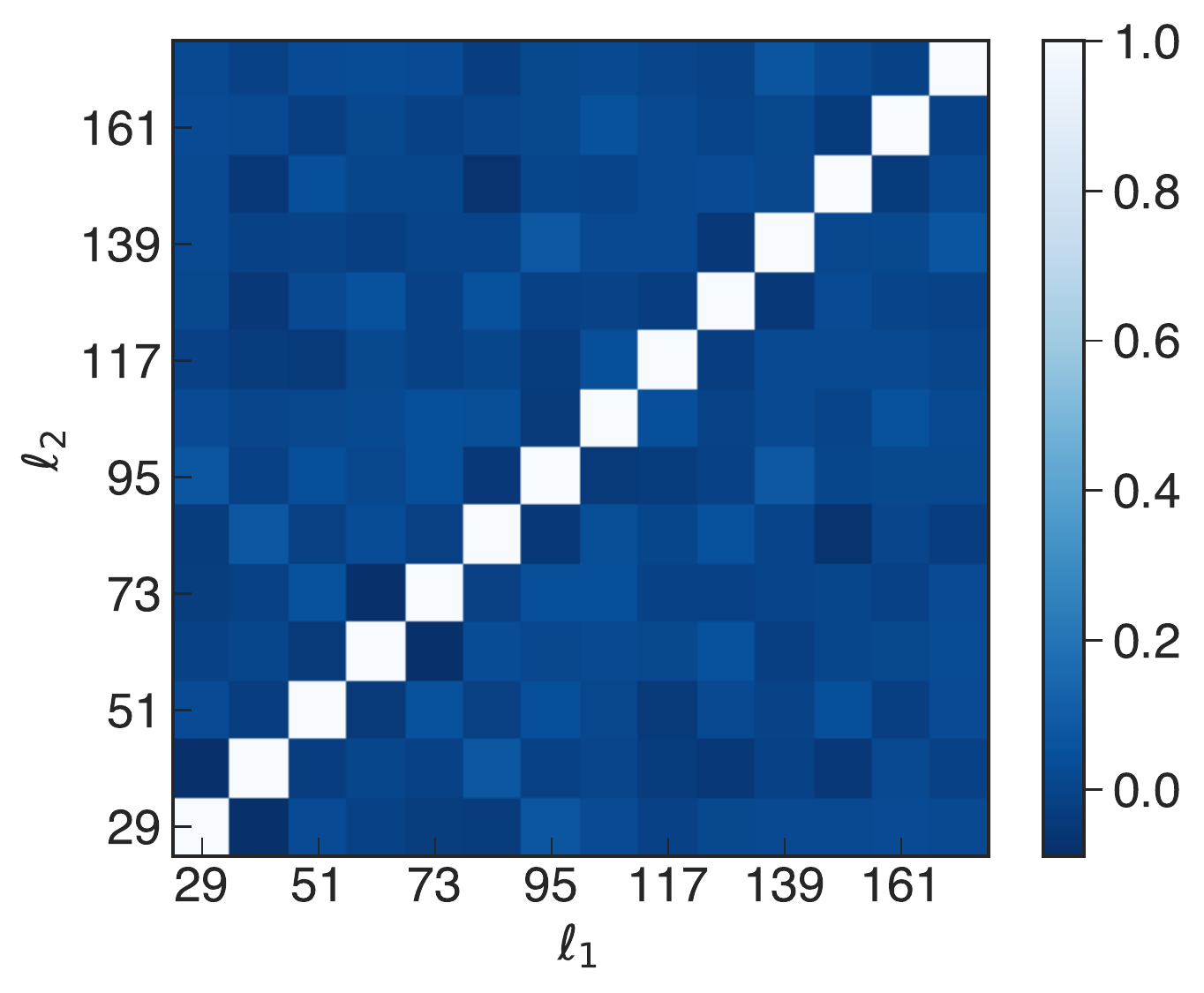}
\includegraphics[width=0.33\textwidth]{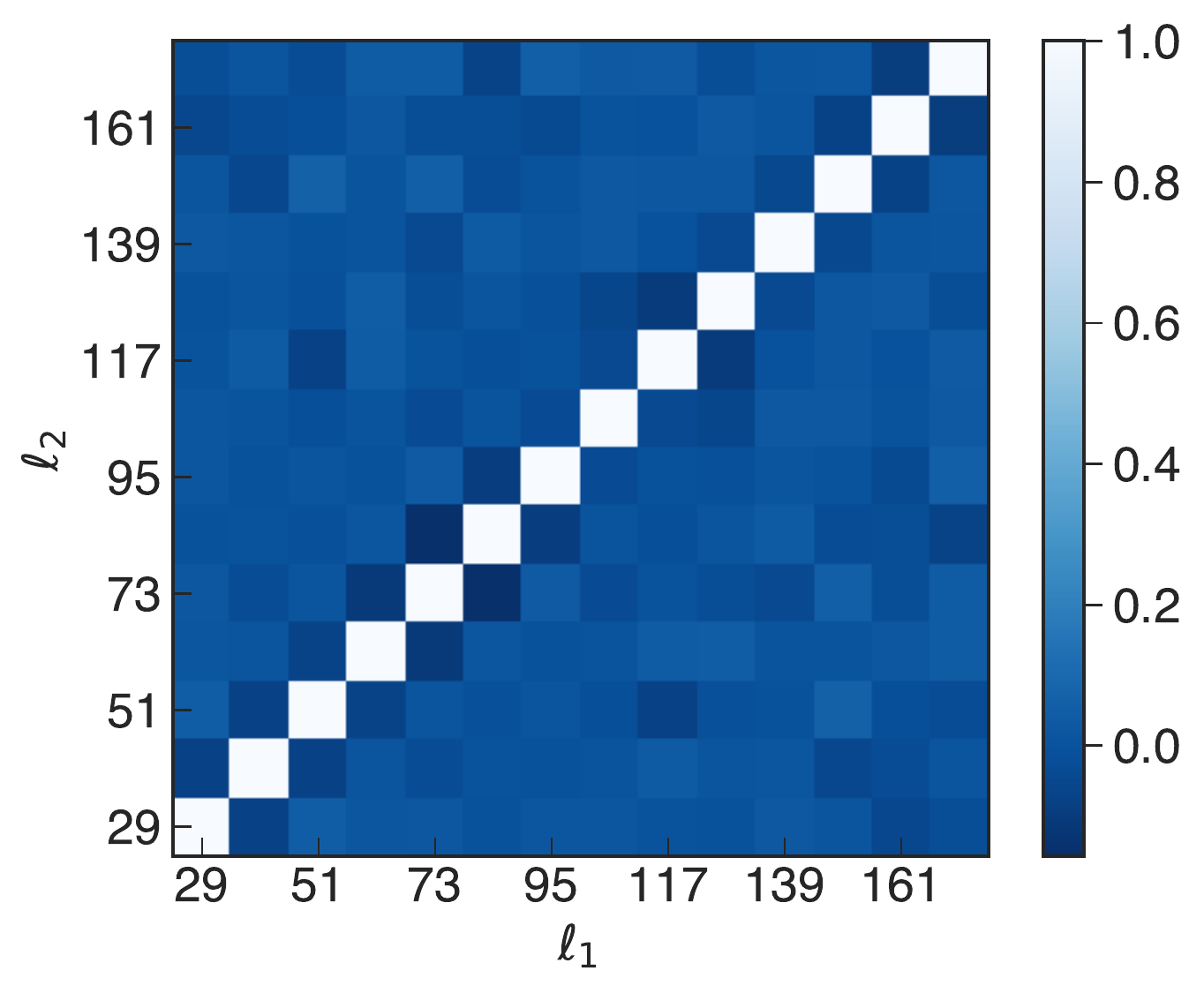}
\includegraphics[width=0.33\textwidth]{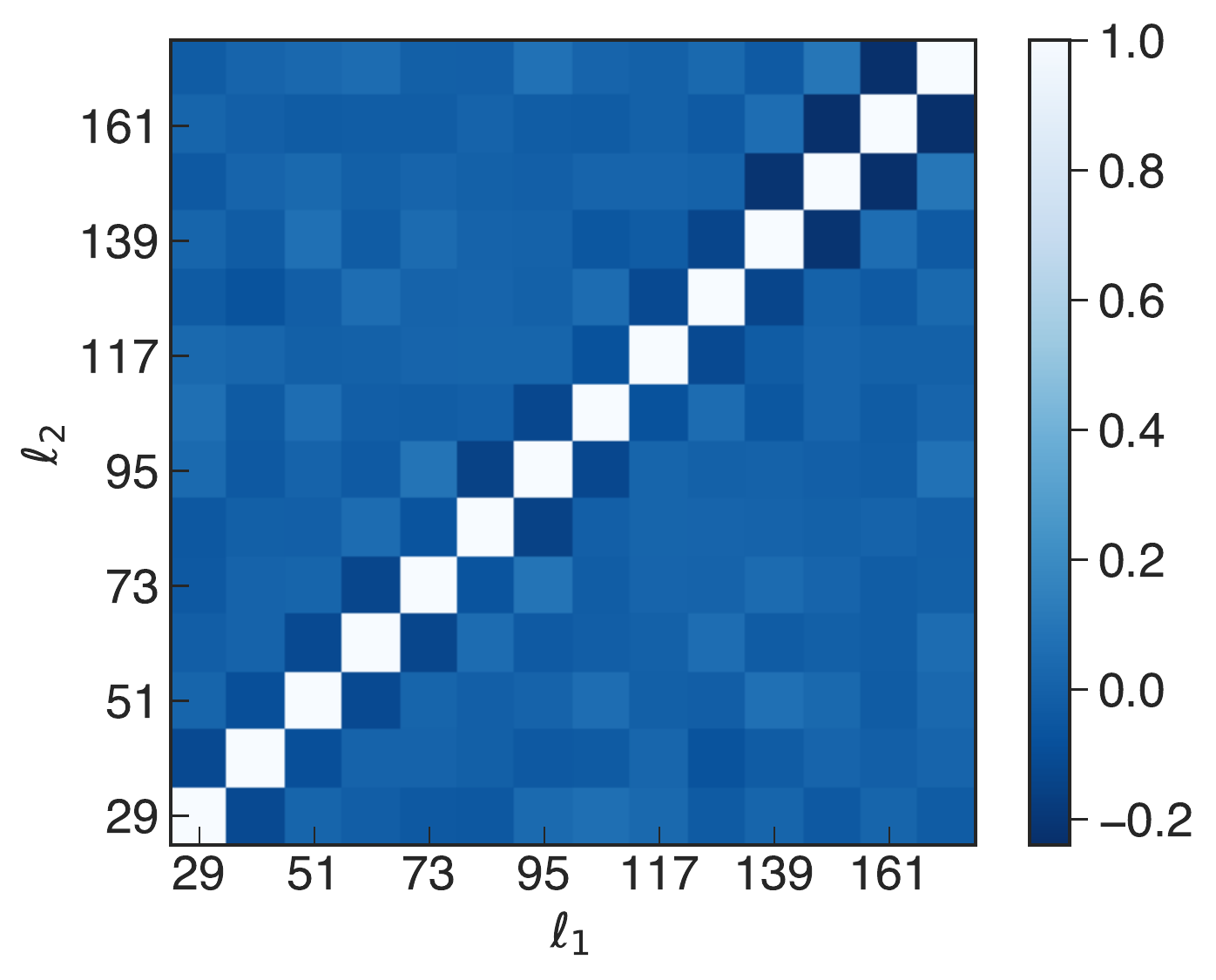}
\includegraphics[width=0.33\textwidth]{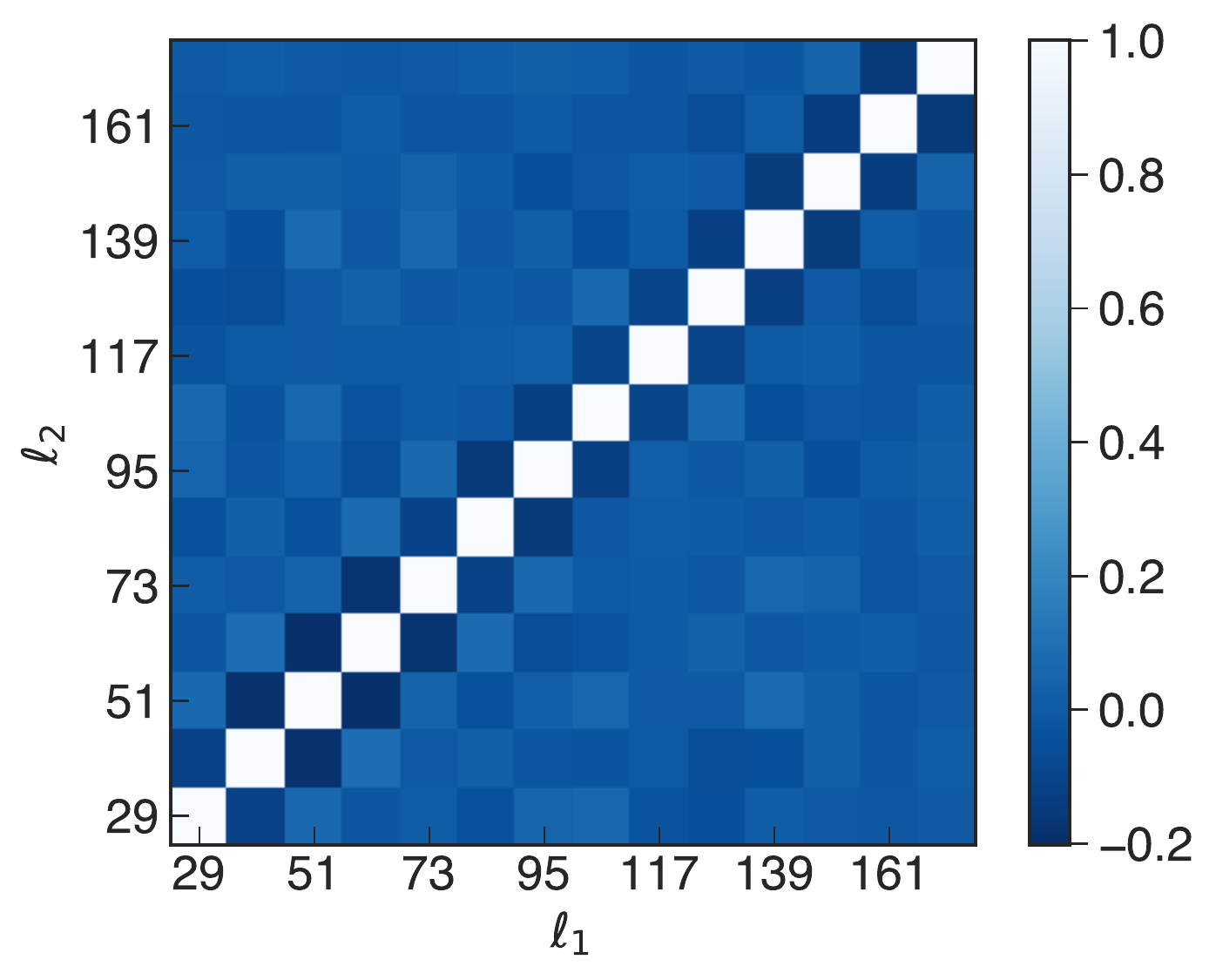}
\includegraphics[width=0.33\textwidth]{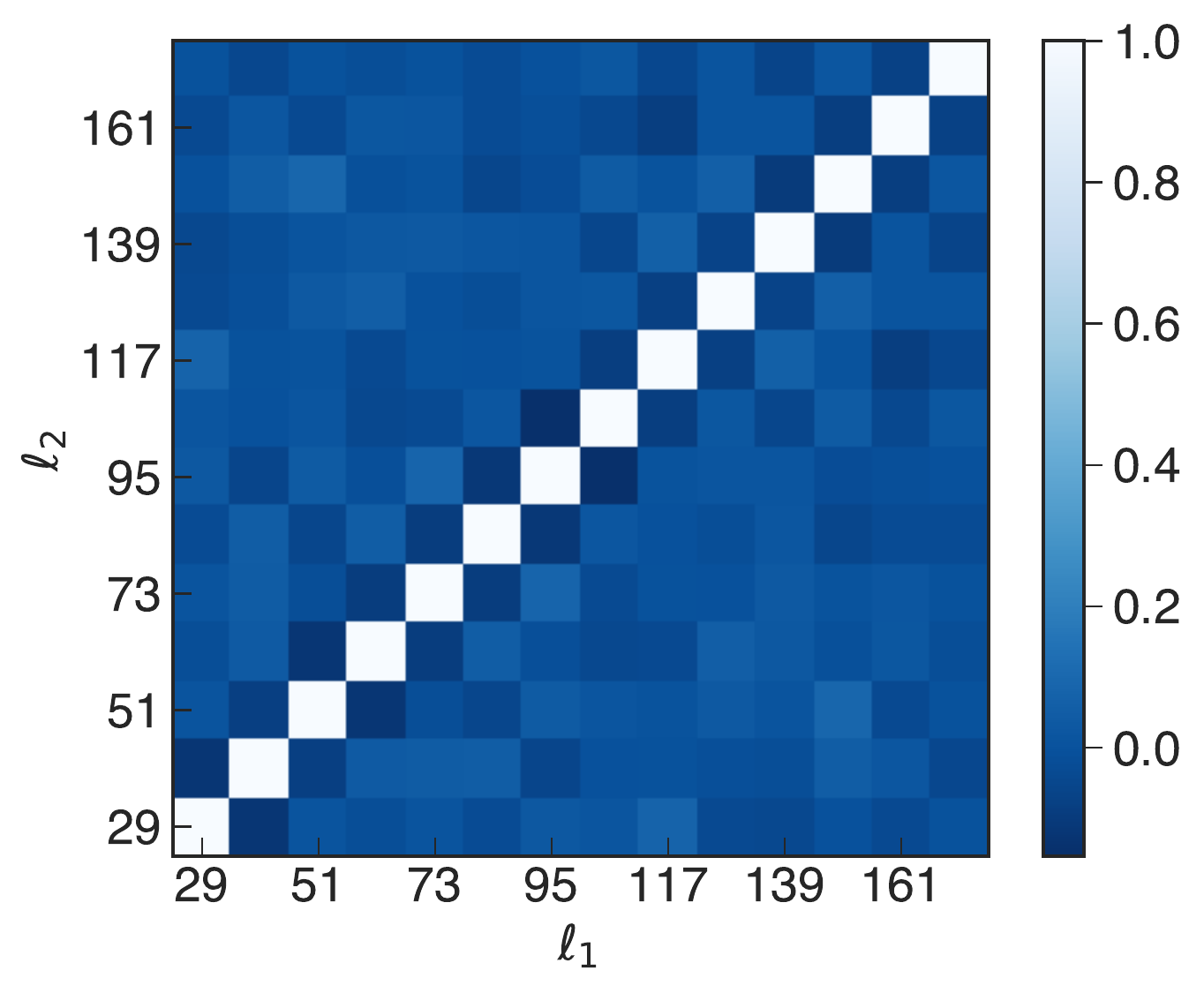}
\caption{Same with Fig.\ref{fig:norm_cov_12}, but here we consider the inhomogeneous noise case for ground-based experiment.}
\label{fig:norm_cov_13}
\end{figure*}
%************************************************************

The corresponding results are shown in Fig. \ref{fig:realex_12}. We find that all the QML-based methods, as well as the PCL method, give unbiased estimates for the $B$-mode band powers for both $r=0$ and $r=0.05$. Let us focus on the $r=0.05$ situation first. Comparing the QML results with that of PCL method, we find that all the QML methods have near-optimal error bars in the entire multipole range, while for the PCL method, uncertainties increase rapidly with increasing angular scale. The QML methods have significant advantages on large scale for $r=0.05$. For $r=0$, QML methods still keep their excellent performance on large scale, and in addition, we find that, even for small scale, the QML methods outperforms the PCL method.

In Fig \ref{fig:norm_cov_12}, we also show the normalized covariance matrices for the power spectra estimates for the three QML methods for this case. Similarly, we find that the covariance matrices are nearly diagonal, which indicates that the correlations between different bands are negligible.  

%\vfill\null
\subsection{Computational performance}
\label{sub:computation}

In section \ref{sec:tests_and_applications}, we discussed that the computational complexity of the QML estimator is $O(N^3_d)$ problem, where $N_d$ is the length of the data vector. We also discussed that $N_d$ for the QML-SZ method or QML-TC method is half of the $N_d$ for the polarization part of the `reshaped' classical QML estimator for polarization. This reduces the computational requirements for the new QML methods. In tables \ref{tab:perf_tab_72} and \ref{tab:perf_tab_12}, we summarize the computational parameters for the three QML estimators for the space-based and ground-based experiment case, respectively. We run all our computation on an Intel Xeon E2620 2.10 GHz workstation, and list \texttt{NSIDE}, $N_d$, $\ell_{\rm max}$, RAM (in gigabytes), computation time for a single computation.

%*************************************   table 1  ***************************************************
%
\begin{table}[h]
\caption{Performance comparison for different estimators in space-based experiment example. \label{tab:perf_tab_72}}
\centering
\begin{tabular}{l c c c c c}   
\tableline
 Estimator & \text{NSIDE} & $N_d$ & $\ell_{max}$ & RAM (GB) & Time ($s$)  \\ 

\tableline
QML 					& 16 & 4454 & 47 & 29.2  & $\sim$ 510 \\ 
QML-SZ 					& 16 & 2131 & 47 & 3.6   & $\sim$ 34  \\ 
%
%QML-TC                  & 16 & 2227 & 47 & 3.8   & $\sim$ 37  \\ 
%
\tableline
\end{tabular}

\end{table}
%********************************************table 1*********************************************

As shown in Table \ref{tab:perf_tab_72}, for the QML-SZ estimators, the computation time is only about $1/13$ of that for classic QML estimator. This happens because the data vector size, $N_d$ for the scalar QML method is about half of that for the classic QML method. Additionally both the QML-SZ and the QML-TC methods are based on the scalar mode QML method, its algorithm complexity lower than the polarization mode, thereby the computation is faster.
%But one must also remember that, for QML-SZ and QML-TC methods, there are large computational overheads for the computation of the scalar $\mathcal{B}$ or $B$-mode maps. However, the computation of $\check{\bs C}$ for the classic QML method requires computation of $F^{12}_\ell, F^{22}_\ell$, which increases computation overhead. The error bars for $\ell \le 5$, for the QML-SZ estimator is subobtimal, the QML-TC method have optimal performance throughout the entire multipole range, but both methods have an order of magnitude difference in terms of computational time. 
In Table \ref{tab:perf_tab_12}, we show the same set of parameters for the ground-based experiment case. In this example, we compute all three QML methods at $\texttt{NSIDE}=64$. And as shown in Table \ref{tab:perf_tab_12}, two scalar QML methods save on both computation time and memory requirements. Comparing with the results listed in the Table \ref{tab:perf_tab_72}, we find that as increase of $N_d$ the advantages of two scalar QML methods at computation times more obvious. In the scalar QML methods there is no need to calculate the rotation angle and perform coordinate transformation of \ref{QML_matrix_C}, and the signal covariance matrix is easier to compute with less memory overhead. For the ground-based case, the two scalar QML methods are near-optimal in the multipole range of interest in this case, and their computational requirements mean that they can be applied on higher resolution maps to compute the power spectrum at higher multipoles.
%In this example, we compute the classic QML at a lower resolution than the two scalar QML methods. The memory requirements for this case $\propto N^2_d \times \ell_{\rm max} / \Delta_\ell$ and computation times $\propto N_d^3$. For the ground-based case, the two scalar QML methods are both optimal, and their computational requirements mean that they can be applied to compute higher multipoles.

%*************************************   table 2  ***************************************************
%
\begin{table}[h]
\caption{Performance comparison for different estimators in ground-based experiment example. \label{tab:perf_tab_12}}
\centering
\begin{tabular}{l c c c c c}   
\tableline
 Estimator & \texttt{NSIDE} & $N_d$ & $\ell_{max}$ & RAM (GB) & Time ($s$)  \\ 

\tableline
QML 					& 64 & 13856 &  192 & 157.6  & $\sim$ 15278  \\ 
QML-SZ 					& 64 & 6297 & 192 & 18.8  & $\sim$ 341   \\ 
QML-TC                  & 64 & 6859 & 192 & 21.8 & $\sim$ 431  \\ 
\tableline
\end{tabular}

\end{table}
%********************************************table 2*********************************************

% section more_realistic_examples_satellite_case (end)
%*********************************************************************%
%*********************************************************************%

%\vfill\null 
\section{Discussions and Conclusions}
\label{sec:conclusion}

In this work, we introduce two novel QML estimators for the CMB $B$-mode power spectrum. Both of them are motivated by methods which isolate the CMB $B$-mode polarization information from the $E$-modes and `ambiguous' modes. Our method relies on the ability to construct a scalar map with only the $B$-mode information, which allows us to use a scalar QML estimator to obtain the CMB $B$-mode power spectrum. This reduces the computational requirements (both the memory requirement and computation time) in comparison to that in the traditional QML estimator for CMB polarization. From the space-based experiment example, we find that, at the same resolution, the new scalar QML methods give us more than 10 times improvement in computation time and more than 8 times reduction in memory requirement.

%Both of them are motivated by the methods to isolate the CMB $B$-mode polarization information from the $E$-modes and `ambiguous' modes. Our method relies on the ability to construct a scalar map with just the $B$-mode information, which allows us to use a scalar QML estimator to obtain the CMB $B$-mode power spectrum. This reduces the scale of the computation problem in comparison with that in the traditional QML estimator for CMB polarization, by reducing both the memory requirement and computation time. From the space-based experiment example, we find that for the computation at the same resolution, the new scalar QML methods show $> 8$ times improvement in computation times and $> 7$ times reduction in memory requirement.

The benefit of the computation efficiency is that the new scalar QML methods are a realistic solution for estimating the $B$-mode power from higher resolution maps, which allows us to extend the use of this method to higher multipoles. We have shown the application of them in the ground-based example, where the low computational requirements of the new QML methods allow us to make computations at a higher resolution, so we can get band power estimates to larger multipole range. %Thus, we demonstrate that the new methods extend the computational feasibility up to higher multipoles. 

In our tests, we find that both estimators give unbiased CMB power spectrum estimates for all cases we considered here. For space-based mission case with large sky surveys, we find that QML-SZ method is sub-optimal for $\ell \le 5$, while it performs near-optimally for rest of the multipole range. From the downgrading tests, we find that this increase in the error bars are likely linked with the effects of \texttt{ud\_grade} on the $\mathcal{B}$-mode map. These errors might be reduced further by making further optimization to the downgrading method, which we will postpone to a future work. 

For the ground-based experiment (small sky patch), the performance of both scalar QML methods is near-optimal in the multipole range of interest. With low computational requirements, we can apply the new methods at higher resolution and obtain the band powers at high multipoles with minimum variance. The performance comparison for the ground-based case shows that both QML-SZ and QML-TC methods make substantial improvements in terms of computational requirements over the traditional QML with some %without any significant 
increase in the error bars. In addition, when comparing with the  CL method, we find that for $r=0$, QML methods have smaller errors in the entire multipole range of analysis for both homogeneous and inhomogeneous noise. While for $r=0.05$, the QML methods have an obvious advantage red{at large scales}
%for $\ell<80$
in the inhomogeneous noise case.
%But we did not see this advantage of homogenouse noise case, PCL method and QML methods all can reconstruct $B$-mode power spectrum perfect in the entire multipole range of analysis.

In this work, we also perform idealized tests, considering the effect of mask, downgrading the input maps and impact of the noise level for the QML methods. From these tests, we conclude that the scalar QML methods are suitably adapted to application considering complex sky masks, and/or different noise levels. We also find that the downgrading process would require further optimization to improve the performance of the QML-SZ method at low multipoles. We should mention that, in this paper, we have not considered other complications in the CMB observations, like correlated noise, foreground residuals, timestream filtering effects, and so on, which would certainly need to be tested for applicability in real data. 
%However, we do not expect these effect to be showstoppers as traditional QML method has been applied with these complications. 
We postpone these tests and optimizations of our novel methods to a future work.

In conclusion, we can summarize this work as a combination of constructing the pure $B$-mode polarization maps and constructing the scalar QML estimator. Recent proposals of isolating the $B$-mode information without ambiguous modes allow us to reduce the computational requirements of the problem without sacrificing on the size of error-bars. Thus, we can extend the use of minimum variance power spectrum estimator for $B$-modes to higher resolution maps. These novel QML estimators for $B$-mode power spectrum will hopefully be useful for future CMB $B$-mode experiments. 

%==================correct ======================

% At the end of this paper, we should mention that only the instrumental noises of detectors as the contamination are considered in our discussion. The more realistic cases should also include the effects of various foreground emissions and systematical errors, which will be considered in the future works.

\acknowledgments{We would like to thank Jacques Delabrouille, Maria Salatino, Pengjie Zhang and Xinmin Zhang
for the helpful discussions and comments. This work is supported by NSFC No.11773028, 11633001, 11653002, 11603020, 11903030, 11621303, 11653003, 11773021, 11890691, the National Key R$\&$D Program of China (2018YFA0404504, 2018YFA0404601, 2020YFC2201600),  the Fundamental Research Funds for the Central Universities under Grant Nos: WK2030000036 and WK3440000004, the Strategic Priority Research Program of the Chinese Academy of Sciences Grant No. XDB23010200, the 111 project, the CAS Interdisciplinary Innovation Team (JCTD-2019-05), and the China Manned Space Program through its Space Application System.}

\appendix 
\section{Miscellaneous mathematical relations}
\label{apdx:imp_relations}

For an arbitrary function ${}_sf(\hat n)$ with spin $s$, we can define \citep{Newman1966}:
\begin{eqnarray}
\eth {}_sf(\hat n) &\equiv& -\sin^s \theta \left(\frac{\partial}{\partial \theta} + \frac{i}{\sin \theta}\frac{\partial}{\partial \phi}\right) \sin^{-s}\theta {}_s f(\hat n) , \\
\bar{\eth} {}_s f(\hat n) &\equiv& -\sin^{-s} \theta \left(\frac{\partial}{\partial \theta} - \frac{i}{\sin \theta}\frac{\partial}{\partial \phi}\right) \sin^s\theta {}_sf(\hat n).
\end{eqnarray}
Thus, the spin-weighted spherical harmonics are obtained by applying the spin-raising and spin-lowering operators ($\eth$ and $\bar{\eth}$) on the standard (spin-0) spherical harmonics:
\begin{eqnarray} 
	{}_s Y_{\ell m}=\frac{1}{N_{l,s}} \eth^sY_{\ell m}, \qquad  {}_{-s} Y_{\ell m}=\frac{(-1)^s}{N_{l,s}} \bar{\eth}^sY_{\ell m}.
\end{eqnarray}
They have the property: ${}_s Y_{\ell m}^* = (-1)^{s+m} {}_{-s}Y_{\ell (-m)}$.
The functions $F^{10}_\ell$, $F^{12}_\ell$ and $F^{22}_\ell$ below Eq.(\ref{QML_matrix_M}) are given by
\begin{eqnarray}
% \label{QML_component_matrix_M}
F^{10}_\ell(z) &=&  2\frac{\nonumber\frac{\ell z}{1-z^2}P_{\ell-1}(z)-(\frac{\ell}{1-z^2}+\frac{\ell(\ell-1)}{2})P_\ell(z)}{[(\ell-1)\ell(\ell+1)(\ell+2)]^{1/2}} \nonumber \\[0.1cm] 
F^{12}_\ell(z) &=&  2\frac{\frac{(\ell+2)z}{1-z^2}P^2_{\ell-1}(z)-(\frac{\ell-4}{1-z^2}+\frac{\ell(\ell-1)}{2})P^2_\ell(z)}{(\ell-1)\ell(\ell+1)(\ell+2)} \nonumber\\[0.1cm]
F^{22}_\ell(z) &=&  4\frac{(\ell+2)P^2_{\ell-1}(z)-(\ell-1)z P^2_\ell(z)}{(\ell-1)\ell(\ell+1)(\ell+2)(1-z^2)}, \nonumber
\end{eqnarray}
where the $P_\ell$ and $P^2_\ell$ denote associated Legendre polynomials $P_\ell^m$ for the cases $m = 0$ and $m = 2$.
We can use the property of spin raising and lowering operators on Eqs. \ref{pureee} and \ref{purebb}, such that it assumes the form \citep{ferte2013}:
\begin{eqnarray}
\tilde{\mathcal E}_{\ell m} &=& -\frac{1}{2} \int d\hat{n} \bigg{[} P_{+}\bigg{(} \left( \bar{\eth} \bar{\eth} W \right) Y_{\ell m}^{\ast} + 2N_{\ell,1} \left(\bar{\eth} W \right)  {}_{1}Y_{\ell m}^{\ast} 
 + N_{\ell,2} W {}_{2}Y_{\ell m}^{\ast}\bigg{)}+P_{-}\bigg{(}\left({\eth}{\eth}W \right)Y_{\ell m}^{\ast}-2N_{\ell,1}\left({\eth} W\right) {}_{-1}Y_{\ell m}^{\ast}
+N_{\ell,2} W {}_{-2}Y_{\ell m}^{\ast}\bigg{)}\bigg{]},  \\
\tilde{\mathcal B}_{\ell m} &=&-\frac{1}{2i}\int d \hat{n} \bigg{[}
P_{+}\bigg{(} \left(\bar {\eth} \bar{\eth}W\right) Y_{\ell m}^{\ast} +2N_{\ell,1} \left(\bar{\eth } W\right)  {}_{1}Y_{\ell m}^{\ast} + N_{\ell,2} W {}_{2}Y_{\ell m}^{\ast}\bigg{)} -P_{-}\bigg{(}\left({\eth}{\eth}W \right)Y_{\ell m}^{\ast}-2N_{\ell,1}\left({\eth} W\right ) {}_{-1}Y_{\ell m}^{\ast} 
+N_{\ell,2} W {}_{-2}Y_{\ell m}^{\ast}\bigg{)}\bigg{]}, \label{eq:pureB_implement}
\end{eqnarray}
where
\begin{eqnarray}
\bar{\eth}W &=&-\frac{\partial W}{\partial \theta} - \frac{i}{\sin \theta}\frac{\partial W}{\partial \phi},  \nonumber\\
 \bar{\eth} \bar{\eth}W  &=& -\cot \theta\frac{\partial W}{\partial \theta} +\frac{\partial^2 W}{\partial \theta^2} - \frac{1}{\sin \theta^2}\frac{\partial^2 W}{\partial \phi^2} 
-\frac{2i \cot \theta}{\sin \theta} \frac{\partial W}{\partial \phi} + \frac{2i}{\sin\theta}\frac{\partial^2 W}{\partial \theta \partial \phi}.  \nonumber 
\end{eqnarray}

\section{Template cleaning residuals}
\label{apdx:QML-TC_method}

\begin{figure}[th]
\centering
\includegraphics[width=0.48\textwidth]{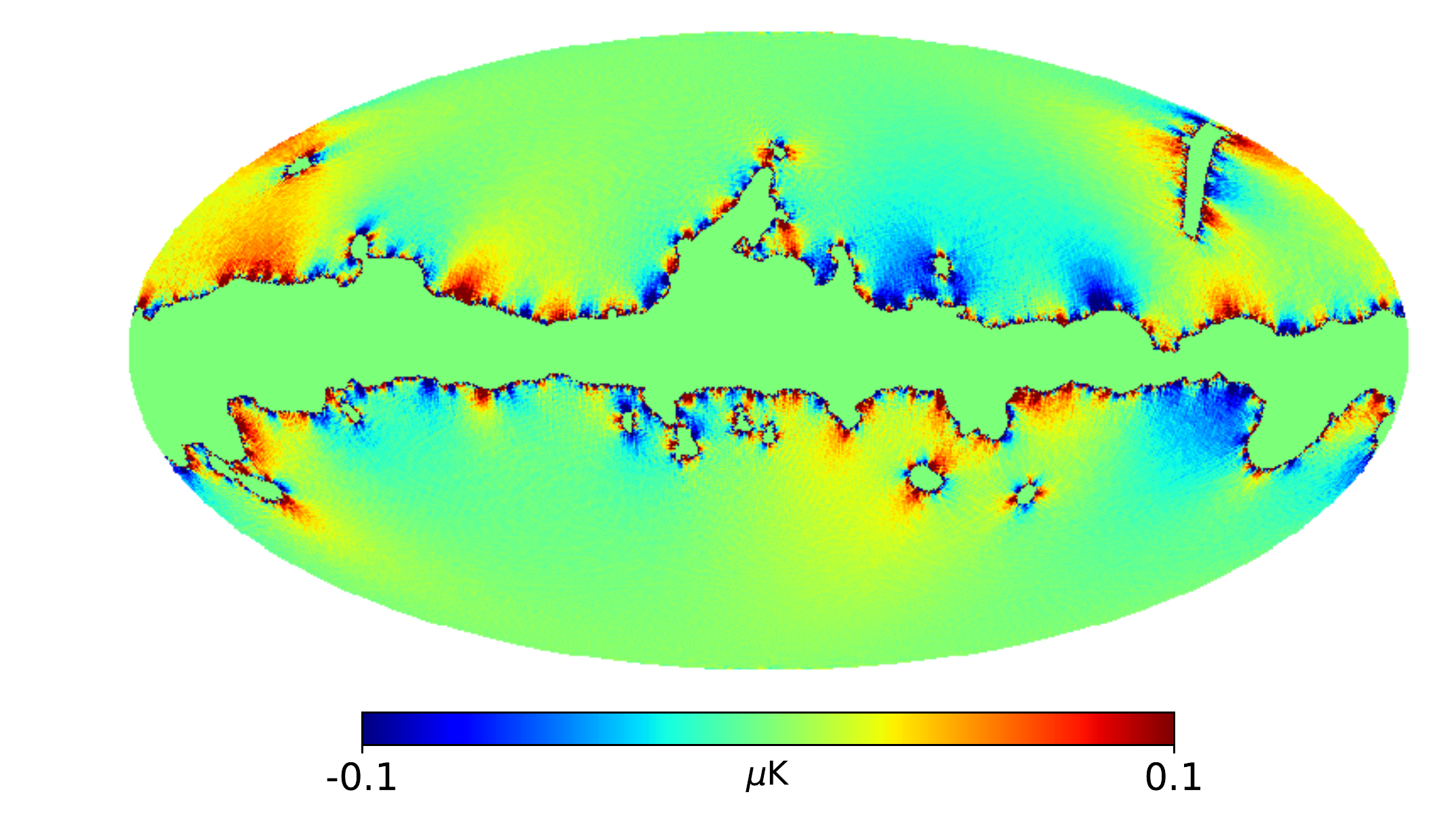} \qquad 
\includegraphics[width=0.48\textwidth]{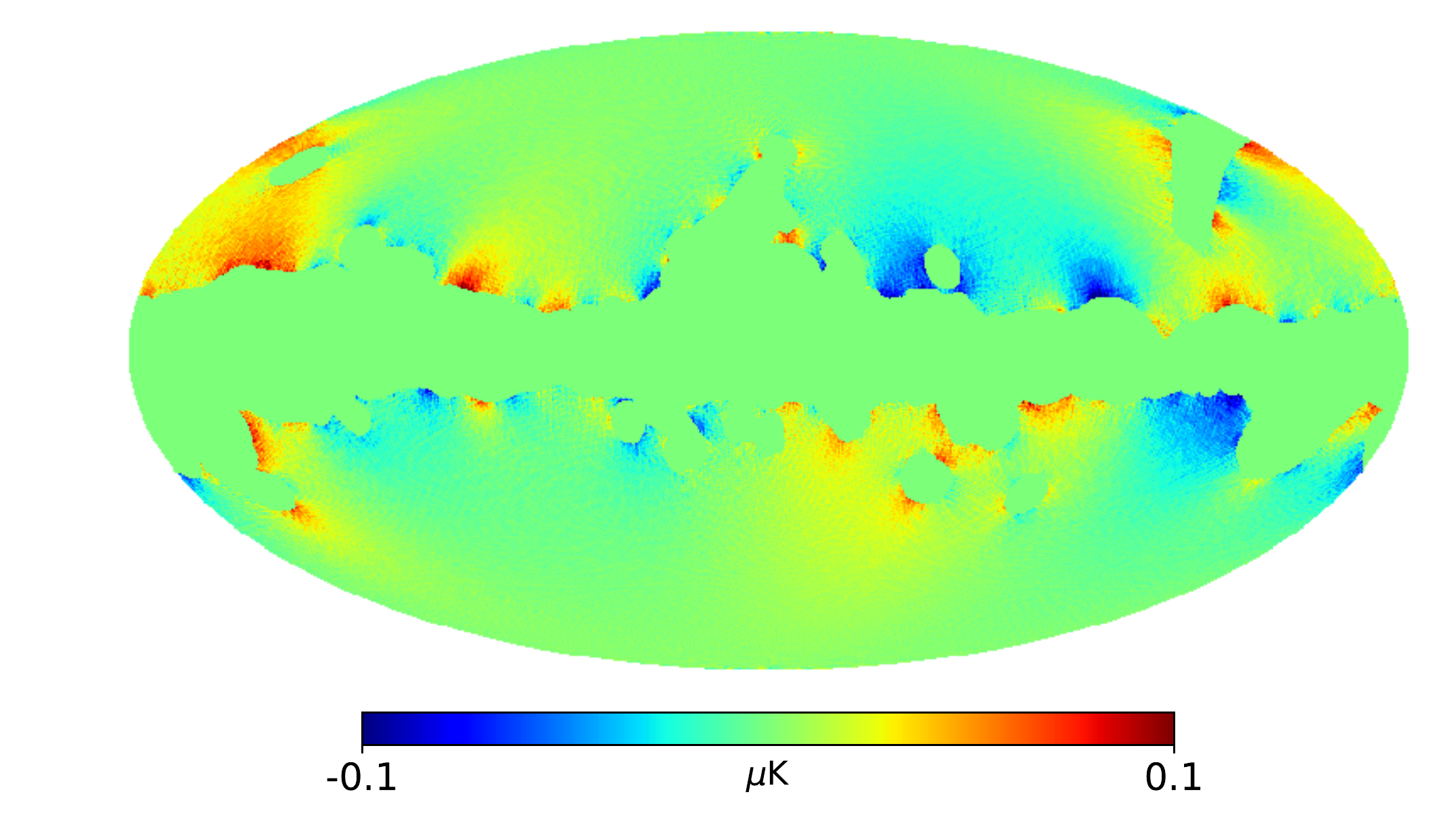}
\includegraphics[width=0.48\textwidth]{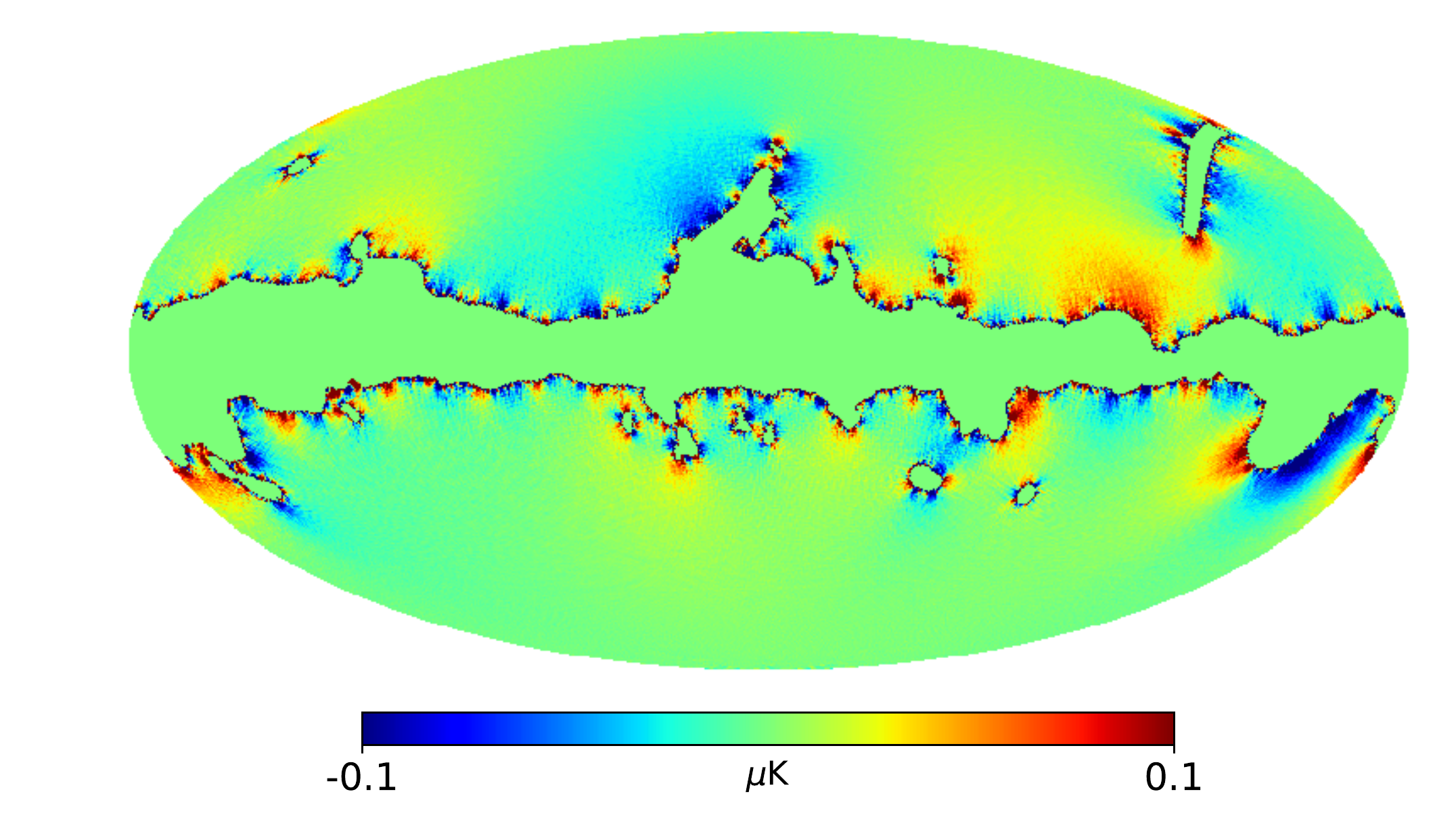} \qquad 
\includegraphics[width=0.48\textwidth]{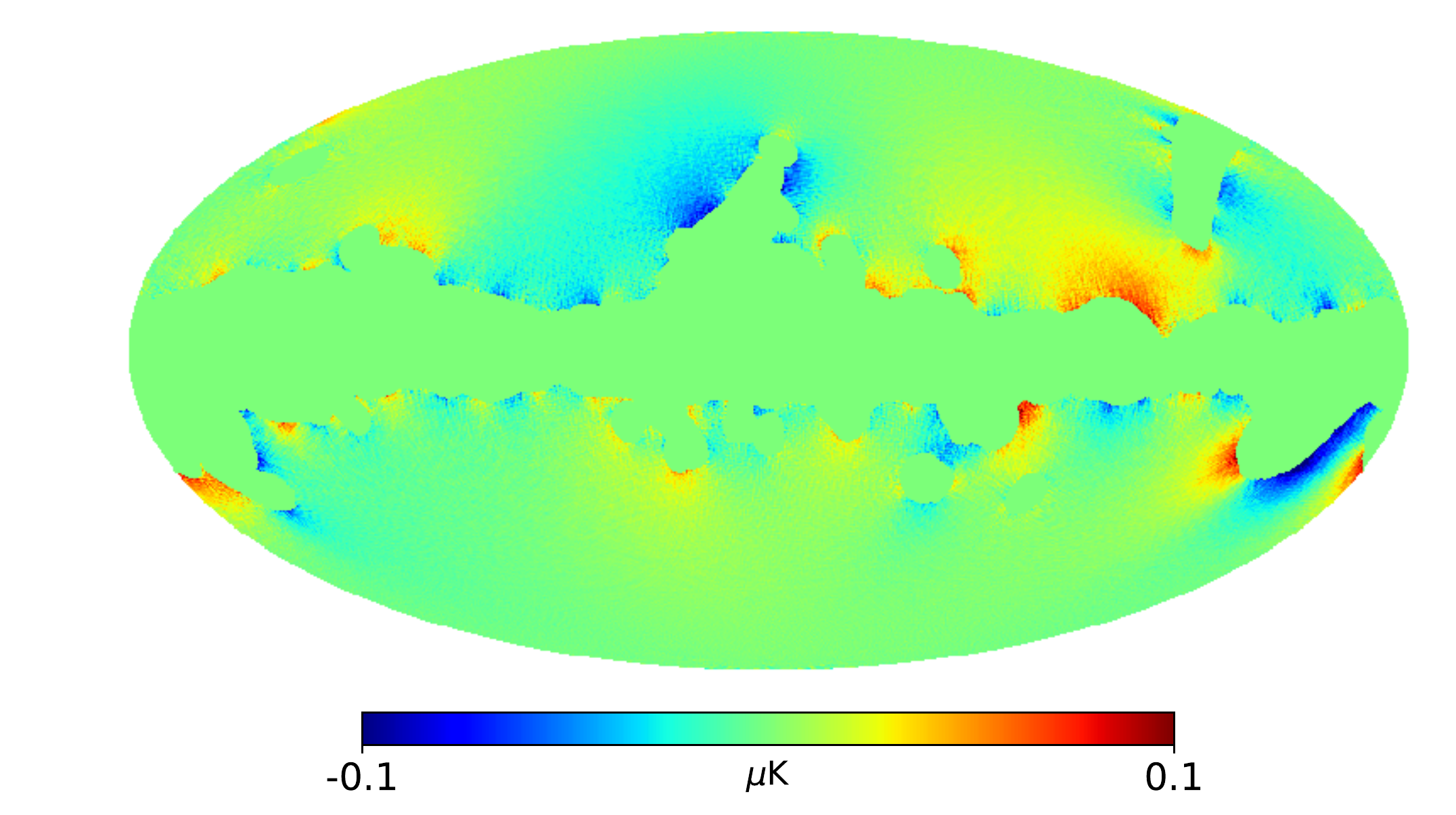}
 \caption{Residual $B$-maps after template cleaning. The left panel shows the results with Planck mask and the right panel shows the results with $3^\circ$ removed from the edge. The upper panel shows the results for an input cosmological model with $r = 0.05$, while the lower panel shows the results for $r = 0$. }
\label{fig:residual_map} 
\end{figure}

We have stated in section \ref{sec:realistic_examples} that for satellite experiment cases we do not use QML-TC method due to the presence of residuals after cleaning the maps by the template cleaning method. After cleaning with the leakage template, the cleaned $B$-mode maps have residuals that cannot be ignored for unbiased recovery of the power at the largest scales. In Fig.\ref{fig:residual_map} we show the residuals for template cleaned $B$-maps with the Planck mask and without noise. From Fig. \ref{fig:residual_map}, we can see that most residuals are largely limited to the boundary of the observed patch. These may be removed by additionally masking $3^\circ$ inside from the boundary of the mask. On removing the edge we can see most of the residual is removed leaving behind residual contaminations on the large angular scales. The presence of these residuals mean that the QML-TC method does not give unbiased power spectrum estimates. In the $r=0.05$ case the power spectra estimates are biased for $\ell<4$ and unbiased everywhere else. When $r=0$, we found it more challenging to obtain unbiased power spectrum estimates for the low multipoles. We have tried various lengths of cuts from the edge of the mask, however, our results did not improve significantly. Thus the large angular scale $B$-mode residuals from template cleaning make it difficult to obtain unbiased power spectrum estimates at low multipoles. For this reason the QML-TC method is unsuitable for use in the satellite experiment cases, where we hope to recover the power on the largest angular correctly.

%*********************************************************************%
%**********************************************************************%

We conclude that the QML-TC method needs further optimization for application in satellite experiments. It may be possible to modify the template cleaning procedure to reduce or remove the residuals or we can even look to account for the residual in the QML-TC pipeline. However, we will postpone any such modifications to a future work.

\bibliographystyle{aasjournal}
\bibliography{biblio}

\end{document}